\documentclass{aa} 

\usepackage{graphicx}	
\usepackage{amsmath}	
\usepackage{amssymb}	
\usepackage{caption}
\usepackage{lscape}
\usepackage{verbatim}
\usepackage{hyperref}
\usepackage{txfonts,xcolor}
\usepackage{rotating}
\usepackage{pdflscape}
\usepackage[colorinlistoftodos]{todonotes}

\usepackage{scalerel}
\usepackage{tikz}
\usetikzlibrary{svg.path}

\definecolor{orcidlogocol}{HTML}{A6CE39}
\tikzset{
  orcidlogo/.pic={
    \fill[orcidlogocol] svg{M256,128c0,70.7-57.3,128-128,128C57.3,256,0,198.7,0,128C0,57.3,57.3,0,128,0C198.7,0,256,57.3,256,128z};
    \fill[white] svg{M86.3,186.2H70.9V79.1h15.4v48.4V186.2z}
                 svg{M108.9,79.1h41.6c39.6,0,57,28.3,57,53.6c0,27.5-21.5,53.6-56.8,53.6h-41.8V79.1z M124.3,172.4h24.5c34.9,0,42.9-26.5,42.9-39.7c0-21.5-13.7-39.7-43.7-39.7h-23.7V172.4z}
                 svg{M88.7,56.8c0,5.5-4.5,10.1-10.1,10.1c-5.6,0-10.1-4.6-10.1-10.1c0-5.6,4.5-10.1,10.1-10.1C84.2,46.7,88.7,51.3,88.7,56.8z};
  }
}
\newcommand\orcidicon[1]{\href{https://orcid.org/#1}{\mbox{\scalerel*{
\begin{tikzpicture}[yscale=-1,transform shape]
\pic{orcidlogo};
\end{tikzpicture}
}{|}}}}

\usepackage[utf8]{inputenc}
\usepackage{subcaption}
\usepackage{multirow}
\usepackage{enumitem}

\def\arcmin{\hbox{$^\prime$}}
\def\arcsec{\hbox{$^{\prime\prime}$}}
\begin{document} 


\title{New JWST redshifts for the host galaxies of CDF-S XT1 and XT2: understanding their nature}

   \author{
        J. Quirola-V\'asquez\inst{1,2,3,\orcidicon{0000-0001-8602-4641}}
            \and
        F. E. Bauer\inst{2,3,4,\orcidicon{0000-0002-8686-8737}}
            \and
        P.~G. Jonker\inst{1}
            \and
        A. Levan\inst{1}
            \and
        W. N. Brandt\inst{5,6,7}
            \and
        M. Ravasio\inst{1}
            \and
        D. Eappachen\inst{8,\orcidicon{0000-0001-7841-0294}}
            \and
        Y. Q. Xue\inst{9,10,\orcidicon{0000-0002-1935-8104}}
            \and
        X. C. Zheng\inst{11}
}

    \institute{
    Department of Astrophysics/IMAPP, Radboud University, P.O. Box 9010, 6500 GL, Nijmegen, The Netherlands \email{j.quirolavasquez@astro.ru.nl}
        \and
    Instituto de Astrofísica, Pontificia Universidad Católica de Chile, Casilla 306, Santiago 22, Chile  
         \and
    Millennium Institute of Astrophysics (MAS), Nuncio Monse$\tilde{\rm n}$or S\'otero Sanz 100, Providencia, Santiago, Chile
        \and
    Space Science Institute, 4750 Walnut Street, Suite 205, Boulder, Colorado 80301, USA
        \and
    Department of Astronomy \& Astrophysics, 525 Davey Laboratory, The Pennsylvania State University, University Park, PA 16802, USA
        \and
    Institute for Gravitation and the Cosmos, The Pennsylvania State University, University Park, PA 16802, USA
        \and
    Department of Physics, 104 Davey Laboratory, The Pennsylvania State University, University Park, PA 16802, USA
        \and
    Indian Institute of Astrophysics, Koramangala, Bengaluru-560034, Karnataka, India
        \and
    CAS Key Laboratory for Research in Galaxies and Cosmology, Department of Astronomy, University of Science and Technology of China, Hefei 230026, China
        \and
    School of Astronomy and Space Science, University of Science and Technology of China, Hefei 230026, China
        \and
    Key Laboratory for Research in Galaxies and Cosmology, Shanghai Astronomical Observatory, Chinese Academy of Sciences, 80 Nandan Road, Shanghai 200030, PR China
}

  \abstract
   {}
   {CDF-S XT1 and XT2 are considered two ``canonical'' extragalactic fast X-ray transients (FXTs). In this work, we report new constraints on both FXTs, based on recent \emph{JWST} NIRCam and MIRI photometry, as well as NIRspec spectroscopy for CDF-S XT2 that allow us to improve our understanding of their distances, energetics, and host galaxy properties compared to the pre-{\it JWST} era.
   }   
   {We use the available \emph{HST} and \emph{JWST} archival data to determine the host properties and constrain the energetics of each FXT based on spectral energy distribution (SED) photometric fitting.}
   {The host of CDF-S XT1 is now constrained to lie at $z_{\rm phot}{=}2.76_{-0.13}^{+0.21}$, implying a host absolute magnitude $M_{R}{=}-19.14$~mag, stellar mass $M_{*}{\approx}2.8{\times}10^8$~$M_\odot$, and star formation rate SFR${\approx}0.62$~$M_\odot$~yr$^{-1}$. These properties lie at the upper end of previous estimates, leaving CDF-S XT1 with a peak X-ray luminosity of $L_{\rm X,peak}{\approx}2.8{\times}10^{47}$~erg~s$^{-1}$. We argue that the best progenitor scenario for XT1 is a low-luminosity gamma-ray burst (GRB), although we do not fully rule out a proto-magnetar association or a jetted tidal disruption event involving a white dwarf and an intermediate-massive black hole.
   In the case of CDF-S XT2, \emph{JWST} imaging reveals a new highly obscured component of the host galaxy, previously missed in \emph{HST} images, while NIRspec spectroscopy securely places the host at $z_{\rm spec}{=}3.4598{\pm}0.0022$. The new redshift implies a host with $M_{R}{=}-21.76$~mag, $M_{*}{\approx}5.5{\times}10^{10}$~$M_\odot$, SFR${\approx}160$~$M_\odot$~yr$^{-1}$, and FXT $L_{\rm X,peak}{\approx}1.4{\times}10^{47}$~erg~s$^{-1}$.
   The revised energetics, similarity to X-ray flash event light curves, small host offset, and high host SFR favor a low-luminosity collapsar progenitor for CDF-S XT2. Although a magnetar model is not ruled out, it appears improbable.}
   {While these \emph{HST} and \emph{JWST} observations shed light on the host galaxies of XT1 and XT2, and by extension, on the nature of FXTs, a unique explanation for both sources remains elusive. Rapid discovery, for instance, with the \emph{Einstein Probe} satellite, and contemporaneous multiwavelength detections of FXTs remain essential for clarifying the nature of FXTs.} 

\date{Received August 7, 2024; accepted January 31, 2025}

\maketitle



\section{Introduction} \label{sec:intro}


Extragalactic fast X-ray transients (FXTs) are singular bursts of X-ray photons (in the ${\sim}$0.3--10~keV band) with durations of seconds to hours \citep{Heise2010}. Our knowledge of FXTs has increased substantially during the last two decades via the identification and characterization of sources detected by the \emph{Swift}-XRT \citep[e.g.,][]{Soderberg2008}, \emph{Chandra X-ray Observatory} \citep[e.g.,][]{Jonker2013,Glennie2015,Bauer2017,Xue2019,Lin2022,Eappachen2022,Eappachen2023,Quirola2022,Quirola2023}, the \emph{X-ray Multi-mirror Mission - Newton telescope} \citep[\emph{XMM-Newton}, e.g.,][]{Novara2020,Alp2020,DeLuca2021,Eappachen2024}, or the recently launched \emph{Einstein Probe Telescope} \citep[e.g.,][]{Liu2025,Levan2024,Gillanders2024,van_Dalen2024}. 
Despite these efforts, the relative numbers of confirmed FXTs remain low (${\approx}$34 FXTs), and hence, their nature and physical processes are still uncertain.  Their properties strongly overlap with the theoretical parameter space occupied by core-collapse supernova (CC-SN) shock breakout events \citep[SBOs; e.g.,][]{Soderberg2008,Alp2020,Sun2022,Scully2023}, white dwarf-intermediate-mass black hole jetted tidal disruption events \citep[WD-IMBH TDEs; e.g.,][]{Jonker2013,MacLeod2014,Glennie2015,Peng2019,Maguire2020,Saxton2021}, gamma-ray bursts \citep[GRBs; e.g.,][]{Zhang2013,Bauer2017,Xue2019,Sarin2021,Ai2021,Lin2022,Levan2024,Wichern2024}, and binary neutron star mergers \citep[BNS; e.g.,][]{Bauer2017,Xue2019,Quirola2024}.

Two ``canonical'' FXTs have been detected in the \emph{Chandra Deep Field-South} \citep[CDF-S;][]{Luo2017,Xue2017}, denoted as CDF-S XT1 \citep{Luo2014,Bauer2017} and CDF-S XT2 \citep{Zheng2017,Xue2019}, enabling detailed studies due to the large amount of multiwavelength ancillary data available there. 
%
Several attempts have been made to identify a unique progenitor \citep[e.g.,][]{Bauer2017,Xue2019,Peng2019,Sun2019,Lu2019,Xiao2019,Sarin2021,Quirola2024}, but it has not been possible for either object.
Moreover, despite constraints from a wide variety of powerful telescopes spanning radio to X-ray wavelengths \citep[e.g., Chandra, HST, MUSE, VLT, ALMA, Spitzer;][]{Xue2016,Guo2013,Skelton2014,Aravena2016,Inami2017,Herenz2017}, the available data have not permitted detailed characterization of their host galaxies. For instance, in the case of CDF-S XT1, despite the \emph{Hubble Space Telescope} (HST) observations of the host, its photometric redshift shows large uncertainties \citep[$z_{\rm photo}\approx$0.4-3.2;][]{Bauer2017}, which have left large uncertainties on its energetics and hence its progenitor. 
The revolutionary launch of the \emph{James Webb Space Telescope} \citep[JWST;][]{Gardner2006,Gardner2023} has brought new life to various deep survey fields in the $0.6-25~\mu$m range, with aims to study the formation of the first galaxies (even in the first 300~Myr) and (metal-free) stellar populations, and the formation and evolution of massive black holes (MBHs). 
In the CDF-S, for instance, the JWST Advanced Deep Extragalactic Survey (JADES) \citep{Bunker2024,Rieke2023,Eisenstein2023a,Eisenstein2023b} provides Near-Infrared Camera \citep[NIRCam;][]{Rieke2023b} and Mid-Infrared Instrument \citep[MIRI;][]{Wright2023} imaging as well as Near-Infrared Spectrograph \citep[NIRSpec;][]{Boker2023} spectroscopy. These new data can help to shed light on the nature of FXTs like CDF-S XT1 and XT2 and, by extension, allow us to understand better the overall FXT population, which is especially relevant for the ongoing detections being made by the \emph{Einstein Probe Telescope} \citep{Yuan2022}.


In this paper, we combine new JWST data with archival data from other telescopes to study the host galaxies of FXTs CDF-S XT1 and XT2. Our main goals are to (re)characterize their host galaxies, refine their distances and energetics, and ultimately revise our interpretations of both FXTs. The paper is organized as follows. $\S$\ref{sec:data} and $\S$\ref{sec:SED} introduce the JWST data that we are using in this work and the spectral energy distribution (SED) models and packages taken into account for deriving host properties, respectively. In $\S$\ref{sec:CXO}, we explain the methods and results of the \emph{Chandra} X-ray analysis under the new distances. Finally, in $\S$\ref{sec:results} and $\S$\ref{sec:conclusions}, we discuss the results and offer some conclusions in light of the nature of these transients, respectively.
Throughout the paper, a concordance cosmology with parameters $H_0{=}$70~km~s$^{-1}$~Mpc$^{-1}$, $\Omega_M{=}$0.30, and $\Omega_\Lambda{=}$0.70 is adopted. Magnitudes are quoted in the AB system. Unless otherwise stated, all errors are quoted at a 1$\sigma$ confidence level.

\begin{figure*}[h!]
    \centering
    \includegraphics[scale=0.35]{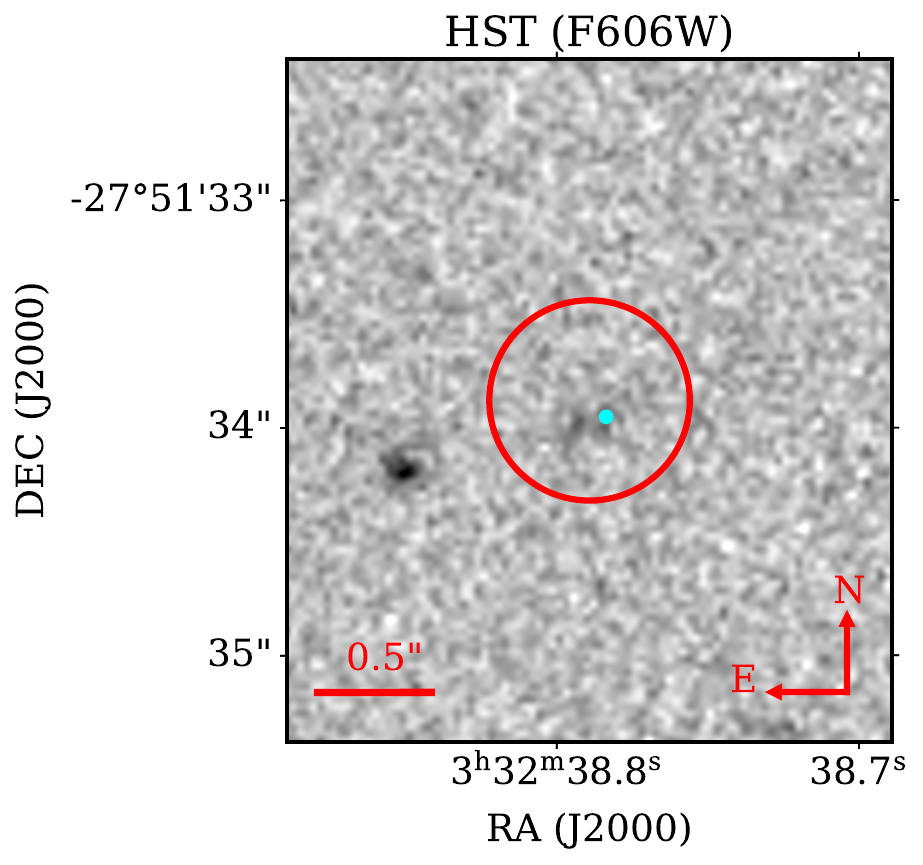}
    \includegraphics[scale=0.35]{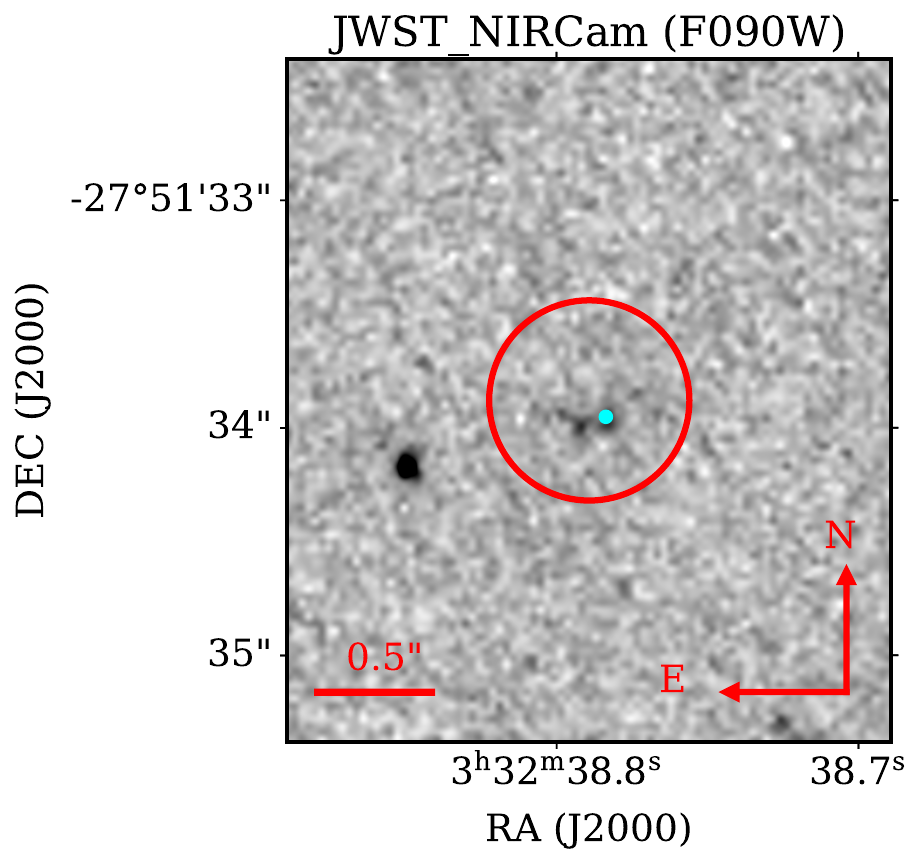}
    \includegraphics[scale=0.35]{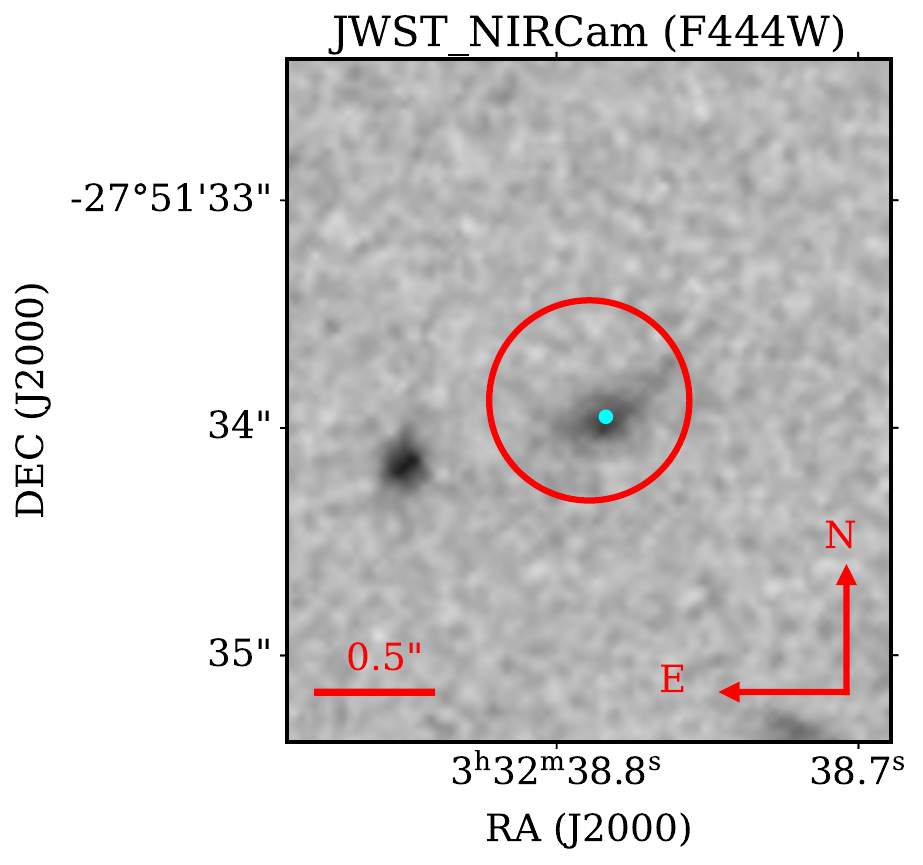}
    \caption{HST and JWST imaging of CDF-S XT1 at three different filters (one per panel) for illustration. The red circle depicts the 2$\sigma$ X-ray position uncertainty of CDF-S XT1, while the cyan dot shows the position of the host according to JADES. Figure~\ref{fig:imaging_XRT_141001_AP} shows the complete version of this figure.
    }
    \label{fig:imaging_XRT_141001}
\end{figure*}

\begin{figure*}[h!]
    \centering
    \includegraphics[scale=0.36]{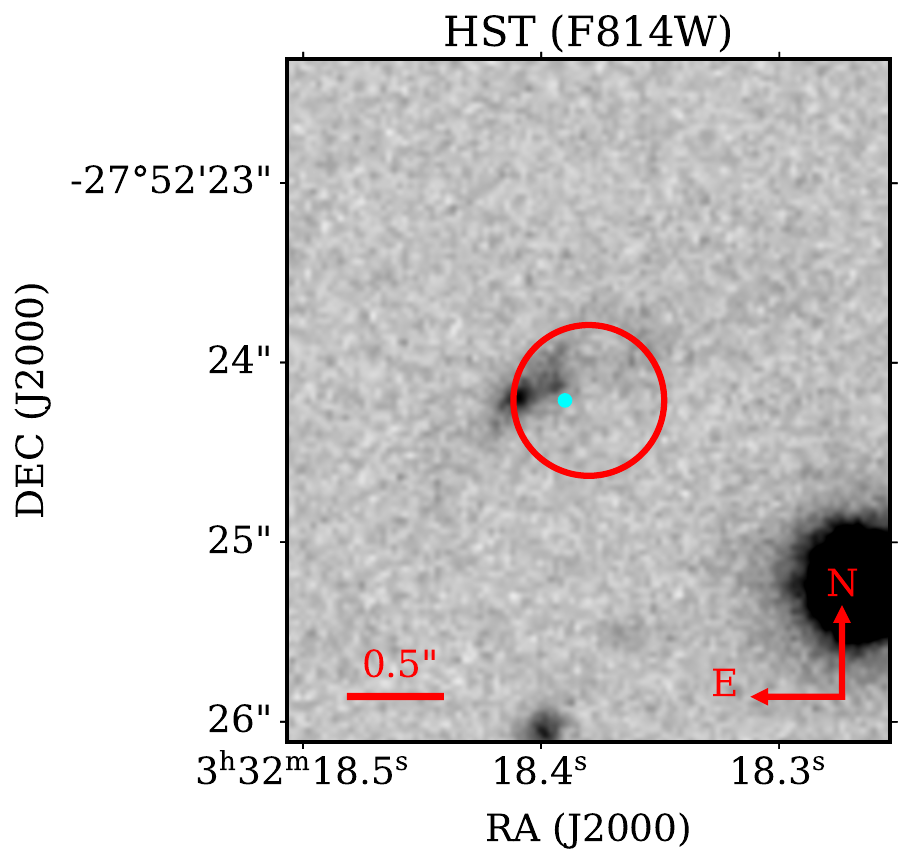}
    \includegraphics[scale=0.36]{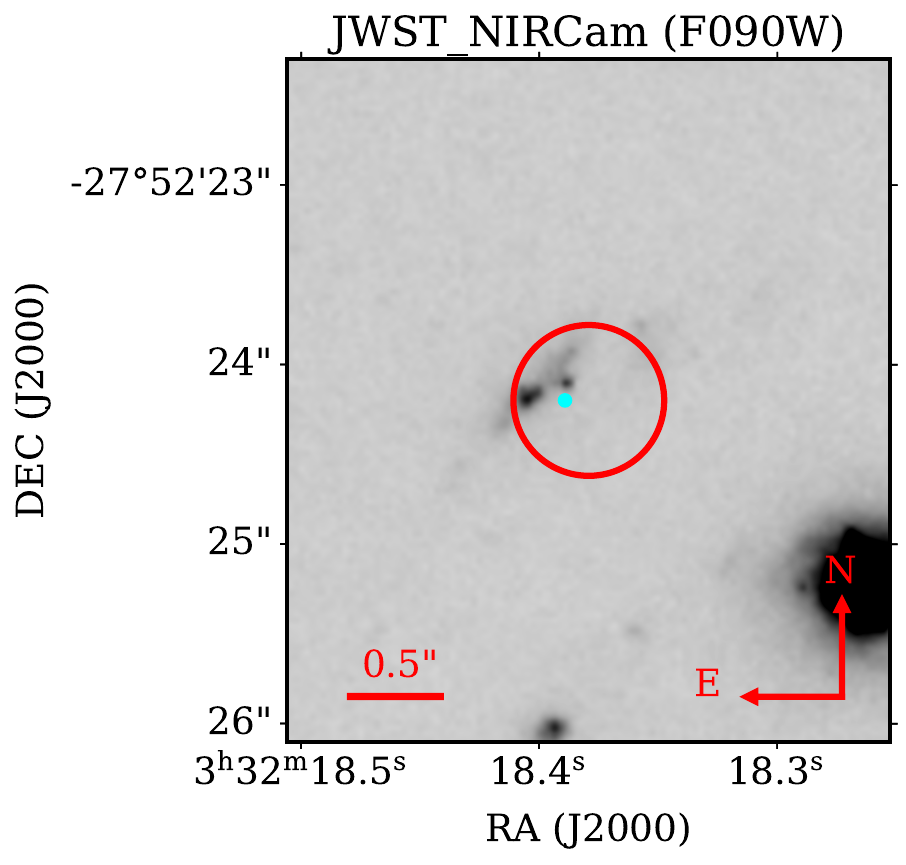}
    \includegraphics[scale=0.36]{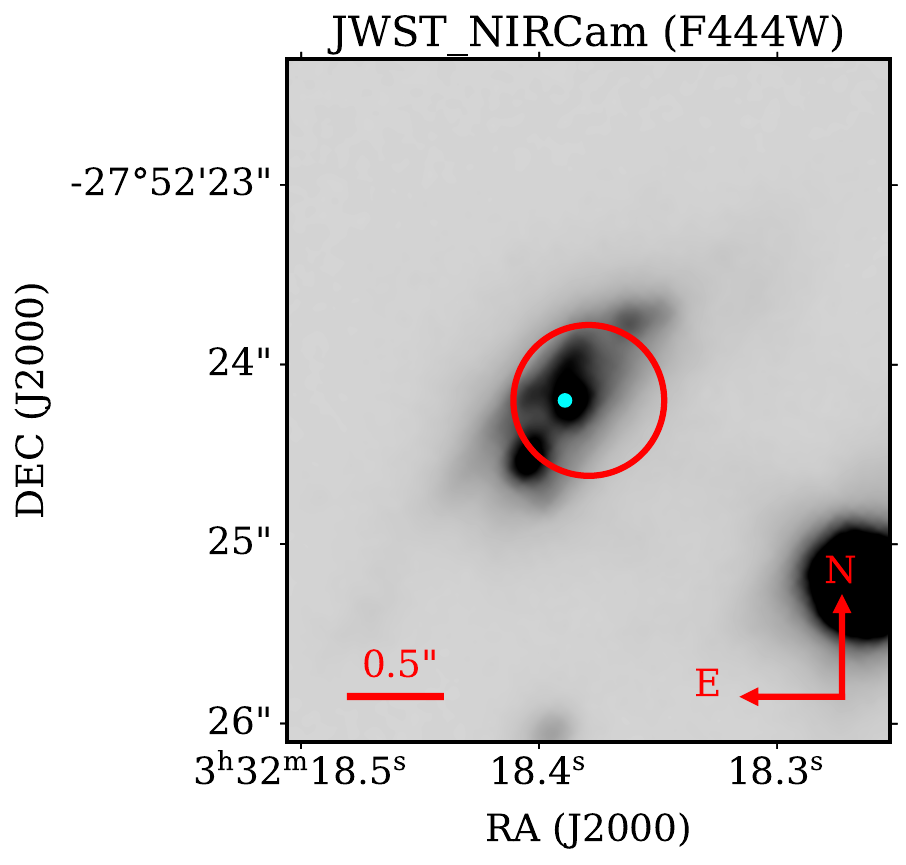}
    \caption{HST and JWST imaging of CDF-S XT2 at three different filters (one per panel) for illustration. The red circle depicts the 2$\sigma$ X-ray position uncertainty of CDF-S XT1, while the cyan dot shows the position of the host according to JADES. Figure~\ref{fig:imaging_XRT_150322_AP} shows the complete version of this figure.}
    \label{fig:imaging_XRT_150322}
\end{figure*}

\begin{table*}
    \centering
    \caption{X-ray Properties taken from the literature of the two extragalactic FXTs considered in this work.}
    \begin{tabular}{ccccccccccc}
    \hline\hline
    Target & ObsID & Exp.~(ks) & Date & RA & DEC & Pos. Unc. & $T_{90}$~(ks) & Ref. \\ 
    (1) & (2) & (3) & (4) & (5) & (6) & (7) & (8) & (9) \\ \hline
    XT1 & 16454 & 49.5 & 2014-10-01 & 53.161550 & -27.859353 & 0\farcs26 & $5.0^{+4.2}_{-0.3}$ & (1),(2) \\
    XT2 & 16453 & 74.7 & 2015-03-22 & 53.076583 & -27.873389 & 0\farcs11 & 10.3$_{-6.3}^{+9.0}$ & (3),(4) \\
    \hline
    \end{tabular}
    \tablefoot{
    \emph{Column 1:} X-ray transient identifier. 
    \emph{Column 2:} \emph{Chandra} observation ID detection.
    \emph{Column 3:} Exposure time of observation.
    \emph{Column 4:} Detection date. 
    \emph{Columns 5 and 6:} Right Ascension and Declination in J2000 equatorial coordinates, in units of degrees, of the FXTs from literature. 
    \emph{Column 7:} Estimated 1$\sigma$ X-ray positional uncertainty, in units of arcseconds. 
    \emph{Column 8:} $T_{90}$ duration. 
    \emph{Column 9:} References: (1) \citet{Bauer2017}, (2) \citet{Quirola2022}, (3) \citet{Xue2019}, (4) \citet{Quirola2023}.
    }
    \label{tab:x_ray_prop}
\end{table*}

\section{Data}\label{sec:data}

This work aims to update and improve the host-galaxy characterization for CDF-S XT1 and XT2 using newly available {\it JWST} data in conjunction with archival data products and, by extension, refine the properties of the transients themselves. 
Below, we explain the photometric and spectroscopic data used in this work\footnote{The data used in this work is available in the next link: \href{https://zenodo.org/records/14844380}{https://zenodo.org/records/14844380}.}. 

\subsection{JWST-NIRCam and HST-WFC3/ACS}\label{sec:nircam}

We consider here the reduced NIRCam imaging data from the public medium and deep tiers of the JADES Origins Field \citep[JOF;][]{Eisenstein2023a,Bunker2024,Rieke2023,DEugenio2024}, and in particular, we incorporate nine JWST filters: F090W, F115W, F150W, F200W, F277W, F335M, F356W, F410M, and F444W. We additionally consider images from the \emph{Hubble Legacy Field} \citep[HST/ACS F435W, F606W, F775W, and F814W and HST/WFC3 F105W, F125W, F140W, and F160W mosaics;][]{Illingworth2016,Whitaker2019}, which have been uniformly assessed and assimilated into the JADES photometric catalogs. 
The JADES JWST images were reduced with the JWST Calibration Pipeline \citep[including some modifications which are explained in detail by][]{Rieke2023}. They were aligned to the HST F160W image (or F850LP when F160W was lacking), which was itself registered to Gaia DR2 \citep{Gaia2018,Rieke2023}. Source positions were determined based on the \texttt{SEP} \citep{Barbary2018} implementation of the \texttt{SExtractor} \citep{Bertin1996} package, applied to a combination of F150W and F200W images. Given the elongated shapes of both host galaxies, we adopted the Kron photometry derived by JADES, where the Kron radius was established using \texttt{Photutils} \citep{Bradley2023}. Photometry from the HST images was derived in a consistent manner. JADES also provides photometric-redshift estimates for the resulting source catalog using the template-fitting code \texttt{EAZY} \citep[which we use for comparison;][]{Brammer2008}.  

The host galaxies of XT1 and XT2 are identified with IDs 180371 and 283332, respectively, in the JADES catalog. For CDF-S XT1, the revised position of its host galaxy is $\alpha$$_{\rm J2000}=$03$^{\rm h}$32$^{\rm m}$38$^{\rm s}$.78 and $\delta$$_{\rm J2000}=$-27$^\circ$51$\arcmin$33$\arcsec$.96, 
while the host galaxy of CDF-S XT2 is now centered at $\alpha_{\rm J2000}=$03$^{\rm h}$32$^{\rm m}$18$^{\rm s}$.39 and $\delta_{\rm J2000}=$-27$^\circ$52$\arcmin$24$\arcsec$.21. Figures~\ref{fig:imaging_XRT_141001} and \ref{fig:imaging_XRT_150322} depict a sub-sample of HST and JWST images of the fields of CDF-S XT1 and XT2, respectively. The \emph{red circles} denote the 1-$\sigma$ positional uncertainty of the \emph{Chandra} detections, while the cyan dot represents the host centroid (see $\S$\ref{sec:CXO} for more details). Tables~\ref{tab:XT1_photometry} and \ref{tab:XT2_photometry} list the JWST and HST host photometry of XT1 and XT2, respectively. 

\subsection{JWST-NIRSpec}\label{sec:nirspec}


Recently, \citet{Barrufet2024} shows the first results from the Cycle-1 JWST-NIRSpec program (\#2198, Barrufet \& Oesch PIs) that focuses specifically on determining the potential quiescent or dusty nature for a sample of 23 "HST-dark" galaxies, using low-resolution NIRSpec-MSA PRISM spectra (with resolving power of $R\sim100$), covering wavelengths of 0.6--5.3~$\mu$m. The host galaxy of CDF-S XT2 turned out to be one of these targets.
The observations consist of two NIRSpec pointings (allowing the targeting of 140 galaxies) with three slitlets per target and dithered along the three shutters. Each spectrum thus received a total exposure time of 2451~s. The spectra were reduced and extracted with \texttt{msaexp2}, a JWST NIRSpec MSA data tool.
A robust spectroscopic redshift of $z=3.4588$ was derived using the \texttt{msaexp} tool \citep{Brammer2019,Heintz2024} for the host of CDF-S XT2, which surprisingly contradicts the redshift of $z=0.738$ originally reported by \citet{Balestra2010} and quoted by \citet{Xue2019}. A detailed discussion surrounding the cause of this large shift appears in Appendix~\ref{sec:redshift_discre}. Figure~\ref{fig:Spec_XRT_150322} shows the NIRSpec spectra of XT2's host, with the most important emission lines identified. Unfortunately, NIRSpec did not observe the host galaxy of CDF-S XT1.
To confirm the reported redshift by \citet{Barrufet2024}, we fitted the emission lines H$\alpha\lambda$6563, [Si III]$\lambda\lambda$9069,9531, and [He I]$\lambda$10,830 using multiple Gaussian functions, and obtained the best-fitting central wavelengths and their associated errors. The derived redshift is $z_{\rm spec}=3.4598{\pm}0.0022$, consistent with the reported value in the literature \citep{Barrufet2024}.

\subsection{JWST-MIRI}\label{sec:miri}

JADES provides some redder mid-infrared (MIR) coverage via coordinated parallels with the MIRI instrument, featuring ${\sim}9$~arcmin$^2$ with 43 hours of exposure at 7.7~$\mu$m and twice that area with 2--6.5 hours of exposure at 12.8 and 15.0~$\mu$m. The fields were designed to overlap with some NIRCam mosaics \citep[especially deep MIRI F770W imaging;][]{Eisenstein2023a,Eisenstein2023b}. That being said, the public MIRI data have not been incorporated into any published catalogs as yet, so we developed our own method to subtract the photometry of the host galaxies. Unfortunately, the MIRI coverage only extends to CDF-S XT2. Using the MAST JWST Search tool\footnote{https://mast.stsci.edu/search/ui/\#/jwst}, we download all available public, product level 3 MIRI imaging in the filters F770W, F1280W, and F1500W (\#1180, Eisenstein PI). The level 3 data are already calibrated, astrometrically, and photometrically corrected, and finally, combined. An elliptical aperture was defined from the Kron parameters reported by JADES to perform comparable aperture photometry. The ellipse was centered in the XT2's host position, with semi-major and minor axes of 8 and 5 pixels, respectively, while an empty background region 4$"$ away was chosen to avoid contamination from the nearby AGN-elliptical galaxy \citep[CANDELS \#4210;][]{Santini2015}. The subtraction was done using the package \texttt{Photutils} \citep{Bradley2023}. Table~\ref{tab:XT2_photometry} lists the JWST-MIRI photometry of XT2 host galaxy.

\begin{figure}
    \centering
    \includegraphics[scale=0.6]{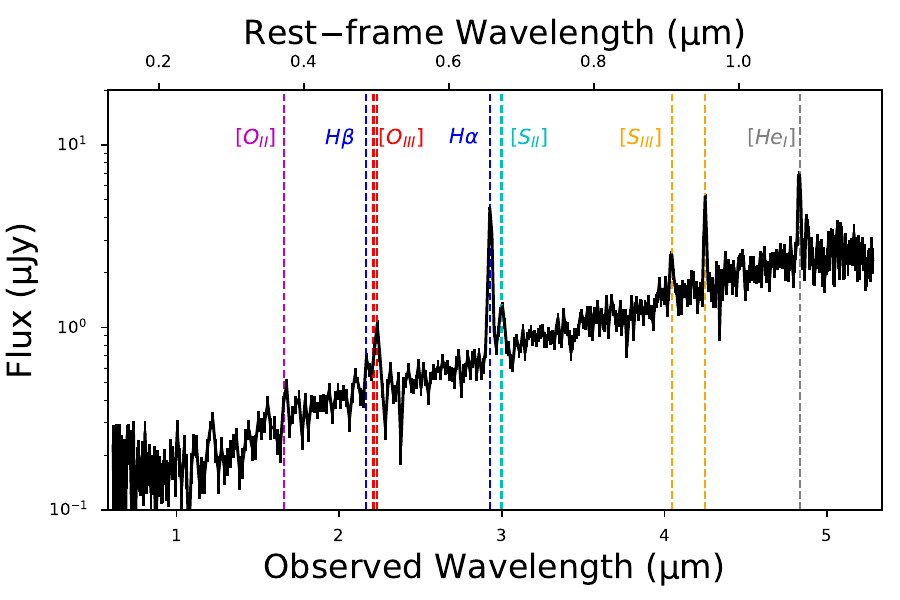}
    \caption{JWST spectrum of the host galaxy of CDF-S XT2 in the observed and rest frames. Dashed vertical lines depict the most important emission lines. According to the emission lines, this galaxy is consistent with a redshift of $z_{\rm spec}{=}3.4598$.}
    \label{fig:Spec_XRT_150322}
\end{figure}

\begin{figure*}
    \centering
    \includegraphics[scale=0.56]{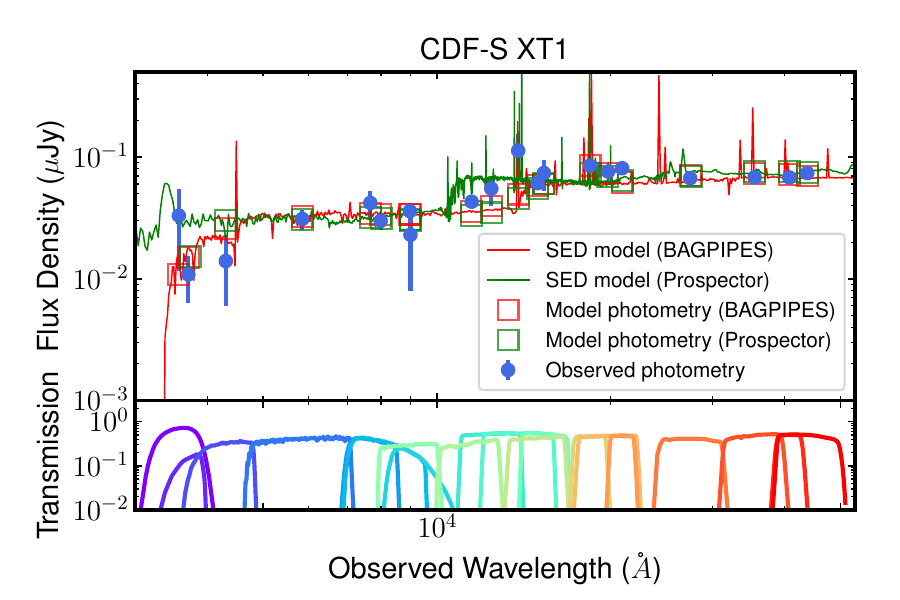}
    \includegraphics[scale=0.56]{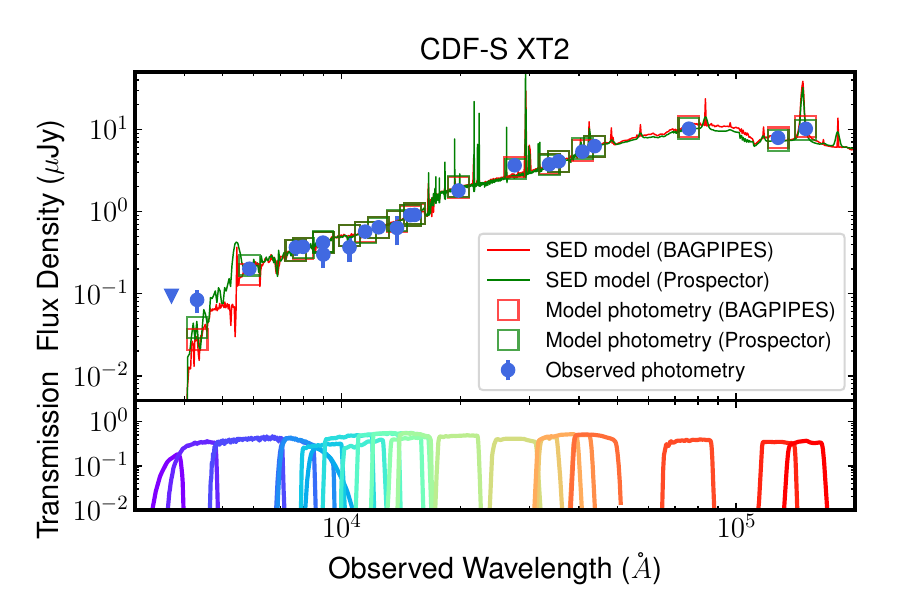}
    \caption{Best-fitting SED models from \texttt{Bagpipes} and \texttt{Prospector} for the hosts associated with CDF-S XT1 (left) and XT2 (right). The photometric data and their 1-$\sigma$ uncertainties are marked by the \emph{blue markers}, while the \emph{triangle} depicts 3-$\sigma$ upper limits. Moreover, the relative transmission functions of the different filters are in the bottom panels (\emph{colored curves}).
    }
    \label{fig:SED_fitting}
\end{figure*}

\begin{table}
    \centering
    \caption{Comparison of host parameters obtained from the literature and by our SED fitting using the \texttt{Bagpipes} \citep{Carnall2018} and \texttt{Prospector} \citep{Leja2017} packages.}
    \scalebox{0.85}{
    \begin{tabular}{llllll}
    \hline\hline
    Target & $z$ &  Log(M$_*/M_\odot$) & SFR/($M_\odot$/yr) & $A_V$ (mag). &  Ref. \\
    (1) & (2) & (3) & (4) & (5) & (6)  \\ \hline
    \multicolumn{6}{c}{Parameters obtained from the literature} \\ \hline
    XT1 & 2.23$_{-1.84}^{+0.98}$ & $7.99{\pm}0.20$ & $1.15{\pm}0.04$ & 0.021 & (1)  \\ 
    XT2 & 0.738 & 9.07 & 0.81 & 0.02 &  (2) \\ \hline
    \multicolumn{6}{c}{Parameters derived using \texttt{Bagpipes}} \\ \hline
    XT1 & $ 2.7 _{- 0.08 }^{+ 0.14 }$ & $ 8.35 _{- 0.11 }^{+ 0.11 }$ & $ 0.85 _{- 0.22 }^{+ 0.30 }$ & $ 0.34 _{- 0.14 }^{+ 0.12 }$ & (3) \\
    XT2 & $3.4598$ & $ 10.75 _{- 0.01 }^{+ 0.01 }$ & $ 156.92 _{- 2.19 }^{+ 2.19 }$ & $ 1.41 _{- 0.01 }^{+ 0.01 }$ & (3) \\ \hline
    \multicolumn{6}{c}{Parameters derived using \texttt{Prospector}} \\ \hline
    XT1 & $ 2.87 _{- 0.24 }^{+ 0.16 }$ & $ 8.50 _{- 0.07 }^{+ 0.06 }$ & $ 0.44 _{- 0.12 }^{+ 0.16 }$ & $ 0.02 _{- 0.02 }^{+ 0.05 }$ & (3) \\
    XT2 & $3.4598$ & $ 10.68 _{- 0.01 }^{+ 0.01 }$ & $ 158.93 _{- 2.70 }^{+ 2.31 }$ & $ 1.25 _{- 0.01 }^{+ 0.01 }$ & (3) \\
    \hline
    \end{tabular}
    }
    \tablefoot{
    \emph{Column 1:} Target. 
    \emph{Column 2:} Host galaxy redshift. 
    \emph{Columns 3 and 4:} Logarithmic of the host galaxy stellar mass and SFR, respectively. 
    \emph{Column 5:} Dust attenuation. 
    \emph{Column 6:} References: (1) \citet{Bauer2017}, (2) \citet{Xue2019}, and (3) This work.}
    \label{tab:SED_para}
\end{table}

\begin{figure*}
    \centering
    \includegraphics[scale=0.6]{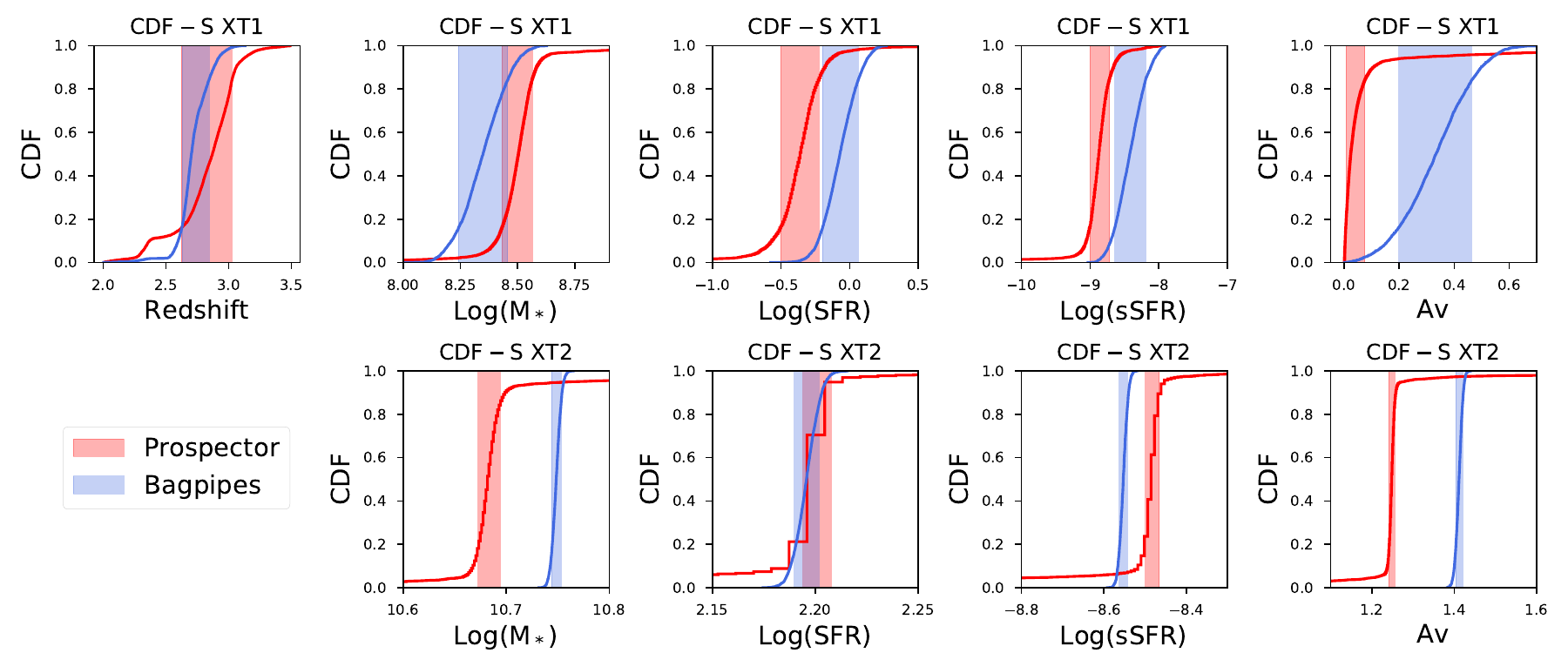}
    \caption{CDFs of the posterior distributions of CDF-S XT1 (top panels) and XT2 (bottom panels) for photometric redshift (just CDF-S XT1), $\log(M_*/M_\odot)$, $\log(\rm SFR)$ in $M_\odot$~yr$^{-1}$, $\log(\rm sSFR)$ in yr$^{-1}$, and $A_V$ in mag. The different colors depict the posterior distribution from \texttt{Prospector} (red) or \texttt{Bagpipes} (blue), respectively, while the color regions depict the 68\% confidence level.
    }
    \label{fig:post_dist}
\end{figure*}

\section{Spectral energy distribution fitting} \label{sec:SED}

We constrain the host properties of both FXTs through SED fitting of their HST+JWST photometry. We consider a star-formation history (SFH) model described by an exponentially decreasing function \citep[SFH $\propto e^{-t/\tau}$, which is probably the most commonly applied SFH model;][]{Simha2014,Carnall2019} with a timescale parameter defined by the e-folding factor $\tau$\footnote{The model assumes that star formation jumps from zero to its maximum value at some time, after which star formation declines exponentially with a timescale $\tau$.}, constructed using the Kroupa initial mass function \citep[IMF;][]{Kroupa2002}, and including the effects of nebular emission and line strengths through a nebular emission model \citep[using the latest version of the \texttt{Cloudy} photoionization code;][]{Byler2017,Ferland2017}. Moreover, to model the dust attenuation in the SEDs, we used the theoretical framework developed by \citet{Calzetti2000}, where $A_V$ is the dust attenuation and is a free parameter. The free parameters of the model and their priors are listed in Table~\ref{tab:SED_model}. In the case of CDF-S XT1, due to the fact that we do not know its distance, the redshift ($z$) is a free parameter determined from the SED fitting. Meanwhile, for CDF-S XT2, the redshift was fixed to the spectroscopic value obtained by the JWST of $z_{\rm spec}=3.4598$. The other fitted free parameters are the mass formed ($M_F$)\footnote{$M_F$ is different than the current stellar mass because stellar mass loss during evolution (e.g., AGN emission and SN explosions) depends on metallicity, SFH, and IMF.}, age of the galaxy at the time of observation ($t_{\rm age}$), the SFH e-folding factor $\tau$, dust attenuation ($A_{\rm V,dust}$), gas ionization parameter ($\log(U)$) and stellar metallicity $\log({Z/Z_\odot})$ (see Table~\ref{tab:SED_model}).

We use two SED fitting packages to derive host properties: $i)$ \texttt{Prospector} \citep{Leja2017,Leja2018,Johnson2021} and $ii)$ \texttt{Bagpipes} \citep[Bayesian Analysis of Galaxies for Physical Inference and Parameter EStimation;][]{Carnall2018,Carnall2019}.
For fitting the available photometric data, both SED packages use the same set of free parameters and prior distributions (see Table~\ref{tab:SED_model}). Moreover, after fitting the SED and constraining the SFH, we derive commonly used physical properties of galaxies to enable direct comparisons with other transients (see Table~\ref{tab:SED_model}). For instance, we get the current star formation rate (SFR) using the posteriors on $t_{\rm age}$, $\tau$, and $M_F$ through the equation
\begin{equation}
    {\rm SFR}(t){=}M_F{\times}\left(\frac{e^{-t/\tau}}{\int_0^te^{-t/\tau}dt}\right).
\end{equation}
We also use the posterior distributions of $t_{\rm age}$ and $\tau$ to derive the posterior on the mass-weighted age ($t_m$) in Gyr as
\begin{equation}
    t_m{=}t_{\rm age}-\frac{\int_0^{t_{\rm age}}t{\times}{\rm SFR}(t)dt}{\int_0^{t_{\rm age}}{\rm SFR}(t)dt},
\end{equation}
which is more physically meaningful than $t_{\rm age}$ (which simply measures the time at which star formation started). 
Figure~\ref{fig:SED_fitting} shows the best-fit SED models obtained by \texttt{Prospector} and \texttt{Bagpipes}; both are consistent with the HST+JWST photometric detections and upper limits.

Table~\ref{tab:SED_para} depicts the host galaxy properties of both FXTs, as well as a comparison with the literature \citep{Bauer2017,Xue2019}. For each host, we report the median and 68\% credible interval of the posteriors in several relevant stellar population properties. Moreover, Fig.~\ref{fig:post_dist} shows the cumulative distribution function (CDF) of posteriors of CDF-S XT1 (top panels) and XT2 (bottom panels) obtained by \texttt{Bagpipes} and \texttt{Prospector}. In the case of CDF-S XT1, the ratios between mean values of host parameters such as SFR, stellar mass, and sSFR from \texttt{Prospector} and \texttt{Bagpipes} lie between factors of ${\approx}$0.4 and 1.4; some parameters (photometric redshift and stellar mass) match at 68\% confidence level (see Fig.~\ref{fig:post_dist}), while others remain somewhat less consistent. Meanwhile, for CDF-S XT2, the posterior CDFs cover a relatively small region in the parameter space (see Fig.~\ref{fig:post_dist}), with ratios between \texttt{Prospector} and \texttt{Bagpipes} mean values being ${\sim}$0.9 and 1.1; aside from the SFR prediction, the two SED-fitting code CDFs are inconsistent at high confidence.
Given the small but significant discrepancies between results, we adopt the mean values from 5,000 randomly selected posteriors selected equally from \texttt{Prospector} and \texttt{Bagpipes}. Finally, it is important to mention that the redshifts of both FXTs are consistent with the photometric redshifts reported by JADES using \texttt{EAZY} \citep{Brammer2008}.

\begin{figure*}
    \centering
    \includegraphics[scale=0.55]{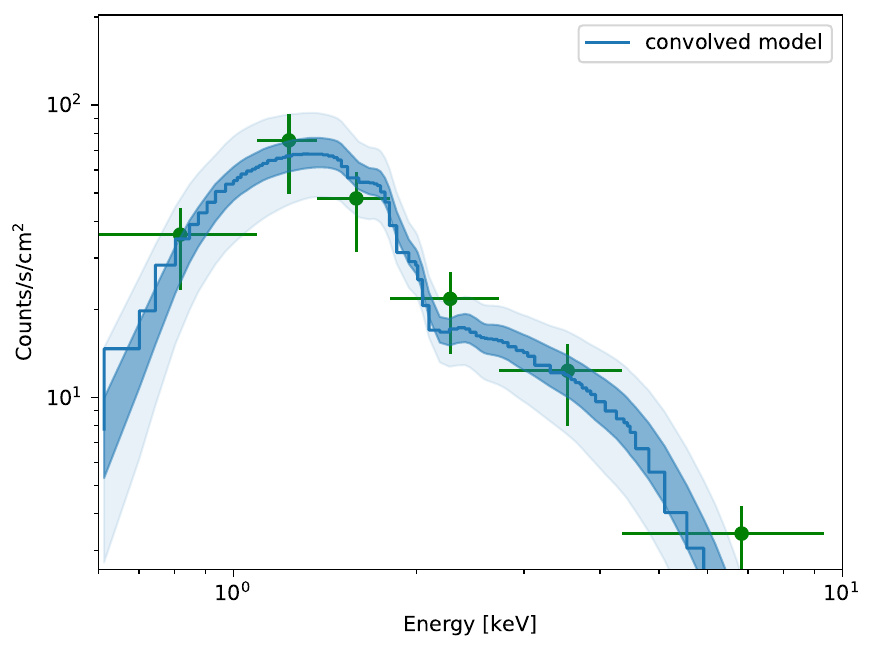}
    \includegraphics[scale=0.55]{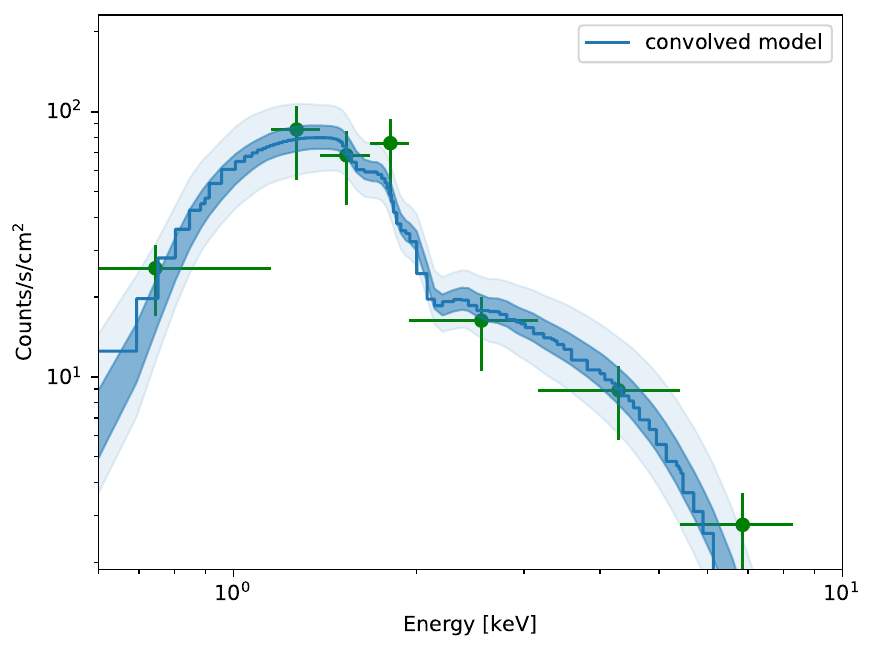}
    \caption{BXA fit of the \texttt{phabs*zphabs*pow} model to the \emph{Chandra} X-ray data of CDF-S XT1 (left) and XT2 (right), where the darker (lighter) blue-filled bands enclose 68\% (98\%) of the posterior. The Galactic hydrogen column density and redshift are fixed parameters during the fitting process (see Table~\ref{tab:spectral_para}).}
    \label{fig:spec_model}
\end{figure*}

\section{Revised \textit{Chandra} X-ray constraints}\label{sec:CXO}

We recompute the X-ray positions and spectral parameters (which depend on the new redshifts) of CDF-S XT1 and XT2. 
We reprocess and analyze the data from the two respective ObsIDs where the transients are detected with the {\sc CIAO 4.16} software developed by the \emph{Chandra} X-Ray Center employing CALDB version 4.11.1. We identify all X-ray sources on the ACIS-I detector and determine their positions and associated uncertainties using the {\sc CIAO} source detection tool \texttt{wavdetect} \citep{Freeman2002}, which employs a series of ``Mexican Hat'' wavelet functions to account for the varying point spread function (PSF) size across the detector. To improve the \emph{Chandra} astrometric solution, we cross-match the positions of the X-ray sources with those of their respective counterparts in the JADES catalog \citep{Eisenstein2023a} located within the {\it Chandra} 90\% error regions. For this, we use the \texttt{wcs\_match} script in {\sc CIAO}. The astrometrically corrected position for CDF-S XT1 is RA$_{\rm J2000.0}=$03$^{\rm h}$32$^{\rm m}$38$^{\rm s}$.79,  Dec$_{\rm J2000.0}=$-27$^\circ$51$\arcmin$33$\arcsec$.89, with a 1$\sigma$ uncertainty position of 0\farcs22.
Meanwhile, for CDF-S XT2 its position is RA$_{\rm J2000.0}=$03$^{\rm h}$32$^{\rm m}$18$^{\rm s}$.38, Dec$_{\rm J2000.0}=$-27$^\circ$52$\arcmin$24$\arcsec$.21,  with a 1$\sigma$ uncertainty position of 0\farcs21.
The quoted uncertainty of both transients considers the X-ray positional errors derived by \texttt{wavdetect} and the alignment dispersion between \emph{Chandra} observations and the reference catalog.

We extract the spectra with the \texttt{specextract} package in {\sc CIAO}. For this, we include the X-ray counts within a circular region centered at the X-ray position with a radius corresponding to an encircled energy fraction of ${\approx}$98\% \citep[$R_{98}$; 5\farcs2 and 4\farcs9 for CDF-S XT1 and XT2, respectively, and using the method of][]{Vito2016} given their instrumental off-axis angle (4\farcm3 and 4\farcm5 for CDF-S XT1 and XT2, respectively). For the background regions, we consider an annulus region (centered on the transient's position) with an internal and external radius of $R_{98}$ and $R_{98}+20$~pixels, respectively\footnote{Formally, in the case of CDF-S XT2, the nearby AGN CANDELS \#4210 \citep{Santini2015} will contaminate our aperture and background, however, it should contribute on average only $\approx$0.5 photons within this aperture within the ${\approx}70$~ks exposure; as this represents only ${\approx}0.4$\% of the total photons, we opt not to mask it out, since it would remove a much larger percentage of photons associated with CDF-S XT2.}.
To obtain the X-ray spectral parameters, we use the Bayesian X-ray Astronomy package \citep[BXA;][]{Buchner2014}, within the fitting environment of \texttt{XSPEC} version 12.14 \citep{Arnaud1996}, and considering the Cash statistic \citep{Cash1979} to account for the low number of detected counts for the FXTs under study. We fit an absorbed power-law model (\texttt{phabs*zphabs*pow} model in \texttt{XSPEC}) to the binned spectra (requiring at least one photon per bin), where \texttt{phabs} and \texttt{zphabs} describe the Galactic and the combined intrinsic plus host galaxy absorption, respectively. For the extraction of the X-ray spectral parameters, we fix the Galactic hydrogen column density ($N_{H,\rm Gal}$) to $2.0\times$10$^{20}$~cm$^{-2}$ \citep[taken from][]{Kalberla2005, Kalberla2015}, while the combined intrinsic plus host galaxy hydrogen column density ($N_H$) is a free parameter in our fit. Redshifts remain fixed parameters in the fitting process, adopting values of $2.76$ (mean photometric redshift from \texttt{Prospector} and \texttt{Bagpipes} combined) and $3.4598$ (spectroscopic redshift from JWST) for CDF-S XT1 and XT2, respectively. The absorbed fluxes are computed using the standard \texttt{XSPEC} tasks, while the unabsorbed fluxes are derived by the \texttt{XSPEC} convolution model \texttt{cflux}. Figure~\ref{fig:spec_model} depicts the best BXA fitted model 
of both transients (with CDF-S XT1 and XT2 in the top and bottom panels, respectively), while Table~\ref{tab:spectral_para} represents the best X-ray spectral parameters of both FXTs obtained by the posterior distribution of BXA.

\begin{table*}
    \centering
    \caption{Results of the 0.5--7 keV \hbox{X-ray} spectral fits for CDF-S XT1 and XT2. }
\scalebox{1.0}{
    \begin{tabular}{llllllllll}
    \hline\hline
    FXT & $z$ & $N_{\rm H,Gal}$ & $N_H$ & $\Gamma$ & $\log{\rm Norm}$ & Abs. Flux & Unabs. Flux & C-stat(dof) & $\ln \mathcal{Z}$ \\
    (1) & (2) & (3) & (4) & (5) & (6) & (7) & (8) & (9) & (10) \\ \hline
    CDF-S XT1 &  2.76 & $0.02$ & $2.08^{+2.67}_{-1.51}$ &  $1.74^{+0.24}_{-0.21}$ & $-5.14^{+0.09}_{-0.08}$ & $4.77_{-0.49}^{+0.34}$ & $4.92_{-0.46}^{+0.49}$ & $79.76(91)$ & $-48.002{\pm}0.276$ \\
    CDF-S XT2 &  3.4598 & $0.02$ & $6.43^{+3.46}_{-2.43}$ & $1.86{\pm}0.20$ & $-5.06{\pm}0.08$ & $3.84_{-0.39}^{+0.42}$ & $5.82_{-0.51}^{+0.54}$ & $74.64(95)$ & $-44.460{\pm}0.228$ \\ \hline
    \end{tabular}
    }
    \tablefoot{
    \emph{Column 1:} Target. 
    \emph{Column 2:} Redshift assumed. 
    \emph{Columns 3 and 4:} Galactic and intrinsic column density absorption ($\times$10$^{22}$) in units of cm$^{-2}$, respectively. The former is kept fixed during the fit. 
    \emph{Column 5:} Photon index from the power-law model. 
    \emph{Column 6:} Normalization parameter (in units of photons~keV$^{-1}$~cm$^{-2}$~s$^{-1}$). 
    \emph{Columns 7 and 8:} Absorbed and unabsorbed 0.3--10~keV fluxes ($\times$10$^{-14}$) in units of erg~cm$^{-2}$~s$^{-1}$. 
    \emph{Column 9:} C-stat value and the number of degrees of freedom in parentheses. 
    \emph{Column 10:} Log-evidence values ($\ln \mathcal{Z}$) for each model, with 1-$\sigma$ confidence errors from the posterior distributions obtained by BXA.}
    \label{tab:spectral_para}
\end{table*}

\begin{figure*}
    \centering
    \includegraphics[scale=0.8]{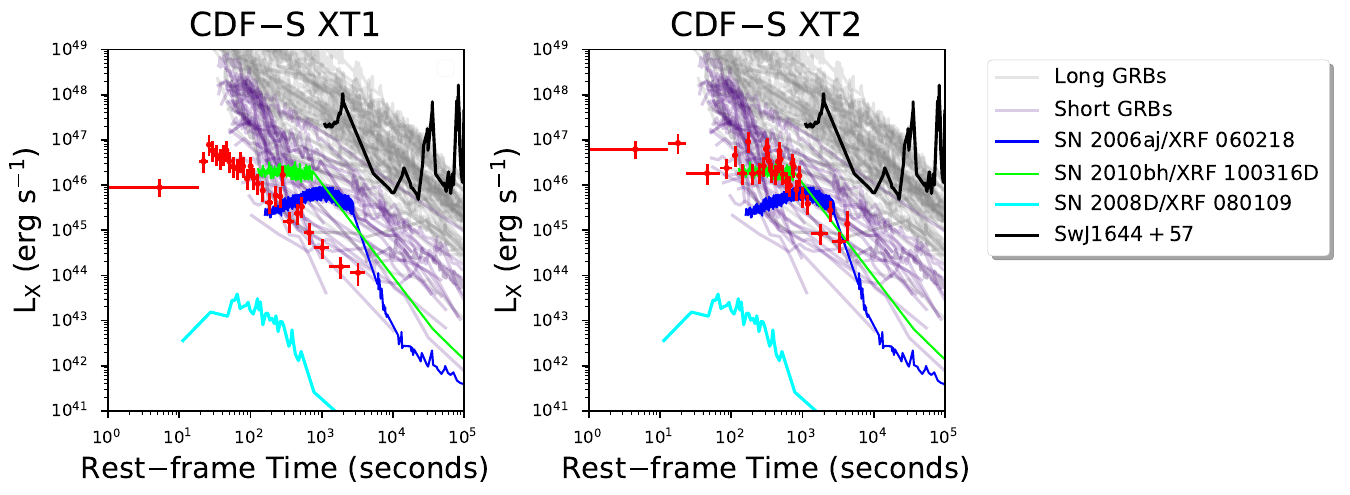}
    \vspace{-0.1 cm}
    \caption{0.3--10 keV light curves of CDF-S XT1 and XT2 in luminosity units (marked in red). Also shown are the \hbox{X-ray} afterglow light curves of 64 LGRBs plus 32 SGRBs \citep{Bernardini2012,Lu2015}, as well as several individual transients are overplotted: the shock-breakout supernova XRF\,080109-SN\,2008D \citep[\emph{cyan line}, 27~Mpc;), the X-ray flashes XRF\,060218-SN\,2006aj (\emph{blue line}, 145~Mpc;) and XRF\,100316D-SN\,2010bh (\emph{green line},263~Mpc;][]{Barniol2015,Starling2011,Modjaz2009,Evans2009,Soderberg2008,Evans2007,Campana2006}, and the relativistically beamed TDE \emph{Swift}\,J1644+57 \citep[\emph{black line}, $z{=}0.3543$;][]{Bloom2011,Levan2011}. The reference time for both FXTs is the arrival time of the first photon to the detector, while for GRBs, \emph{Swift}\,J1644+57, and X-ray flashes their reference time are based on their initial gamma-ray detections.}
    \label{fig:LC_comparison}
\end{figure*}

\begin{figure*}
    \centering
    \includegraphics[scale=0.45]{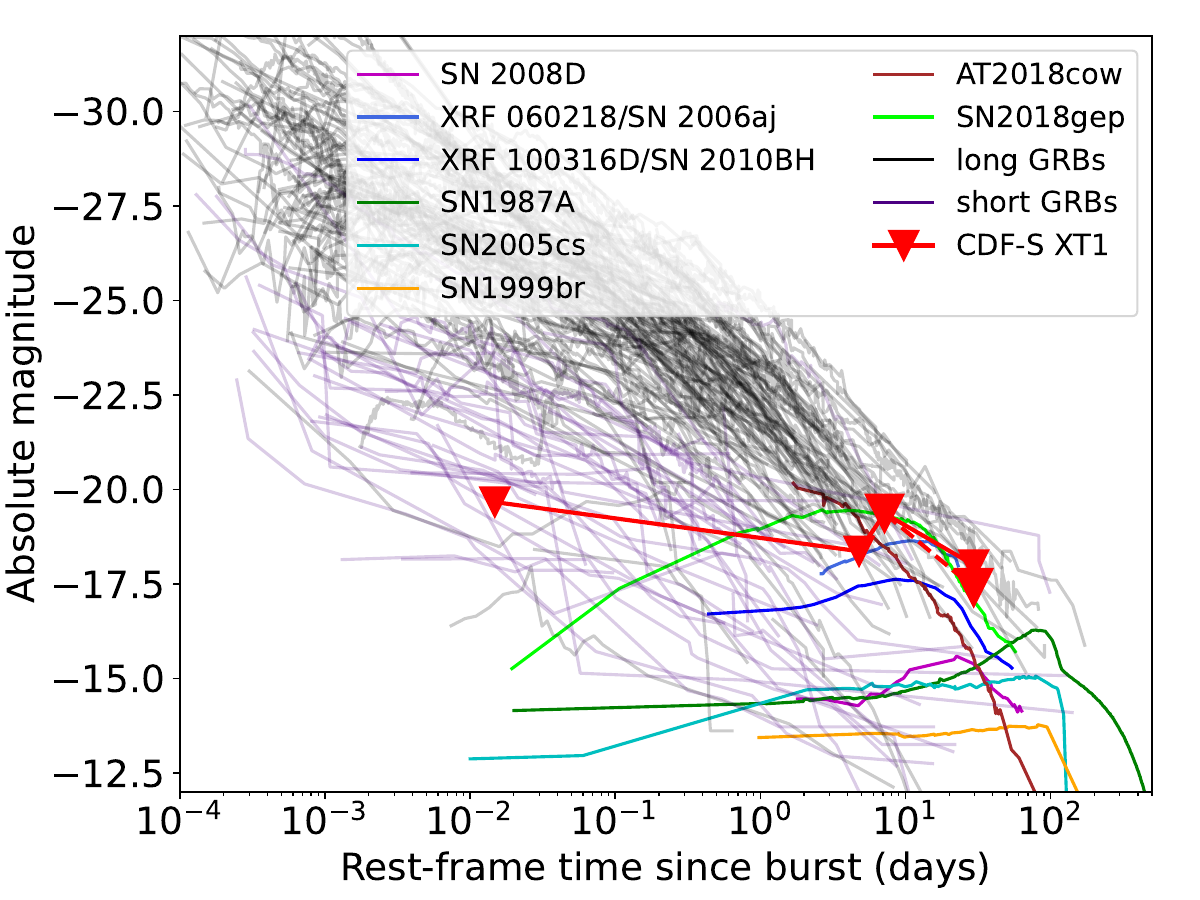}
    \includegraphics[scale=0.45]{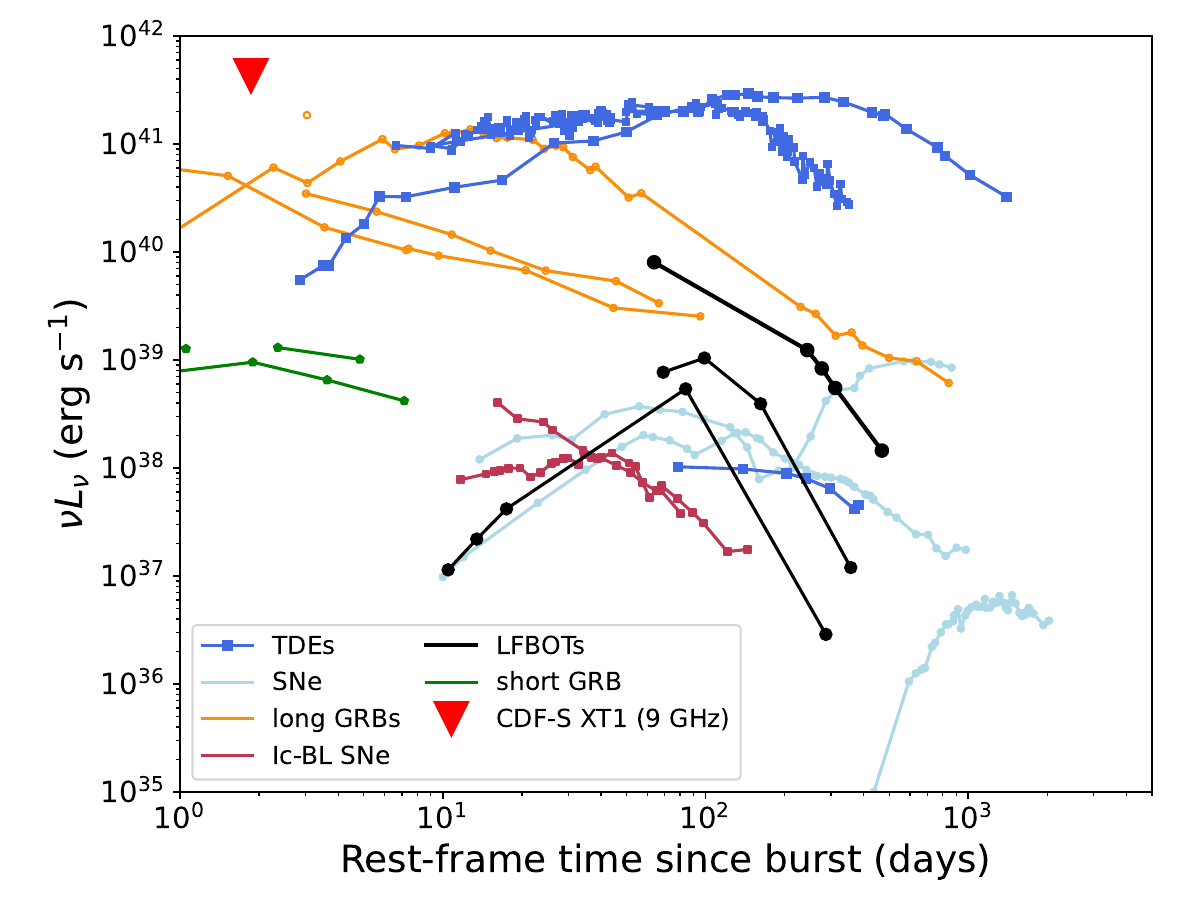}
    \vspace{-0.5 cm}
    \caption{Left: $R$-band optical light curves of CDF-S XT1 (red) compared to long and short GRBs \citep{Kann2006,Kann2010,Kann2011,Nicuesa2012}, the SBO SN~2008D \citep{Soderberg2008}, the low-luminosity events XRF 060218-SN 2006aj and XRF 100316D-SN 2010BH, the relativistic Ic-BL SN~2018gep \citep{Ho2019}, LFBOT AT2018cow \citep{Xiang2021}, and Type II-pec SN~1987A (50 kpc), SN 1999br (7.1 Mpc) and Type II-P SN 2005cs (8.6 Mpc). The {\it HST} F110W constraint at 111~days is displayed for the range $m_R{-}m_{\rm F110W}\approx$0.4--1.0~mag typical of GRBs \citep{Kann2006,Kann2010,Kann2011}.
    Right: 9~GHz upper limit of CDF-S XT1 (red) compared to low-frequency (1-10 GHz) light curves of different classes of energetic explosions such as TDEs \citep{Zauderer2011,Berger2012,Zauderer2013,Alexander2016,Eftekhari2018}, SNe exploding in dense circumstellar medium \citep{Soderberg2005,Salas2013} and normal SNe \citep{Weiler1986}, relativistic Ic-BL SNe \citep{Kulkarni1998,Soderberg2010}, LGRBs \citep[GRB~030329A, GRB~111209A, and GRB~130427A;][]{Berger2003,Hancock2012,Perley2014,van_der_Horst2014}, and luminous fast blue optical objects \citep[LFBOTs;][]{Margutti2019,Coppejans2020,Ho2020b}.}
    \label{fig:kann_plot}
\end{figure*}

\section{Results and interpretations}\label{sec:results}

Below, we report the host-galaxy parameters obtained from the SED fitting, revised X-ray spectral parameters, and possible interpretations to explain the progenitors of CDF-S XT1 and XT2. Finally, we derive the event rate density of both FXTs.

\subsection{CDF-S XT1}

The serendipitous detection of this FXT by \emph{Chandra} (ObsID 16454, ${\approx}$50~ks exposure) in the CDF-S was first reported by \citet{Luo2014} and analyzed in detail by \citet{Bauer2017}.
XT1 shows a light curve with a fast rise (with a duration of ${\approx}100$~s) followed by a power-law decay \citep[$\propto t^{-1.5}$;][]{Bauer2017} without a clear spectral evolution.
As explained above, we consider HST + JWST data to recompute the parameters of both transients and their host galaxies (see Table~\ref{tab:SED_para}). Firstly, the random alignment probability of XT1 and the host galaxy is 0.467\%, reconfirming an association. The new data improve the photometric redshift from $z_{\rm photo}{=}2.23^{+0.98}_{-1.84}$ \citep{Bauer2017} to $2.76_{-0.13}^{+0.21}$, refining the peak luminosity ($L_{\rm X,peak}$) and projected physical offset ($\delta R$) of CDF-S XT1 to be $L_{\rm X,peak}{\approx}2.8{\times}10^{47}$~erg~s$^{-1}$ and $\delta R=1.2{\pm}2.4$~kpc, respectively.
Figure~\ref{fig:LC_comparison} compares the X-ray light curves with other X-ray transients such as SN SBOs (XRF\,080109-SN\,2008D), X-ray flashes (XRF\,060218-SN\,2006aj and XRF\,100316D-SN\,2010bh), GRB afterglows, and jetted TDEs (\emph{Swift}\,J1644+57). 

The source position of CDF-S XT1 was serendipitously observed ${\approx}80$ minutes after the X-ray trigger by the VIsible MultiObject Spectrograph (VIMOS) instrument mounted in the VLT \citep{Bauer2017}, but no contemporaneous optical transient was detected at the location of the CDF-S XT1 ($M_R{>}-21.1$~mag). This field was imaged on three subsequent occasions, by VLT/FORS2, Gemini-South/GMOS-S, and {\it HST}/WFC3 at ${\approx}$18, 27, and 111 days after the X-ray trigger, respectively, with no counterpart detection to $M_R{>}-19.8$, $-20.8$, and ${-}18.0$ (or ${-}17.4$)~mag\footnote{Assuming a color dependence of $m_R{-}m_{\rm F110W}\approx$0.4--0.7 mag based on GRB
afterglow power-law spectral slopes in the range of $-0.6$ to $-1.1$
\citep{Kann2010,Kann2011} and 0.4–1.0 mag for CCSNe between
$z=0.0$ and 1.0 \citep{Poznanski2007,Drout2011,Bianco2014}.}, respectively. 
Finally, the Australian Telescope Compact Array (ATCA), using five different frequencies (2.1, 5, 9, 17, and 19~GHz), observed the source position ${\approx}7$~days after the X-ray onset, all resulting in non-detections \citep[$L_\nu{\lesssim}4{\times}10^{31}-10^{32}$~erg~s$^{-1}$~Hz$^{-1}$;][]{Bauer2017}.
Unfortunately, the gamma-ray instruments \emph{Swift}-BAT and \emph{Fermi} were not covering the position of CDF-S XT1 during the transient trigger. However, the field was covered by the Interplanetary Network, with no counterpart detected \citep[peak flux upper limit of $1.4\times10^{-7}$~erg~cm$^{-2}$~s$^{-1}$ in the energy range $25-150$~keV;][]{Bauer2017}, i.e., an upper limit of $L_{\rm \gamma,peak}<4.8\times10^{52}$~erg~s$^{-1}$ (at $z=2.76$) in the energy range of $1-10^4$~keV\footnote{Considering a $k$-correction factor of $k\approx5.2$, assuming a power-law function given by the photon index $\Gamma=1.74$.}.

Inside the CDF-S XT1 error position, an extended source is located with an absolute magnitude of $M_R=-19.14$~mag considering a redshift of $z=2.76$. At this distance, the angular half-light radius of ${\approx}0\farcs13$ \citep[defined from the JWST F200W image using \texttt{Sextractor};][]{Bertin1996} converts to a projected half-light radius of ${\sim}1.1$~kpc. The luminosity-size relation of the XT1 host shows that the galaxy is not consistent with dwarf galaxies and globular clusters \citep[see for instance,][]{Taylor2018,Simon2019}.
The best-fitted host-galaxy parameters of CDF-S XT1 are $\log(M_*){=}8.44_{-0.14}^{+0.09}$, $\log(\rm SFR){=}-0.21_{-0.21}^{+0.21}$ and $\log(\rm sSFR){=}-8.68_{-0.25}^{+0.38}$, which permit the classification of the host as a star-forming galaxy based on its location relative to regarding the galaxy main sequence \citep{Santini2017}, and following the standard star-forming scheme classification of \citet{Tacchella2022}. A comparison between host-galaxy properties of different transients and both FXTs is presented in Fig.~\ref{fig:host_comparison}. Below, we analyze the potential scenarios associated with CDF-S XT1.

\subsubsection{CDF-S XT1: SN SBO scenario}

One possible scenario is that the X-ray burst could be related to an SN SBO, such as the most studied SBO XRF\,080109 \citep[serendipitously discovered during \emph{Swift}-XRT observations of SN 2007uy in NGC 2770;][]{Soderberg2008}. This was initially rejected by \citet{Bauer2017} because the optical/NIR constraints excluded a substantial portion of SN phase space. With the revised redshift, the peak luminosity of CDF-S XT1 now appears to be ${\approx}$3--4~dex higher than XRF\,080109 (with $L_{\rm X,peak}{\sim}10^{43}$~erg~s$^{-1}$, see Fig.~\ref{fig:LC_comparison}) and the SBO theoretical limits \citep[$L_{\rm X,peak}{\approx}10^{42}-10^{44}$~erg~s$^{-1}$;][]{Waxman2017,Goldberg2022}, allowing us to discard SBOs as a potential progenitor of XT1 more conclusively.

\subsubsection{CDF-S XT1: standard GRB afterglow scenario}

Another scenario for CDF-S XT1 is associated with the \hbox{X-ray} afterglow of a GRB. The \emph{Swift} and \emph{Fermi} satellites did not cover the area of CDF-S XT1 at the time of the burst \citep[][and reference inside]{Bauer2017}. However, the field was covered by the Interplanetary Network, with no counterpart detected \citep{Bauer2017}, excluding any association with a strong GRB but still allowing for faint GRBs or orphan afterglows \citep{Ghirlanda2015}. 
Figure~\ref{fig:LC_comparison} compares CDF-S XT1 and the X-ray afterglows of on-axis short and long GRBs \citep[SGRBs and LGRBs, hereafter;][]{Evans2007,Evans2009,Bernardini2012,Lu2015}. It is clear that CDF-S XT1's luminosities lie well below the range of afterglows for normal LGRBs and near the lower bound of known SGRBs. Although the decay rate after the peak matches those of normal X-ray afterglows \citep[$F_X^{\rm GRBs}{\propto}t^{-1.2}$;][]{Zhang2006,Evans2009,Racusin2009}, XT1's initial fast rise phase is not consistent with the typical X-ray afterglows of on-axis GRBs observed at such early times. 
In Fig.~\ref{fig:kann_plot} (left panel), we compare $R$-band light curves derived from a database of optical measurements of $\sim160$ GRBs with known redshifts \citep{Kann2006,Kann2010,Kann2011,Nicuesa2012}, nearby SNe (SN~1987A, SN~1999br, and SN~2005cs), low-luminous events XRF~060218-SN~2006aj and 100316D-SN 2010BH, the SN SBO SN\,2008D-XRF\,080109, the relativistic Ic-BL SN~2018gep, the luminous fast blue optical objects (LFBOTs) AT2018cow, and the optical upper limits of CDF-S XT1. These upper limits are inconsistent with the majority of long GRBs (consistency remains only with a few long GRBs at $z\lesssim0.2$), especially the earliest upper limit ($\approx80$~min. after the X-ray trigger). Similarly, only a small fraction of short GRB afterglows appear consistent with the upper limits. The strong X-ray and optical limits in combination with the deep gamma-ray upper limit from the Interplanetary Network disfavors a typical on-axis GRB scenario. The lack of a gamma-ray counterpart opens the possibility of off-axis GRBs, which might be accompanied by a weak optical emission \citep{Lazzati2017}. We will explore the off-axis scenario in more detail. Finally, Fig.~\ref{fig:kann_plot} (right panel) shows the 9~GHz upper limit for XT1 compared with other transients. The weak limit fails to exclude any of the objects shown, leaving the nature of XT1 at radio wavelengths largely unconstrained.

One consideration is that the rising and decaying phases of XT1 could mimic an X-ray flaring episode, which has been seen in some afterglows \citep{Chincarini2007,Chincarini2010,Margutti2011}. To compare the pulse shape of XT1 with the X-ray flaring episodes in GRBs' afterglows, we fitted a fast rise exponential decay function \citep[see][for more details]{Norris2005} to the light curve of CDF-S XT1 (in counts rate units and a time bin of 50~sec. to be consistent with \citet{Chincarini2010}), and derived the observed width of the pulse ($w\approx$320~sec.) and its peak time ($\tau_{\rm peak}\approx100$~sec.). The peak time and the width of CDF-S XT1 remain marginally consistent with GRB flares \citep{Chincarini2007,Chincarini2010}, although the spectral slope and decay time slope lie on the hard and slow ends of their respective distributions \citep{Chincarini2007,Chincarini2010}. The ratio $\log(w/\tau_{\rm peak})$ of CDF-S XT1 lies outside the expected transition for GRBs \citep[see][fig.~3]{Chincarini2010}, with similar inconsistencies identified in the $L_{\rm X,peak}-\tau_{\rm peak}$ and $L_{\rm X,peak}-w$ relations \citep{Chincarini2010}. Finally, the total duration of XT1 is longer than GRB X-ray flares (i.e., $\sim$10--300~s). Thus, the GRB flaring episode afterglow scenario is not considered viable.

Based on the rise and decay light curve of the transient, an additional possibility related to the on-axis scenario is an early afterglow emission caused by the deceleration of a GRB jet \citep[e.g.,][]{Sari1999,Molinari2007}, where the early afterglow might peak on a time-scale similar to the deceleration time of the jet \citep[$t_{\rm dec}$;][]{Sari1999}.\footnote{We note that the deceleration jet scenario also predicts an afterglow emission with an optical peak, which is usually used to constrain the initial Lorentz factor of GRBs. High intrinsic extinction from the host galaxy could explain the lack of strong optical emission for CDF-S XT1. Extrapolating an $A_V$ estimate from the hydrogen column density (i.e., $N_H\approx2\times10^{22}$~cm$^{-2}$) yields $A_V=9.4_{-6.8}^{+12.1}$~mag \citep{Guver2009} for the Galactic gas-to-dust ratio. Assuming a gas-to-dust ratio for a Small-Magallanic (SMC) environment \citep[GRBs are well described by an SMC-type extinction curve with moderate to low extinction;][]{Zafar2011} yields $E(B-V)=1.4_{-1.0}^{+1.8}$~mag, which is more consistent with a galaxy with a moderate SFR similar to the host of CDF-S XT1. Both values can easily explain the optical non-detection.}
To explore this scenario, we assume that CDF-S XT1 does not have any contribution from the low-energy tail of the prompt emission of any potentially associated GRB \citep[e.g., due to the long-lasting central engine activity;][]{Ghisellini2007,Genet2007,Ioka2006,Toma2006,Panaitescu2008,Nardini2010} or any flaring episodes caused by the late-time central engine \citep{Margutti2010,Ghirlanda2018}. We compute the deceleration time of the jet \citep[using the formalism of][]{Sari1999,Ghirlanda2018,Zhang_book_2018}, considering that the deceleration is caused by the interaction of the jet with the interstellar medium (ISM\footnote{The ISM is described by the number density medium $n_0$ in units of cm$^{-3}$.}; see Appendix \S\ref{sec:deceleration} for more details about the methodology).
We assume typical values for the kinetic energy ($E_{\rm k}\sim10^{50}-10^{53}$~erg) and the initial Lorentz factor ($\Gamma_0\sim50-1000$) of GRBs, and a number density of the ISM of $n\sim10^{-2}-10$~cm$^{-3}$ to cover the entire parameter space of GRBs. If the origin of CDF-S XT1 is related to the early afterglow, we expect that the deceleration time should be similar to the peak time of the transient \citep{Sari1999}, i.e., $t_{\rm dec}=t_{\rm X,peak}\approx110$~sec. Figure~\ref{fig:early_aft} shows the deceleration time covering a wide parameter space of $\Gamma_0$ and $E_{\rm k}$, where we found that $t_{\rm dec}\approx t_{\rm X,peak}$ in the range $\Gamma_0\approx60-600$. Under the condition $n\leq1$~cm$^{-3}$, we obtain $\Gamma_0\gtrsim100$, which is consistent with the distribution of Lorentz factors of normal GRBs \citep[e.g.,][]{Racusin2011,Ghirlanda2018}; however, at $n>1$~cm$^{-3}$, the early afterglow interpretation favors a lower Lorentz factor $\Gamma_0\approx70-200$. This Lorentz factor range is covered by standard GRBs, which can be excluded based on the gamma-ray upper limit obtained by \emph{Konus Wind} \citep[i.e., $F_{\rm \gamma,peak}<1.4\times10^{-7}$~erg~s$^{-1}$~cm$^{-2}$, in the range 25-150~keV;][]{Tsvetkova2017}, leaving this scenario as an unlikely possibility, and favoring other possibilities such as faint GRBs and/or off-axis GRBs.



Next, we consider an association with off-axis GRBs \citep[e.g.,][]{Granot2002,Ryan2019,Perley2025}. The early rise phase of CDF-S XT1 lasts ${\sim}100$~s, which a slightly off-axis afterglow could explain; note that the short duration and observed luminosity of XT1 immediately rule out comparisons to higher off-axis angles (hence longer rise) scenarios such as GRB~170817A- or SN~2020bvc-like events \citep[e.g.,][]{Nynka2018,DAvanzo2018,Troja2020,Troja2022,Izzo2020}. Notably, XT1's decay rate is comparable to the normal decay of GRBs. \citet{Sarin2021} suggested an association of an X-ray afterglow produced by a relativistic structured jet viewed off-axis with a viewing angle of $\theta_{\rm obs}{\approx}10$~deg and an ultra-relativistic jet core of $\theta_{\rm core}{\approx}4.4$~deg, explaining the early and later optical limits. On the other hand, \citet{Wichern2024}, using off-axis uniform and Gaussian-structured jet afterglow models from \citet{Ryan2019}, explored the link between FXTs and the afterglows of off-axis merger-induced GRBs. They found that in the particular case of CDF-S XT1, the early and high peak X-ray luminosity of CDF-S XT1 are consistent with a relatively on-axis viewing angle, while the deep optical upper limits imply a highly off-axis viewing angle, leading them to conclude that the overall properties of CDF-S XT1 are not self-consistent with any off-axis scenario for a normal merger-induced GRB. Unfortunately, our host galaxy SED-fitting constraints do not shed much additional light on the nature of XT1. Although the moderate SFR is consistent with LGRB and SGRB hosts, its low stellar mass lies at the lower bounds ($\lesssim$10 and 5\% of LGRBs and SGRBs, respectively; having lower stellar masses; see Fig.~\ref{fig:host_comparison}). As the reader can appreciate, this scenario cannot be completely excluded.

Finally, we consider a dirty fireball scenario, which describes GRBs with a higher baryon load \citep[a proton content as small as $\gtrsim10^{-4}$~$M_\odot$;][]{Cenko2013} than typical events. Here, the fireball accelerates more slowly due to the greater inertia of the baryons, producing a low initial Lorentz factor $\Gamma_0\ll100$ \citep{Dermer1999,Huang2002}. This model predicts afterglow emission, but inhibits any high-energy emission via $e^--e^+$ pair production \citep{Dermer1999,Huang2002,Rhoads2003,Ghirlanda2012}. For instance, the transients PTF11agg, AT~2019pim, and AT~2021any \citep[without gamma-ray counterparts;][]{Cenko2013,Xu2023,Perley2025} can be explained by a dirty fireball model considering lower Lorentz factors of $\Gamma_0\sim10-50$. The light curve of a dirty fireball will take longer to rise to a peak, being set by the time it takes the shock to sweep up the material 
\citep[i.e., the deceleration time;][]{Ho2020c}. For a uniform-density medium (assuming $n\sim$1, 10, and 30~cm$^{-3}$), using Eq.~\ref{eq:003}), no outflow (with a kinetic energy of $E_{\rm K}\sim10^{49}-10^{54}$~erg) with $\Gamma_0\lesssim40$ can produce an afterglow that rises to a peak on a time scale similar to the peak time of CDF-S XT1. Thus, we conclude that this scenario does not adequately explain the timing properties of XT1.

\subsubsection{CDF-S XT1: TDE scenario}

Given that the decay time slope of CDF-S XT1 appears fully consistent with the predictions for TDEs \citep[i.e., $\propto t^{-5/3}$;][]{Rees1988,Phinney1989,Burrows2011}, a comparison to this class of transient is warranted. However, the high peak luminosity reached by XT1 implies that only a jetted TDE scenario remains possible. 

In the case of an SMBH-TDE, we note that SwJ1644+57 reaches comparable luminosities to XT1 at multiple points in its light curve \citep[see Fig.~\ref{fig:LC_comparison};][]{Bloom2011,Levan2011}, and presents a radio luminosity \citep{Berger2012} consistent with the radio upper limits of XT1 (see Fig.~\ref{fig:kann_plot}, right panel). However, the fast rise time, hard X-ray flux, lack of strong variability, and overall duration of CDF-S XT1 remain strongly inconsistent with the few known jetted SMBH-TDEs, leading us to exclude this scenario.

Considering instead a jetted WD-IMBH TDE, this could potentially explain not only the high luminosity but also the fast rise to the peak (${\approx}$minutes) and duration of XT1 \citep{Krolik2011,Haas2012,Kawana2018}.
We explored this scenario by adopting a simple model developed by \citet{Peng2019}. Once the tidal forces disrupt the WD, the bound debris stream falls back to the disruption site (fallback timescale, $t_{\rm fb}$). However, the returned debris stream is not immediately digested by the BH, as some time is required to form an accretion disk and swirl inward until it reaches the IMBH (accretion timescale, $t_{\rm acc}$). 
The fast rise of CDF-S XT1 can then be explained in terms of short fallback and accretion timescales of ${\sim}$30 and 70~s, respectively. The low $t_{\rm fb}$ implies a relatively heavy and compact WD. Assuming a WD mass of $M_{\rm WD}=1.4$~M$_\odot$ \citep[and using Eq.~5 of][]{Peng2019}, the BH mass should be $M_{\rm BH}{\sim}6{\times}10^2$M$_{\odot}$.
However, the thermal accretion disk emission would be limited by the Eddington luminosity, producing an X-ray luminosity of ${\sim}10^{42}-10^{44}$~erg~s$^{-1}$, which is ${\gtrsim}3$~dex below the peak luminosity of CDF-S XT1. Nevertheless, many of WD-IMBH TDE systems are also predicted to launch relativistic jets \citep{Strubbe2009,Zauderer2011,DeColle2012}, powered by rapid IMBH spin (which has been estimated in some IMBH candidates, e.g.,\citealp{Wen2021} and \citealp{Cao2023}) and high magnetic flux \citep[e.g., from a magnetic WD;][]{Cenko2012, Brown2015,Sadowski2016} via the \citet{Blandford1977} mechanism. For instance, \citet{MacLeod2016} shows typical peak jet luminosities between${\sim}10^{47}-10^{50}$~erg~s$^{-1}$ and rise timescales of $\sim10^2-10^4$~s, matching with CDF-S XT1. Furthermore, the early VIMOS-VLT $R$-band and radio upper limits are consistent with the expected thermonuclear emission of IMBH TDE and the radio afterglow, respectively \citep[see Figs.~4 and 12 of][]{MacLeod2016}. 

In terms of host-galaxy properties, WD-IMBH TDEs are predicted to occur in irregular dwarf galaxies, globular clusters, and hyper-compact stellar clusters \citep[e.g.,][]{Merritt2009,Jonker2012,Reines2013}, resulting in substantial offsets from the center of their host galaxy. While the host galaxy of XT1 appears consistent with irregular dwarf galaxies, the large positional uncertainty does not permit confirmation of any physical offset.
Overall, a jetted WD-IMBH TDE remains plausible to explain the X-ray light curve shape, duration, and luminosity, the multiwavelength early upper limits, and the host properties of CDF-S XT1. 

\subsubsection{CDF-S XT1: proto-magnetar scenario}

This scenario has been explored previously for CDF-S XT1 \citep[e.g.,][]{Sun2019,Quirola2024}, using the mean photometric redshift reported by \citet{Bauer2017} of ${\approx}2.23$. This model assumed that after a BNS merger, a massive magnetar could be formed \citep{Zhang2001}, emitting isotropic X-ray emission by the internal dissipation of the magnetar wind with a characteristic spin-down luminosity \citep[e.g.,][]{Metzger2014}. For relatively on-axis viewing angles, an observer may detect this magnetar X-ray emission as a plateau superimposed with the expected on-axis gamma-rays and/or short GRB afterglow \citep[e.g.,][]{Rowlinson2010,Rowlinson2013,Lu2015,Gompertz2014}. However, in directions far from the SGRB jet, it may be possible to observe directly the more isotropic X-ray magnetar emission or a version obscured by the neutron-rich ejecta material residing around the remnant \citep[for more details see][]{Yu2013,Sun2017,Quirola2024}. The last case has been proposed to explain CDF-S XT1, wherein the X-ray luminosity of CDF-S XT1 is produced by a magnetar (characterized by a magnetic field, $B_p$, and an initial rotational period, $P_i$), while the initial fast rise is a consequence of time-dependent obscuration produced by the expanding neutron-rich ejecta material (characterized by the mass of the ejecta, $M_{\rm ej}$, and its opacity, $\kappa$). 

We explore this scenario using the new redshift-corrected X-ray light curve of XT1, considering the formalism of \citet[][and references within]{Quirola2024}. The derived magnetar properties ($B_p{\approx}3.2{\times}10^{15}$~G and $P_i{\approx}0.7-2.0$~ms) and ejecta ($M_{\rm ej}{\approx}10^{-4}$~M$_\odot$ and $\kappa{\approx}1$~cm$^2$~g$^{-1}$) are consistent with previous works \citep{Lu2019,Quirola2024}. However, it is important to realize that $P_i$ is close to the ${\sim}1.0$~ms breakup limit of rotating neutron stars \citep[][]{Lattimer2004}, and could trigger the disruption of the massive magnetar. Nevertheless, this limit is unclear because it depends on several poorly-constrained neutron star parameters such as the equation-of-state and loss-energy mechanism \citep[e.g.,][]{Lin202b}. Finally, the low stellar mass of the host of XT1 is uncommon among SGRBs, with just $<5$\% of SGRB hosts having equal or lower stellar masses (see Fig.~\ref{fig:host_comparison}). As such, this scenario for CDF-S XT1 remains viable.


\subsubsection{CDF-S XT1: low-luminousity LGRB scenario}

Another scenario considered to explain FXTs is low-luminosity (LL)-LGRBs, which are typically characterized by longer durations, lower energetics, and lower collimation levels. LL-LGRBs are thought to differ from normal LGRBs due to their partial or complete inability to launch a successful jet \citep{Wang2007,Bromberg2011,Nakar2012}.
Two notable examples of LL-LGRBs are XRF~060218 \citep[$z=0.033$;][]{Pian2006,Campana2006} and XRF~100316D \citep[$z=0.0591$;][]{Starling2011}, which both have similar peak luminosities to CDF-S XT1 (see Fig.~\ref{fig:LC_comparison}). However, the light curve of XT1 has a different shape, especially during the fast-rise phase. 
Some explanations for XRFs suggest that they may be related to shock breakout from choked GRB jets \citep{Campana2006,Bromberg2012,Nakar2012,Irwin2016b}, which produce lower-energy emission \citep[fainter by about four orders of magnitude compared to typical LGRBs, i.e., $10^{46}-10^{48}$~erg~s$^{-1}$;][]{Nakar2015}, which, at high redshift, could mean a fainter or absent of $\gamma$-ray signal. The lack of a $\gamma$-ray counterpart of CDF-S XT1 is consistent with this idea, while the radio upper limits of XT1 remain compatible with the radio luminosity of XRF~060218 and XRF~100316D \citep[see][Fig.~3]{Margutti2013b}.

\begin{figure*}
    \centering
    \includegraphics[scale=0.7]{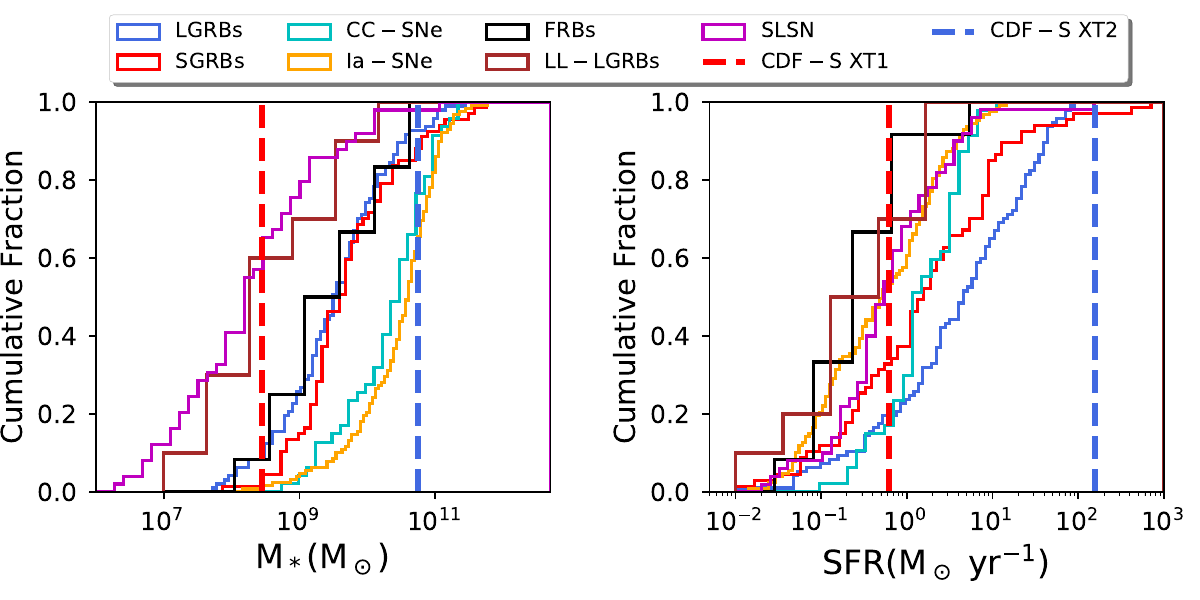}
    \vspace{-0.3 cm}
    \caption{Comparison of the host-galaxy properties of CDF-S XT1 and XT2 with the cumulative distributions of galaxy
    stellar mass (\emph{left panel}) and star-formation rate (\emph{right panel}) for LGRBs \citep{Li2016, Blanchard2016}, 
    SGRBs \citep{Fong2010, Fong2012, Fong2013, Margutti2012b, Sakamoto2013, Berger2013b, Fong2022, Nugent2022}, 
    FRBs \citep{Heintz2020}, 
    CC-SNe and Ia-SNe \citep{Tsvetkov1993,Prieto2008,Galbany2014}, 
    low-luminosity LGRBs \citep[GRB~980425, GRB~020903, GRB~030329, GRB~031203, GRB~050826, GRB~060218, GRB~100316D, GRB~111005A, and GRB~171205A;][]{Christensen2008,Michalowski2014,Levesque2014,Kruhler2017,Wiersema2007,Izzo2017,Wang2018,MichalowskI2018,Arabsalmani2019}, and 
    SLSNe \citep{Schulze2021}. The dashed vertical lines depict the mean values of CDF-S XT1 (red vertical lines) and XT2 (blue vertical lines).}
    \label{fig:host_comparison}
\end{figure*}

Regarding the X-ray evolution, XRF~060218 and XRF~100316D show significant X-ray soft thermal components ($kT{\sim}$0.1--0.2~keV), which cool, and in the case of XRF~060218 also shifts into the optical/UV band at later epochs \citep{Campana2006,Starling2011,Barniol2015}. This behavior is usually interpreted as the break out of a shock driven by a mildly relativistic shell into the dense wind surrounding the progenitor star \citep{Campana2006}. 
However, CDF-S XT1 does not show a softening trend in the X-ray spectra, although this could result from the low number of counts in the FXT rather than intrinsic differences. Furthermore, both XRF~060218 and XRF~100316D were associated with supernovae days later (called SN~2006aj and SN~2010bh, respectively), which also are consistent with the optical upper limits of CDF-S XT1 (see Fig.\ref{fig:kann_plot}, left panel).
The low stellar mass and moderate SFR of XT1's host galaxy are consistent with the host properties of LL-LGRBs (see Fig.~\ref{fig:host_comparison}). Overall, an origin related to LL-LGRB, and thus with massive stars, remains plausible for CDF-S XT1.

\begin{figure*}
    \centering
    \includegraphics[scale=0.68]{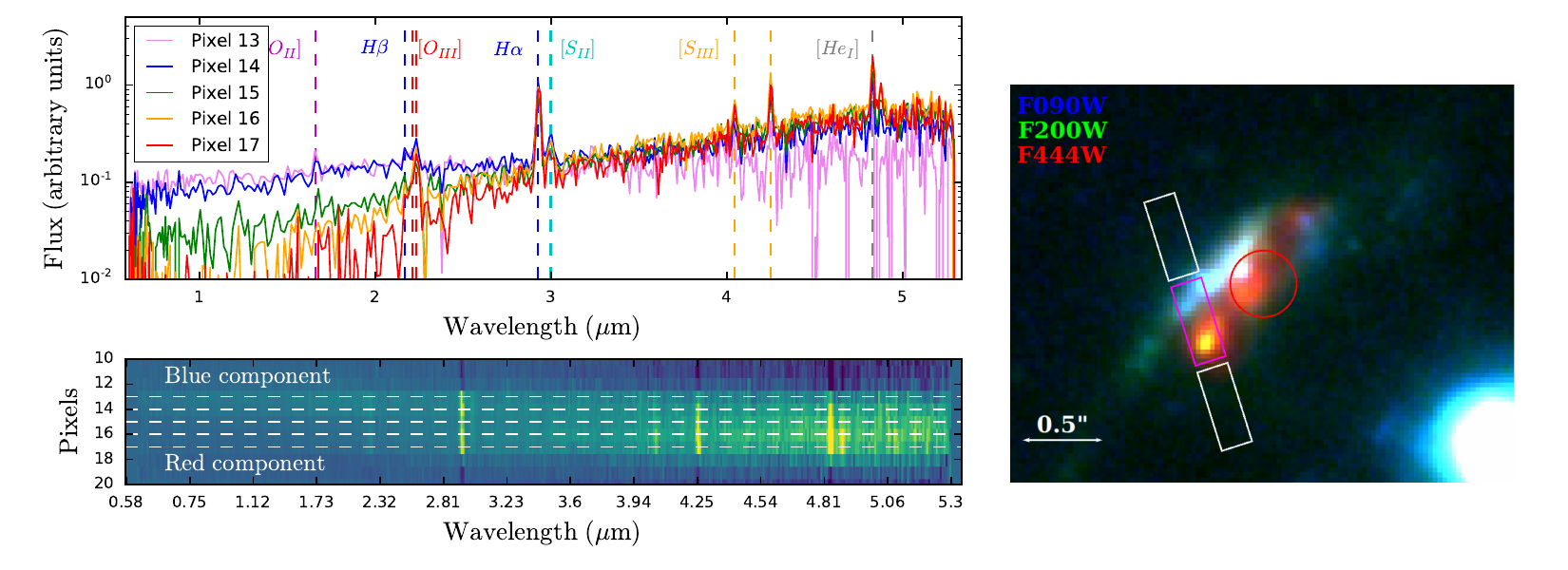}
    \caption{JWST-NIRSpec 2D and 1D spectra per pixel of the host galaxy of CDF-S XT2. Top-left panel: Single-pixel row-extraction spectra through resolved spatial regions of the XT2 host galaxy. 
    Bottom-left panel: 2D spectra in pixel units of the host galaxy of CDF-S XT2. Horizontal dashed lines depict the pixel positions of the single row-extraction spectra shown in the top-left panel. The 2-D is compared to the right panel (North is up). 
    Right panel: RGB (F090W, F200W, and F444W) image of the host of CDF-S XT2, identical in size to Fig~\ref{fig:imaging_XRT_150322}, indicating the NIRSpec slit position.  The magenta rectangle denotes the region of dispersed spectra shown in the bottom-left panel, while the white rectangles are regions to assess the background and the red circle depicts $1-\sigma$ X-ray positional uncertainty of CDF-S XT2.}
    \label{fig:JWST_spectra_pixel}
\end{figure*}

\subsection{CDF-S XT2}

This X-ray transient was also detected serendipitously by \emph{Chandra} (ObsID 16453, ${\approx}$70~ks observation) in the CDF-S field, identified by \citet{Zheng2017} and studied in detail by \citet{Xue2019}. This FXT shows a light curve with a ${\approx}2$~ks plateau phase followed by a power-law decay \citep[$\propto t^{-2}$;][]{Xue2019}, as well as a mild spectral softening evolution (the photon indices before and after the end of the plateau are $\Gamma_{\rm before}{=}1.40{\pm}0.36$ and $\Gamma_{\rm after}{=}2.59_{-0.48}^{+0.51}$, respectively, at 90\% confidence level\footnote{To compute spectral slopes, we split the \texttt{event file} regarding the time when the plateau ends \citep{Quirola2023}, and extract and fit the spectra using the value for the absorption derived from the fit to the full spectrum (see Table~\ref{tab:spectral_para}). We fit both spectral intervals together assuming fixed $N_{\rm H,Gal}$, $N_H$, and redshift.}). This FXT was in the field-of-view (FoV) of \emph{Fermi}-GBM for $\sim1000$~s around the FXT detection time, but no $\gamma$-ray counterpart was found \citep{Xue2019}. Unfortunately, the field was not observed by optical and NIR instruments on dates around the X-ray detection. A host galaxy was identified by \citet{Xue2019} with an apparent magnitude of $m_{\rm F160W}{\approx}24$~AB~mag,\footnote{Source \#4167 in the CANDELS F160W DR1 catalog of \citealp{Guo2013} and \#7699 in the 3D-HST v4.1 catalog of \citet{Skelton2014}.} with a random match probability of only ${\approx}1$\%. 
In light of the new \emph{JWST} data, which includes a revised redshift, we recompute all relevant FXT and host-galaxy parameters below.

\subsubsection{New spectroscopic redshift, energetics, and host properties}

As we explain in \S\ref{sec:nirspec}, the host galaxy of CDF-S XT2 was observed by \emph{JWST}-NIRSpec, resulting in the identification of numerous emission lines and a robust spectroscopic redshift of $z_{\rm spec}{=}3.4598$ (see Fig.~\ref{fig:Spec_XRT_150322}). Although the \emph{JWST}-NIRSpec spectrum by itself is irrefutable proof of the new redshift for the host galaxy, it is important to understand the nature of the previous misidentification. Interested readers can find a forensic assessment and detailed investigation into the origin of the redshift discrepancy in Appendix \S\ref{sec:redshift_discre}. We summarize the most salient details below.

A "robust" spectroscopic redshift of $z=0.7382$ for the host galaxy of CDF-S XT2 was first reported by \citet{Balestra2010}, based on VLT-VIMOS observations taken on 2004-11-17 (Cesarsky PI), which was subsequently adopted in \citet{Xue2019} and thereafter. However, no emission lines appear in a spectral extraction of more recent VLT-Multi Unit Spectroscopic Explorer (MUSE) data \citep[Wisotzki PI;][]{Inami2017,Herenz2017} at the position of XT2's host between 4,500-9,500~\AA~(see Fig.~\ref{fig:old_spec_comp}, as well as \S\ref{sec:redshift_discre} for further technical details), despite similar sensitivities. This, in conjunction with the new high-quality, spatially resolved \emph{JWST} spectra, suggests that the initial VLT-VIMOS spectral extraction was contaminated by the primary target in the slit mask, an X-ray AGN (CANDELS \#4210) in an elliptical galaxy at $z=0.7396$, lying $\sim$2\farcs0 to the southwest. 
Notably, the non-detection of the XT2 host in VLT-VIMOS {\it U}-band imaging (see Fig.~\ref{fig:imaging_XRT_150322_AP}) can be interpreted as a drop-out beyond the Lyman limit, consistent with $z_{\rm spec}=3.4598$. The Lyman-$\alpha$ line is not identified in the MUSE spectra (the MUSE spectrum of XT2 host and the collapsed image of the Ly-$\alpha$ emission line are shown in Figs.~\ref{fig:old_spec_comp} and \ref{fig:slit_image}, respectively). Unfortunately, the available Spitzer-InfraRed Array Camera (IRAC) imaging between 3.6 and 8.0~$\mu$m is not sufficiently informative due to the strong blending between the bright emission from the host of XT2 and the nearby AGN CANDELS \#4210 \citep[IRAC has a pixel scale of ${\sim}1\farcs22$~pix$^{-1}$;][]{Fazio2004}.

Adopting this new spectroscopic redshift, the peak luminosity and projected physical offset of CDF-S XT2 are $L_{\rm X,peak}{\approx}1.4{\times}10^{47}$~erg~s$^{-1}$ and $\delta R=0.96{\pm}2.13$~kpc, respectively. The former has increased by a factor of ${\sim}45$ compared to the value reported by \citet{Xue2019}, while the latter is now consistent with arising from the central, obscured portion of the complex host galaxy (see Figs.~\ref{fig:imaging_XRT_150322} and \ref{fig:imaging_XRT_150322_AP}). On the other hand, the random match probability between CDF-S XT2 and its host is 0.091\%.
The new host galaxy parameters of CDF-S XT2 are $\log(M_*){=}10.74_{-0.06}^{+0.01}$, $\log(\rm SFR){=}2.20_{-0.01}^{+0.01}$ and $\log(\rm sSFR){=}-8.54_{-0.02}^{+0.06}$, indicating that it is a powerful, obscured star-forming galaxy. Compared with the previous values \citep{Santini2015,Xue2019}, the stellar mass and SFR have increased by factors of ${\sim}$45 and 200, respectively.
Comparisons of the X-ray light curves and host-galaxy properties between XT2 and several different transients are depicted in Figs.~\ref{fig:LC_comparison} and ~\ref{fig:host_comparison}, respectively. Below, we first examine the unique properties of the host and then explore potential progenitor scenarios for CDF-S XT2 based on the X-ray and host properties. Notably, the association of XT2 with a dust-obscured, strongly star-forming galaxy lends support to progenitor channels involving massive stars. This is also indirectly supported by the high intrinsic hydrogen column density from X-ray data, which implies a high extinction of $E(B-V)=4.3_{-1.6}^{+2.3}$~mag (considering a SMC-like environment).

\subsubsection{One or two galaxies}

Figure~\ref{fig:imaging_XRT_150322_AP} highlights the strong color gradient observed across the host of CDF-S XT2, as well as the clumpy nature of newly detected emission probed by the NIR-MIR bands of JWST. Figure~\ref{fig:JWST_spectra_pixel} (left panel) shows the 2D dispersed spectra across this color gradient (approximately consistent with the galaxy minor axis), along with a comparison of the 1D single-row extracted spectra of this spatially resolved region. It is clear from both the individual HST, NIRCam, and MIRI filter image cutouts and the spatially-resolved spectra that the host galaxy exhibits a diverse distribution of stellar populations and/or variable dust extinction. In particular, the single-row spectra reveal two fairly distinct components, one bluer (spectra for pixels 13 and 14) and another redder (spectra for pixels 15--17). However, it is clear that both components exhibit the same emission lines at ${\gtrsim}2~\mu$m (the redder component does not show a robust [O$_{II}$] line), i.e., even though both components might arise from different stellar populations, they share a common redshift. We do not see any unidentified emission lines, implying galaxies at two distinct redshifts. The $R\sim100$ spectral resolution of the NIRSpec data does not provide strong constraints ($\lesssim$2500 km~s$^{-1}$ for H$\alpha$) on differences between nearby overlapping galaxies or expected galaxy-scale velocity gradients, which might provide additional insight into the dynamical evolution of the host. Considering the overall morphology of the host, we can now reinterpret the bluer (\emph{HST}) filters, where only one side appears and the emission appears clumpy. In the NIR-MIR filters from \emph{JWST}, on the other hand, the host takes on a somewhat more symmetrical form, albeit also with stronger clumps. The fact that an unambiguous galaxy core in MIRI images at \emph{JWST} suggests that both components belong to the same obscured galaxy. We calculate an offset of ${\sim}0\farcs3$ (i.e., ${\sim}2$~kpc) between the blue and red star-forming components, which also have been seen in other high-redshift dusty galaxies such as GN20 \citep{Carilli2011,Hodge2015,Colina2023} and  A2744-ID02 \citep{Kokorev2023}. These offsets are often interpreted as the result of gravitational encounters or mergers, although that interpretation is not ironclad.

\begin{figure}
    \centering
    \includegraphics[scale=0.7]{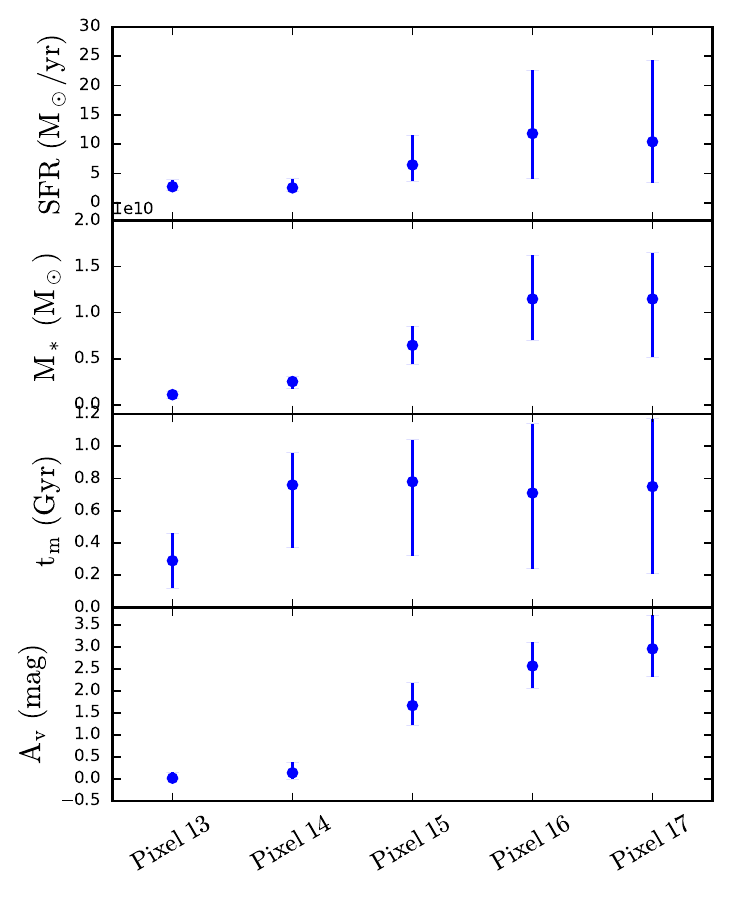}
    \caption{JWST single-pixel spectral parameters obtained by our SED fitting using the \texttt{Bagpipes} package \citep{Carnall2018,Carnall2019} from pixel 13 to pixel 17 assuming the SED model from Table~\ref{tab:SED_model}. The figure depicts the star-formation rate (top panel), stellar mass (2nd panel, in units of $10^{10}$\,$M_{\odot}$), mass-weighted age (third panel), and dust attenuation (bottom panel) per pixel. The uncertainties were computed from the posterior distributions at a 99.9\% confidence level.
    }
    \label{fig:pixel_comparison}
\end{figure}

To demonstrate the different stellar properties of both components, we carry out SED fitting (using \texttt{Bagpipes}) on the JWST single-row spectra from pixels 13 to 17 (the same as Fig.~\ref{fig:JWST_spectra_pixel}) considering the same SFH model and parameters as Table~\ref{tab:SED_model}. Figure~\ref{fig:pixel_comparison} shows the derived SFR, stellar mass and mass-weighted age, and dust attenuation obtained per \emph{JWST} single-pixel spectra with 99.9\% confidence uncertainties. We observe strong gradients in the parameters --- i.e., the SFR, stellar mass, and dust attenuation evolve from lower values in the bluer component to higher for the redder component --- although the errors (at $>$99.9\% confidence) on the individual measurements remain large, due in part to the limited wavelength coverage of the spectra. The variations in both $A_{\rm V}$ and stellar mass are significant, implying that the host galaxy SFR and stellar population age remain relatively consistent across the minor axis, and that the strong color gradient is likely driven by inhomogeneous dust attenuation and a possible clumpy stellar mass distribution.\footnote{Some caution is warranted here to not overinterpret these results, due to known degeneracies between age, mass, and $A_{V}$, which appear as larger parameter error bars associated with the redder spectra.}.
The $A_V$ attenuation in the red component is comparable with the high attenuation in other HST and JWST-dark or optically faint and dark galaxies \citep{Wang2019,Smail2021,Jin2022,Xiao2023,Barrufet2024}.


\subsubsection{CDF-S XT2: SN SBO scenario}

Similar to CDF-S XT1, the peak X-ray luminosity inferred for XT2 in this work is $L_{\rm X,peak}{\sim}10^{47}$~erg~s$^{-1}$. Its peak luminosity is ${\approx}$4~order of magnitudes higher than the expected for SBOs \citep{Waxman2017,Goldberg2022} and the well-known SBO XRF\,080109-SN~2008D \citep[$L_{\rm X,peak}{\sim}10^{43}$~erg~s$^{-1}$, see Fig.~\ref{fig:LC_comparison};][]{Soderberg2008}. As such, we rule out the SBO progenitor scenario.

\subsubsection{CDF-S XT2: GRB afterglow scenario}

Several LGRBs \citep[e.g.,][]{Lyons2010,Beniamini2020} and SGRBs \citep[e.g.,][]{Rowlinson2010,Rowlinson2013,Gompertz2014} show a plateau in their X-ray afterglow. However, only $<$10\% have plateau luminosities ${\lesssim}10^{47}$~erg~s$^{-1}$ (consistent with CDF-S XT2). Some authors claim that the plateau origin is related to some engine activity, which we discuss below. The power-law decay after the plateau remains crudely consistent with the expected X-ray afterglow of GRBs \citep[$\propto t^{-1.2}$;][]{Evans2009}, while the spectral softening in XT2 also has been seen in some GRB afterglows \citep[e.g., GRB~130925A;][]{Zhao2014}. 

It is important to discuss here the implications of the gamma-ray upper limits as they relate to an on-axis scenario. \citet{Xue2019} was unable to find any significant source-like gamma-ray emission signal above the background in the \emph{Fermi}-GBM instrument (using the detectors $n4$, $n5$, and $b0$) around the X-ray trigger, with a $1-10^4$~keV peak flux upper limit of $\approx6\times10^{-7}$~erg~cm$^{-2}$~s$^{-1}$. In light of the new redshift, this implies an isotropic rest-frame peak luminosity of $<1.7\times10^{52}$~erg~s$^{-1}$. This upper limit lies near the $\approx25$ percentile of the distribution of isotropic peak luminosities
for a sample of 150 GRBs (spanning a range of $0.1<z<5$, including 12 short-hard bursts) detected by \emph{Konus-Wind} \citep{Tsvetkova2017}, suggesting a potential association with that population. However, if we consider only the GRBs with similar redshifts (i.e., 12 GRBs with $z>3$), our gamma-ray upper limit lies well outside the distribution, but it could be an observational bias. 
Similarly, comparing the gamma-ray upper limit and the X-ray luminosity with GRBs \citep[e.g.,][]{DAvanzo2012,Rossi2022}, we identify that both parameters are consistent with moderate to low-luminous GRBs. 
This, along with the lack of a gamma-ray detection \citep{Xue2019,Quirola2023} appears poorly consistent with the bright on-axis afterglow scenario. Although it is not possible to discard this scenario completely, we believe it unlikely, leaving us to consider more likely an off-axis one and low-luminous GRBs.
On the other hand, the relatively high luminosity and plateau duration of XT2 effectively exclude high off-axis afterglows such as LGRB SN 2020bvc \citep[with viewing angle $\theta_{\rm obs}{\approx}23$~deg;][]{Izzo2020}, and SGRB GRB 170817A \citep[with viewing angle $\theta_{\rm obs}{\approx}23$~deg;][]{Nynka2018,DAvanzo2018,Troja2020,Troja2022}, both of which are substantially fainter ($L_{\rm X,peak}{\approx}3{\times}10^{41}$~erg~s$^{-1}$) than CDF-S XT2.
Recently, \citet{Wichern2024} explored an association between FXTs and off-axis SGRB afterglows and concluded that it is difficult to reproduce the luminosity, temporal indices, and durations of FXTs in the off-axis SGRB afterglow interpretation in a self-consistent way (i.e., at the same viewing angle), with the closest fits arising from only slightly off-axis angle models.
A mildly off-axis LGRB scenario remains an open possibility to explore, especially given the high star-formation rate of its host galaxy, which reinforces a collapsar origin (over a merger one).

\subsubsection{CDF-S XT2: TDE scenario}

Another potential progenitor channel is TDEs. Based on arguments similar to those of the XT1 case (i.e., luminosity, event duration, and lack of variability), we discard scenarios involving normal and jetted SMBH-TDEs. In the case of WD-IMBH TDEs, the simple model developed by \citet{Peng2019} can mimic the XT2 light curve for fallback and accretion timescale parameters of ${\sim}$2 and 500~s, respectively. However, the high luminosity remains impossible to explain with a thermal accretion model. Thus, the only viable option is a jetted WD-IMBH TDE, which can potentially accommodate the high luminosity \citep{MacLeod2014,MacLeod2016}. Unfortunately, the lack of contemporaneous multiwavelength constraints does not permit us to explore this channel in more detail. One point against a jetted WD-IMBH TDE is the properties of XT2's host, which implies a link to massive stars. In contrast, WDs require considerable time to form and gravitationally couple with IMBHs, which should occur more often in irregular dwarf galaxies, globular clusters, and hyper-compact (older) stellar clusters \citep{Merritt2009,Jonker2012,Reines2013}. As such, we consider the jetted WD-IMBH TDE scenario to remain possible but improbable.

\subsubsection{CDF-S XT2: proto-magnetar scenario}

Similar to CDF-S XT1, the proto-magnetar scenario can be applied to XT2, where its light curve suggests an association with a proto-magnetar emission formed after the BNS merger \citep{Zhang2013,Xue2019,Sun2019}, viewed along a direction without neutron-rich ejecta material in the line-of-sight \citep[or a low amount of ejecta material, e.g., $M_{\rm ej}\lesssim10^{-4}$M$_\odot$;][]{Sun2019,Quirola2024}. Indeed, \citet{Xue2019} considered this interpretation as the most plausible for CDF-S XT2. However, the updated redshift leads to an increase in the peak luminosity by a factor of ${\approx}45$ and a decrease in the plateau duration by a factor of ${\approx}$4.5. Thus, we reexamine this model \citep[following the method developed by][]{Yu2013,Quirola2024}, which requires new magnetic field and initial period parameters of $B_p{\sim}1.3{\times}10^{15}$~G and $P_i{\sim}1$~ms, respectively, to fit the light curve. Although these parameters allow the model to still mimic XT2's light curve reasonably well, the magnetic field is now a factor of ${\sim}1.5$ lower, while the rotation rate is ${\sim}$3 faster. 
Those values could lead to tensions with the forbidden ranges discussed in the literature \citep[e.g.,][]{Sun2019,Quirola2024}, in particular, the initial period necessary to achieve a stable magnetar \citep{Lattimer2004}; however, our current knowledge about neutron star physics does not permit us to rule out this channel. \citet{Xue2019} pointed to the physical offset as a key argument in support of the magnetar model and its association with SGRBs. Nevertheless, the new JWST observations place XT2 near the center of its host galaxy (see Fig.~\ref{fig:imaging_XRT_150322_AP}), and potentially aligned with the strongest star-forming clumps (see Fig~\ref{fig:JWST_spectra_pixel}, right panel). This smaller offset provides less clarity about the progenitor of XT2.
While the host properties (especially the high SFR) remain within the bounds of known SGRB galaxies (see Fig.~\ref{fig:host_comparison}), just a small portion of the SGRB hosts have equal or higher SFR (${\lesssim}5$\%) and stellar mass (${\lesssim}10$\%) compared to XT2's host. We thus view the scenario wherein the FXT arises from a proto-magnetar that formed after the BNS merger with a lower likelihood than before. It may also be interesting to explore a magnetar model related to collapsars (based on the properties of its host).

\subsubsection{CDF-S XT2: low-luminousity-LGRB scenario}

Based on Figure~\ref{fig:LC_comparison}, the X-ray light curve of CDF-S XT2 now appears to share some similarities with X-ray flashes (XRFs). XRFs are interpreted as the shock breakout from choked GRB jets \citep{Campana2006,Bromberg2012,Nakar2012,Barniol2015,Irwin2016b}, leading to the production of less luminous and/or energetic emission compared to typical LGRBs \citep[they lie in the lower left corner in the Amati relation;][]{Amati2006,Willingale2017,Martone2017} and a general absence of any highly relativistic (i.e., $\gamma$-ray) emission.
The lack of a bright $\gamma$-ray counterpart of CDF-S XT2 \citep{Xue2019} would be consistent with this scenario. Comparing X-ray light curves, XRF 100316D produced a plateau with duration and luminosity of ${\approx}1$~ks and ${\approx}10^{46}$~erg~s$^{-1}$, respectively, interpreted as either quasi-spherical breakout emission \citep{Nakar2012} or synchrotron emission from a dissipating Poynting flux-dominated outflow powered by a magnetar ($P_i{\sim}10$~ms and $B_p{\sim}3{\times}10^{15}$~G) once the gravitational collapse occurs \citep{Fan2011}. 
XRF 080109 also exhibited a plateau phase, albeit somewhat longer (${\approx}3$~ks) and less luminous [${\approx}$(3--7)${\times}10^{46}$~erg~s$^{-1}$] than XRF 100316D. 
Notably, XRF 100316D's plateau phase is nearly identical to that of XT2 in duration and luminosity (see Fig.~\ref{fig:LC_comparison}). Unfortunately, the low number of counts of CDF-S XT2 does not permit us to explore the possibility of a thermal component in the X-ray spectra, as observed in XRFs 060218 and 100316D.
In terms of spectral evolution, XRF 060218 shows an X-ray thermal softening evolution ($kT{\sim}0.1-0.2$~keV) \citep{Campana2006,Soderberg2006,Fan2006b}, while XRF 100316D does not present a significant thermal softening at early epochs \citep[although some softening does appear at ${>}10^4$~s;][]{Starling2011,Fan2011,Margutti2013}. CDF-S XT2 also exhibits mild softening, but only at the 90\% confidence level. Both XRFs 060218 and 100316D were found to have associated SNe at later times. There are no contemporaneous constraints for CDF-S XT2, although, at its distance, we would not expect any detection with current facility sensitivities. 

The high SFR of the host galaxy of CDF-S XT2 implies that massive stars are plausible progenitors, thus favoring the LL-LGRB scenario. However, the host parameters of XT2 lie near the upper bound of the distribution among known LL-LGRB hosts (see Fig.~\ref{fig:host_comparison}). This could be a selection effect, as LL-LGRBs produce lower luminosity and/or energy emission, which has restricted detections to the relatively nearby universe to date,
and hence CDF-S XT2 could be the first high redshift LL-LGRB detected. Overall, this scenario remains one of the most plausible ones.

\begin{figure}
    \centering
    \includegraphics[scale=0.6]{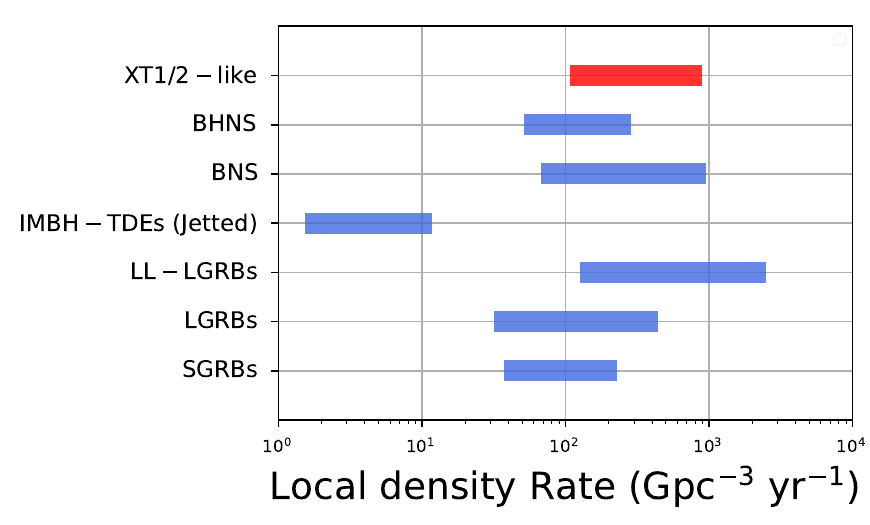}
    \vspace{-0.3 cm}
    \caption{Estimated local density rate comparison between XT1- and XT2-like objects (derived in this work) and 
    the merger rate of neutron star and BH systems \citep[$\rho_0^{\rm BHNS}{=}130_{-69}^{+112}$~Gpc$^{-3}$~yr$^{-1}$;][]{Abbott2021b}, the merger rate of BNS systems \citep[$\rho_0^{\rm BNS}{=}320_{-240}^{+490}$~Gpc$^{-3}$~yr$^{-1}$;][]{Abbott2021a}, jetted WD-IMBH TDEs \citep[$\rho_0^{\rm IMBH}{=}1.8-10$~Gpc$^{-3}$~yr$^{-1}$ 
    ;][]{Malyali2019,Tanikawa2021}, LL-LGRBs \citep[$\rho_{\rm 0,LL-LGRBs}{=}$150~yr$^{-1}$~Gpc$^{-3}$, corrected by $f_b^{-1}{=}1$--14;][]{Liang2007,Zhang_book_2018}, LGRBs \citep[$\rho_{\rm 0,LGRBs}{=}$0.75~yr$^{-1}$~Gpc$^{-3}$, corrected by $f_b^{-1}{=}50$--500;][]{Wanderman2010,Palmerio2021} and SGRBs \citep[$\rho_{\rm 0,SGRBs}{=}$1.75~yr$^{-1}$~Gpc$^{-3}$, corrected by $f_b^{-1}{=}25$--110;][]{Guetta2005b,Wanderman2015}.
    }
    \label{fig:rate_comparison}
\end{figure}

\subsection{Event rate density}

In this section we estimate the event rate density of the XT1- and XT2-like objects (considering all \emph{Chandra} observations between 2000 and 2022), assuming that both FXTs originate from the same progenitor channel based on the similar peak luminosities. A systematic search for FXTs in two decades of \emph{Chandra} data, developed by \citet{Quirola2022} and \citet{Quirola2023}, identified 22 FXTs: $i)$ five associated with nearby galaxies ($\lesssim$100~Mpc) and $ii)$ 17 FXTs with further away galaxies (including CDF-S XT1 and XT2). We exclude the former group, which is likely related to low-luminosity (${\lesssim}10^{39}$~erg~s$^{-1}$) objects such as X-ray binaries, and work only with higher luminosity events \citep{Quirola2022}. For computing the event rate, we consider not only CDF-S XT1 and XT2 but also FXTs with similar features to both FXTs: light curve shape, duration, and optically faint hosts (i.e., $m_r{\gtrsim}24.5$~mag) because of the stronger potential association with similar high redshifts. We identified seven additional FXTs (XRT~030511, XRT~100831 and XRT~110919 from \citealp{Quirola2022} and XRT~140507, XRT~191127, XRT~170901 and XRT~210423 from \citealp{Quirola2023}) that share such similarities with CDF-S XT1 and XT2; therefore, we consider a total of nine XT1- and XT2-like FXTs (i.e., ${\sim}53$\% of the aforementioned \emph{Chandra} distant FXTs).

The event rate density, in general, evolves with cosmic time as $\rho(z)=\rho_0f(z)$ \citep{Sun2015}, where $\rho_0$ is the local event rate density and $f(z)$ is a normalized function that describes the cosmic evolution. The estimated local event density could be derived from $N=\rho_0V_{\rm max}\Omega T/4\pi$ \citep{Zhang2018,Sun2019}, where $V_{\rm max}$ represents the maximum co-moving volume at which XT1- or XT2-like objects could be detected by \emph{Chandra}, $\Omega$ and $T$ depict the \emph{Chandra} FoV and on-sky exposure time, respectively, and $N$ is the number of sources. The factor $\Omega T$ is derived as $\Omega T=\Sigma_i\omega_i t_i$, where $\omega_i$ and $t_i$ are the ACIS-I/S FoV and exposure time per individual observation, respectively, taken from 2000 to 2022 \citep[analyzed by][]{Quirola2022,Quirola2023}.
\emph{Chandra} can detect and identify FXT events as faint as $F_{\rm X,peak}{\sim}2.5{\times}10^{-13}$~erg~cm$^{-2}$~s$^{-1}$ \citep[based on the simulations by][]{Quirola2022}. Adopting a peak luminosity of $L_{\rm X,peak}{\sim}10^{47}$~erg~s$^{-1}$, \emph{Chandra} can then detect XT1 and XT2-like objects to redshifts up to $z_{\rm max}=5.8$, equating to a maximum co-moving volume of $V_{\rm max}{\sim}2300$~Gpc$^3$. This translates to an estimated local density rate of $\rho_0^{\rm XT1/2}{=}352_{-224}^{+398}$~Gpc$^{-3}$~yr$^{-1}$ (at 99\% confidence level).

Figure~\ref{fig:rate_comparison} compares the estimated local volumetric density rate derived for XT1- and XT2-like objects and other transients: 
the merger rates of NS-BH \citep{Abbott2021b} and BNS \citep{Abbott2021a} systems; the rate of 
jetted (beamed) versions of WD-IMBH TDE systems \citep{Malyali2019,Tanikawa2021}, 
the rates of off-axis LL-LGRBs \citep{Liang2007,Zhang_book_2018}, 
LGRBs \citep{Wanderman2010}, and SGRBs \citep{Wanderman2015}. 
To convert beamed objects to off-axis versions, or isotropic objects to beamed ones, we adopt a beaming correction factor, $f_b$, the values of which are taken from the literature as indicated in Fig.~\ref{fig:rate_comparison}. Given the large uncertainties in these beaming factors, which effectively establish the ranges shown in Fig.~\ref{fig:rate_comparison}, we do not factor in other errors. We do not consider SBOs or unbeamed WD-IMBH TDEs here, since they have already been strongly excluded based on their low luminosities \citep[they are shown in][]{Quirola2023}. The rate of XT1+XT2-like objects appears compatible with at least a portion of the phase space of all the possible progenitor channels considered, aside from jetted WD-IMBH TDEs. We caveat here that the predicted rate of IMBH TDEs shows large uncertainties in the literature \citep[e.g.,][]{MacLeod2016,Maguire2020,Malyali2019,Tanikawa2021}, such that rejecting this channel is not possible.

\section{Conclusions and future work} \label{sec:conclusions}

We report new constraints on the FXTs CDF-S XT1 \citep{Luo2014,Bauer2017} and XT2 \citep{Zheng2017,Xue2019}, based on recent JWST NIRCam and MIRI photometry, as well as NIRspec spectroscopy in the case of CDF-S XT2, of their host galaxies. 

The host of CDF-S XT1 is now constrained to lie at $z_{\rm phot}=2.76_{-0.13}^{+0.21}$ \citep[greatly reducing the uncertainty of the previous photometric redshift of $2.23^{+0.98}_{-1.84}$;][]{Bauer2017}, implying a host absolute magnitude $M_R=-19.14$~mag, $M_*\approx2.8{\times}10^8$~M$_\odot$, SFR$\approx0.62$~M$_\odot$~yr$^{-1}$. These properties lie at the upper end of previous estimates \citep{Bauer2017}, leaving CDF-S XT1 with a peak X-ray luminosity of $L_{\rm X,peak}{\approx}2.8{\times}10^{47}$~erg~s$^{-1}$ and projected physical offset of $\delta R=1.2{\pm}2.4$~kpc. At this redshift, the early (${\approx}80$~min post X-ray trigger) upper limit from VLT-VIMOS translates to $M_R{>}-21.1$~mag (which is inconsistent with the majority of on-axis GRB afterglows). We argue that the best progenitor scenario for XT1 is a low-luminosity GRB, where the X-rays are associated with the shock breakout of a choked jet, although we cannot fully rule out a proto-magnetar interpretation (formed after the merger of two neutron stars) and jetted WD-IMBH TDE channels.

In the case of CDF-S XT2, JWST imaging reveals a new highly obscured component of the host galaxy, previously absent from HST imaging, while NIRspec spectroscopy now securely places the host at $z_{\rm spec}=3.4598{\pm}0.0022$. The discrepancy with the previously measured redshift \citep[first reported by][]{Balestra2010} came from a contaminated VLT-VIMOS spectral extraction due to an X-ray AGN elliptical galaxy at  $z=0.7396$ (CANDELS \#4210) lying ${\sim}2''$ to the southwest of XT2's host. The new spectroscopic redshift combined with SED-fitting yields a FXT peak luminosity $L_{\rm X,peak}{\approx}1.4{\times}10^{47}$~erg~s$^{-1}$, host absolute magnitude of $M_R=-21.76$~mag, $M_*\approx5.5{\times}10^{10}$~M$_\odot$, SFR$\approx160$~M$_\odot$~yr$^{-1}$, and mass-weighted stellar age of $0.38{\pm}0.02$~Gyr.
The spatially resolved NIRSpec spectra probe the strong aforementioned \emph{HST}-to-\emph{JWST} color gradient across a projected offset of ${\sim}0.2$~kpc. Based on SED-fitting, the gradient appears to arise primarily from variable dust extinction, potentially as the result of gravitational encounters or mergers. Based on the revised X-ray and host properties, the similarity to X-ray flash event light curves, small host offset, and high, dusty SFR of the host (consistent with the large hydrogen column density observed toward the transient), a low-luminosity collapsar progenitor appears to be a good fit for CDF-S XT2. Nonetheless, based on our current knowledge of the physics of neutron stars, a proto-magnetar model is not ruled out.

Although the current \emph{HST} and \emph{JWST} observations push the boundaries of what is currently possible to understand the hosts of CDF-S XT1 and XT2, they are not enough to obtain a complete understanding of both objects. Finding similar objects to CDF-S XT1 and XT2 with sensitive time-domain observatories such as the new \emph{Einstein Probe} satellite, especially at closer distances where contemporaneous multiwavelength (and perhaps multi-messenger) constraints can be obtained, remains essential to differentiate emission mechanisms and provide the last word on their nature.

\begin{acknowledgements}

We thank Fengwu Sun and Gabriel Brammer for helpful early discussions about available JWST imaging and photometric redshifts for the hosts of CDF-S XT1 and XT2.
The scientific results reported in this article are based on observations made by the {\it Chandra} X-ray Observatory, the ESO Science Archive Facility, and the JADES - JWST Advanced Deep Extragalactic Survey. This research has made use of software provided by the {\it Chandra} X-ray Center (CXC). We acknowledge support from: the European Research Council (ERC) under the European Union’s Horizon 2020 research and innovation programme Grant agreement No. 101095973 (JQ-V, PGJ); the IAU-Gruber foundation fellowship (JQ-V); ANID - Millennium Science Initiative Program - ICN12\_009 (FEB, JQ-V), CATA-BASAL - FB210003 (FEB), FONDECYT Regular - 1241005 (FEB), the Penn State Eberly Endowment (WNB), and the NSFC 12025303 (YQX).

\end{acknowledgements}





\bibliographystyle{aa}
\bibliography{Quirola-Vasquez2019}



\appendix

\section{early afterglow in the on-axis scenario}\label{sec:deceleration}

\begin{figure*}[h!]
    \centering
    \includegraphics[scale=0.29]{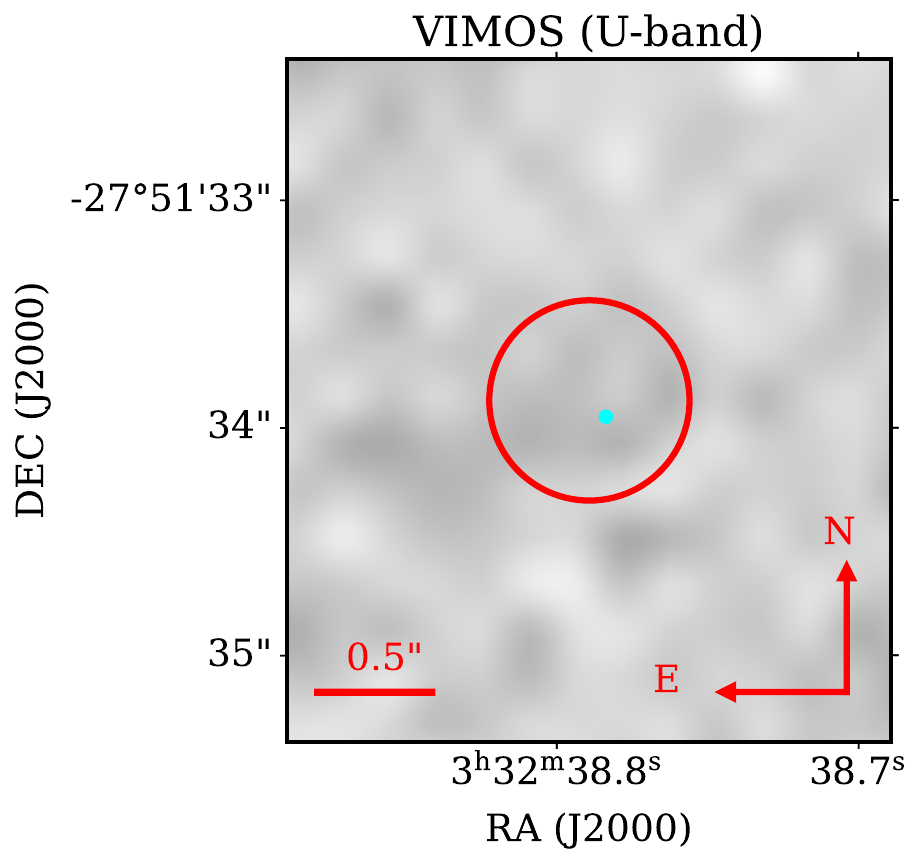}
    \includegraphics[scale=0.29]{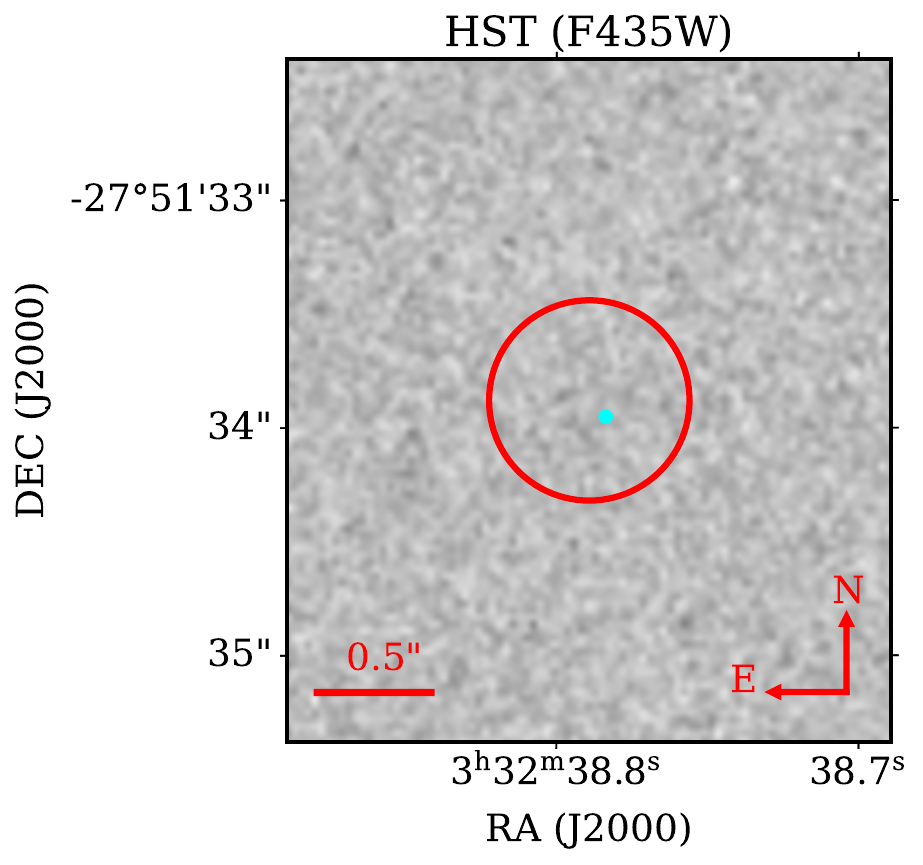}
    \includegraphics[scale=0.29]{XT1_F606W.pdf}
    \includegraphics[scale=0.29]{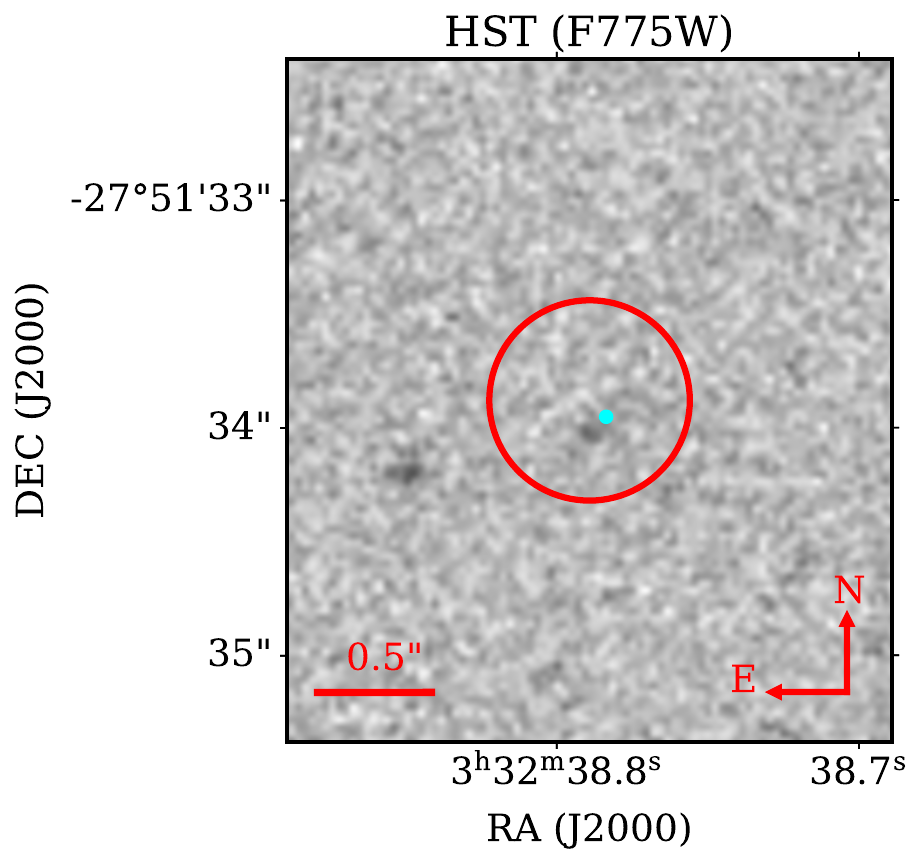}
    \includegraphics[scale=0.29]{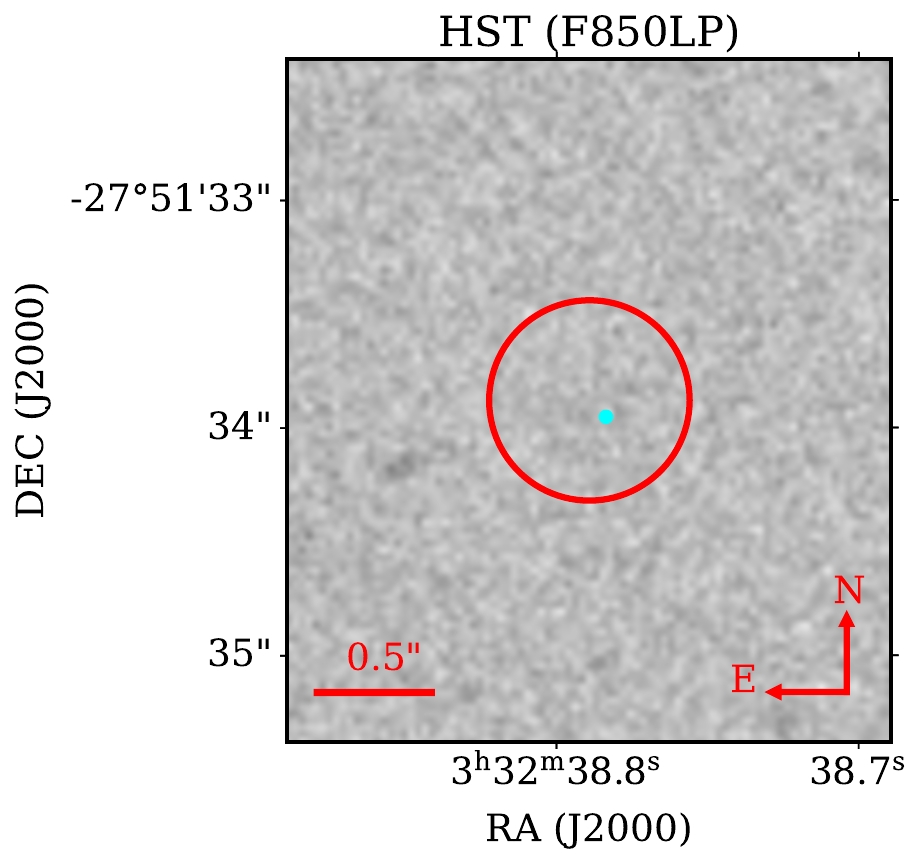}
    \includegraphics[scale=0.29]{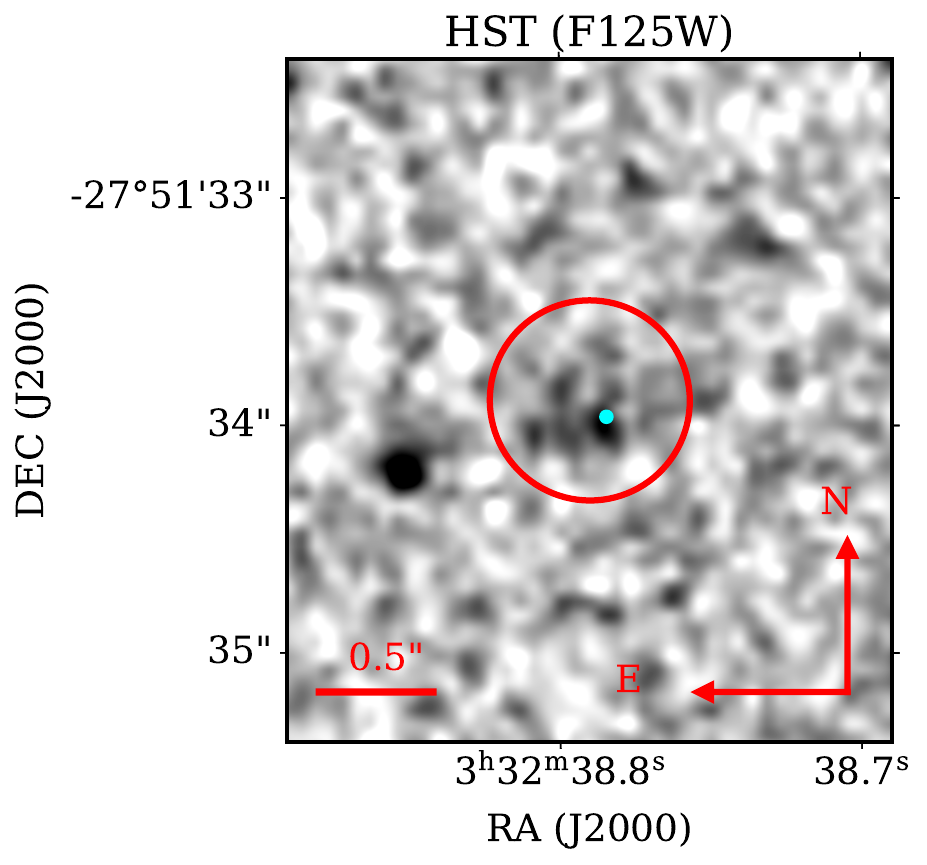}
    \includegraphics[scale=0.29]{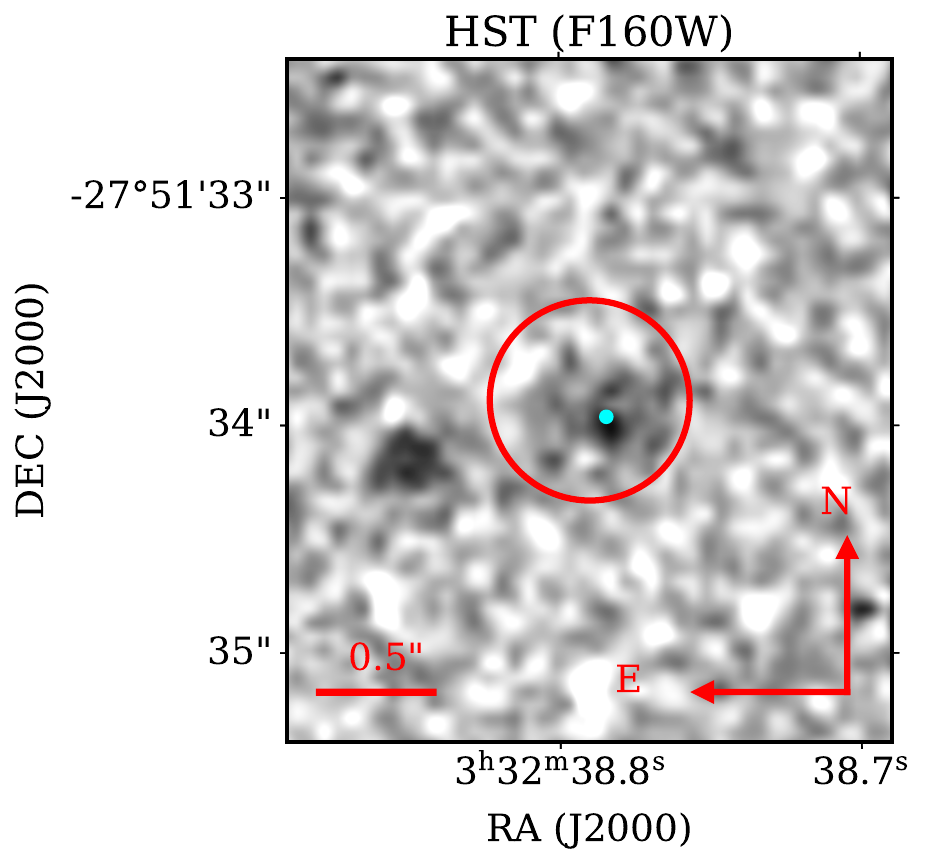}
    \includegraphics[scale=0.29]{XT1_F090W.pdf}
    \includegraphics[scale=0.29]{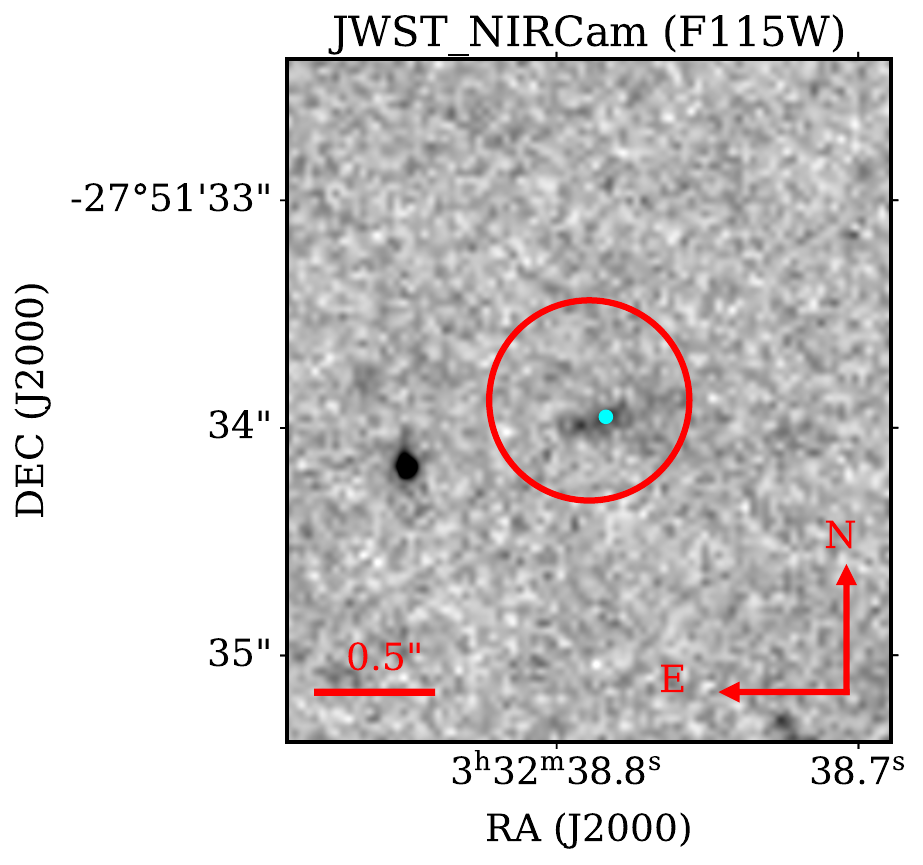}
    \includegraphics[scale=0.29]{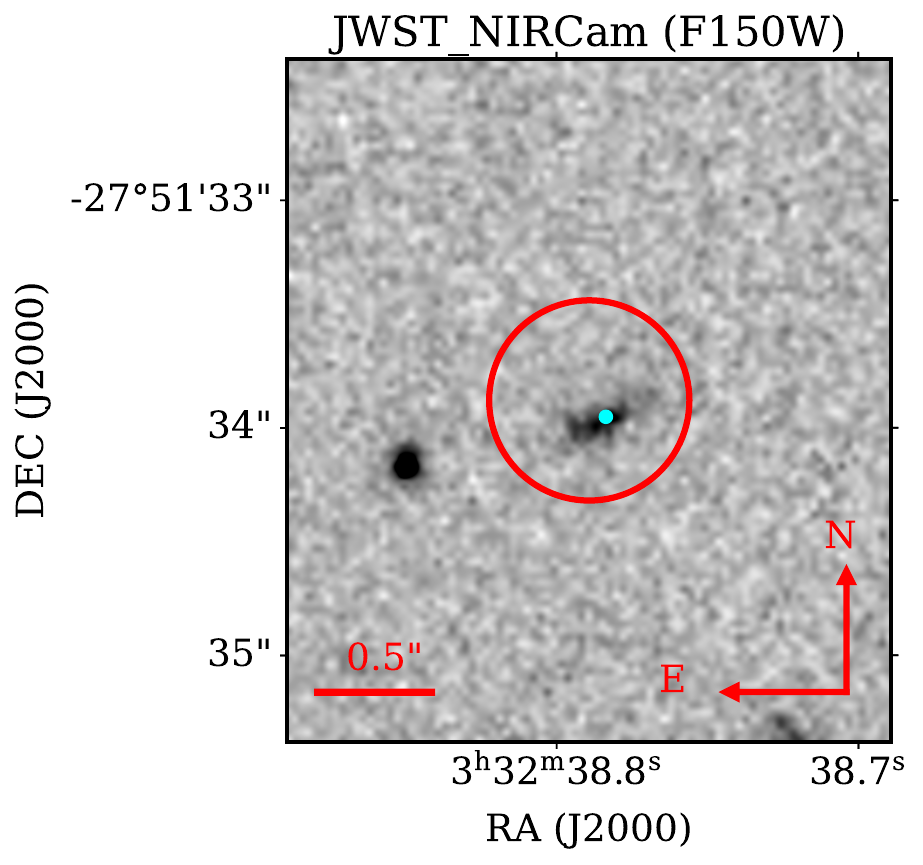}
    \includegraphics[scale=0.29]{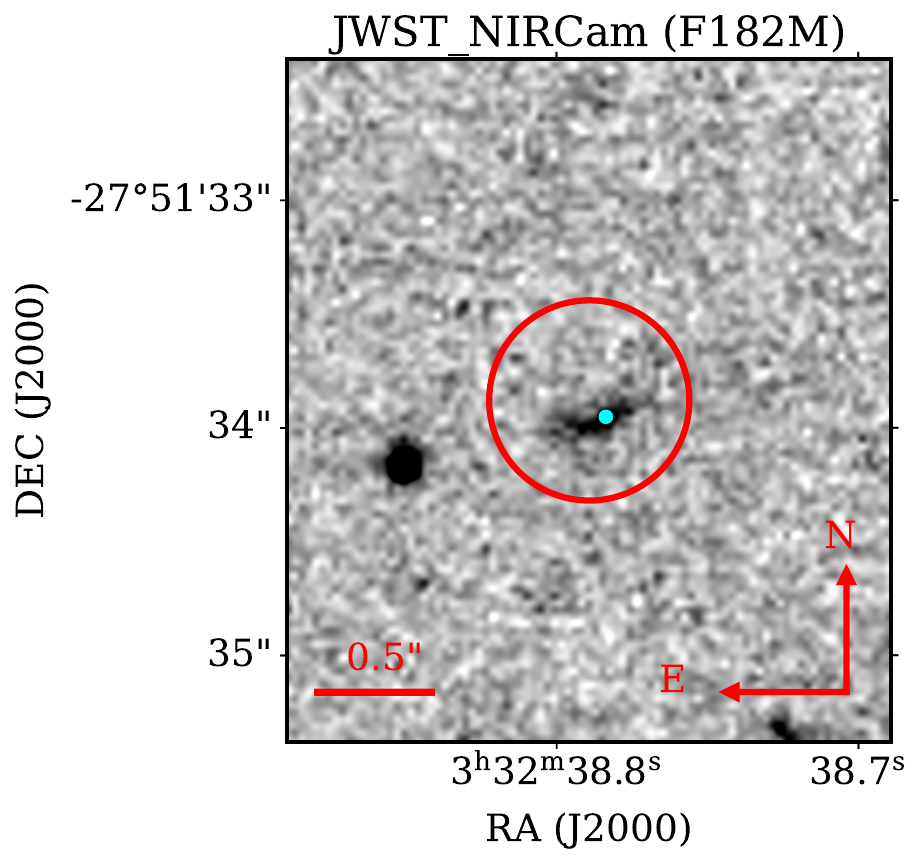}
    \includegraphics[scale=0.29]{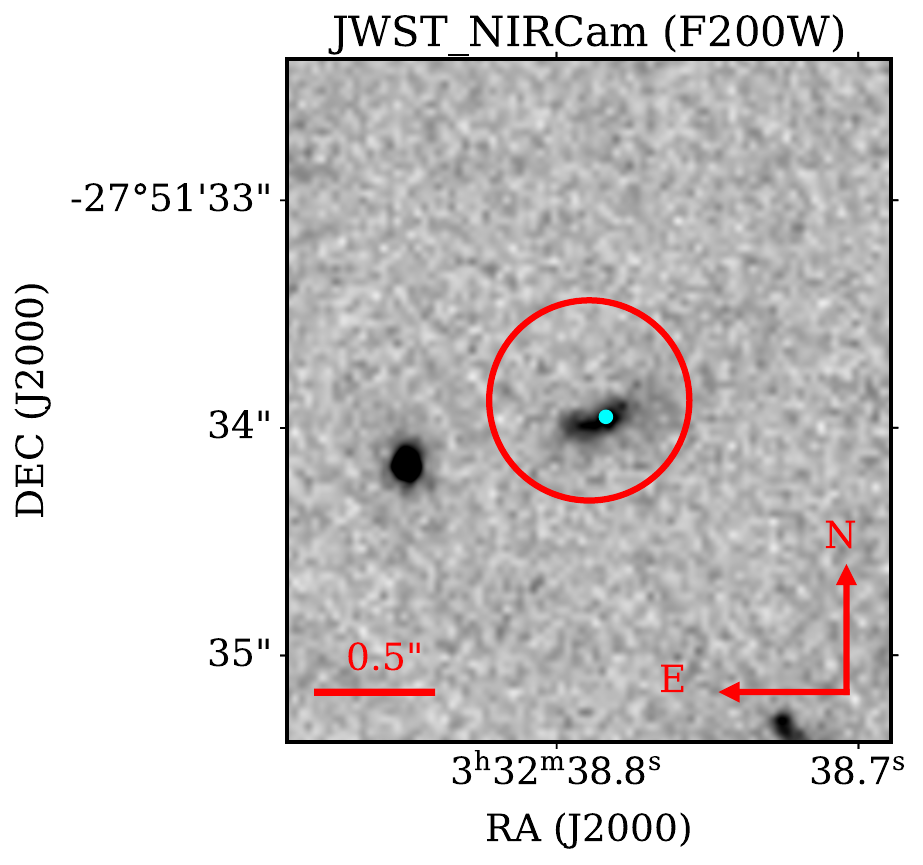}
    \includegraphics[scale=0.29]{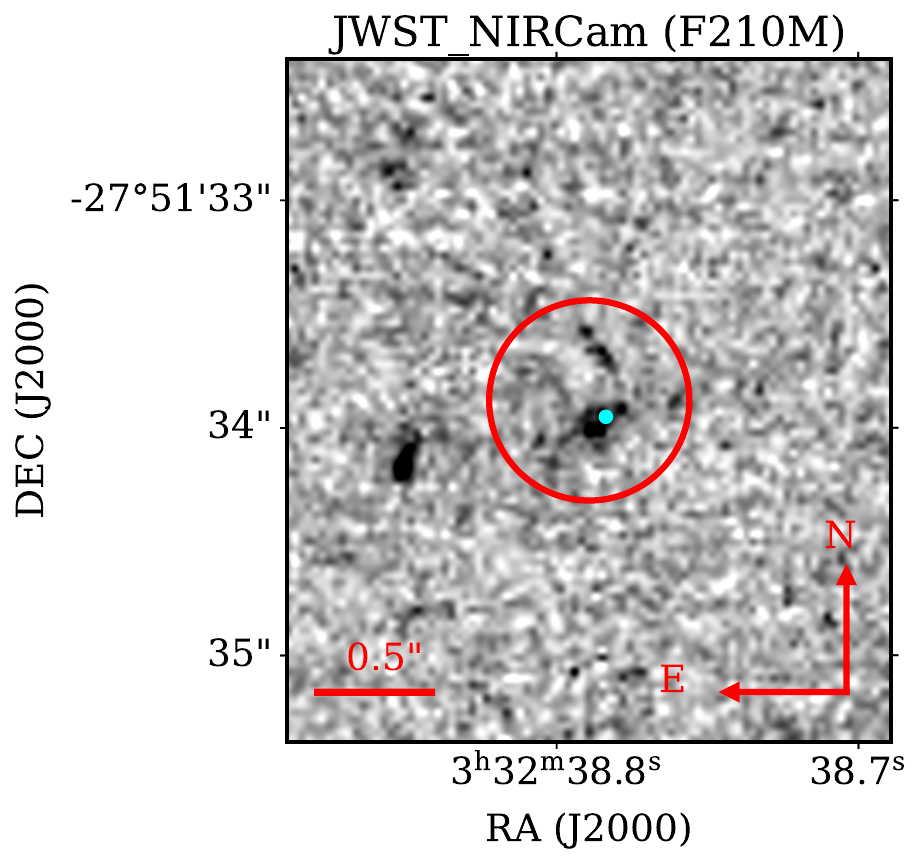}
    \includegraphics[scale=0.29]{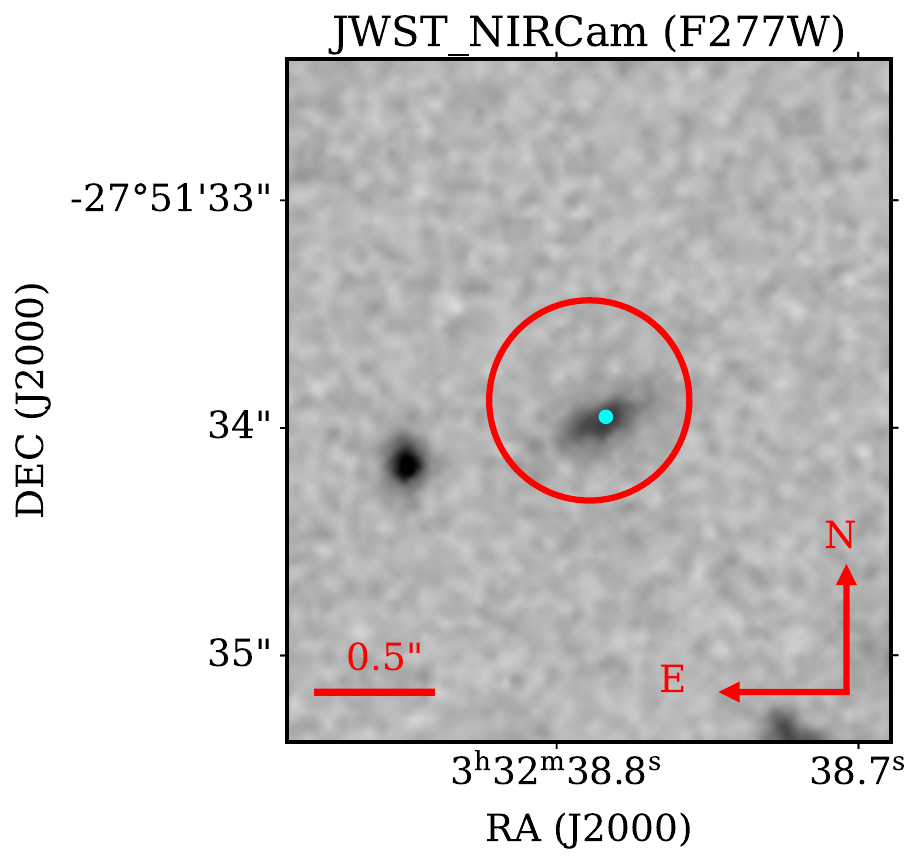}
    \includegraphics[scale=0.29]{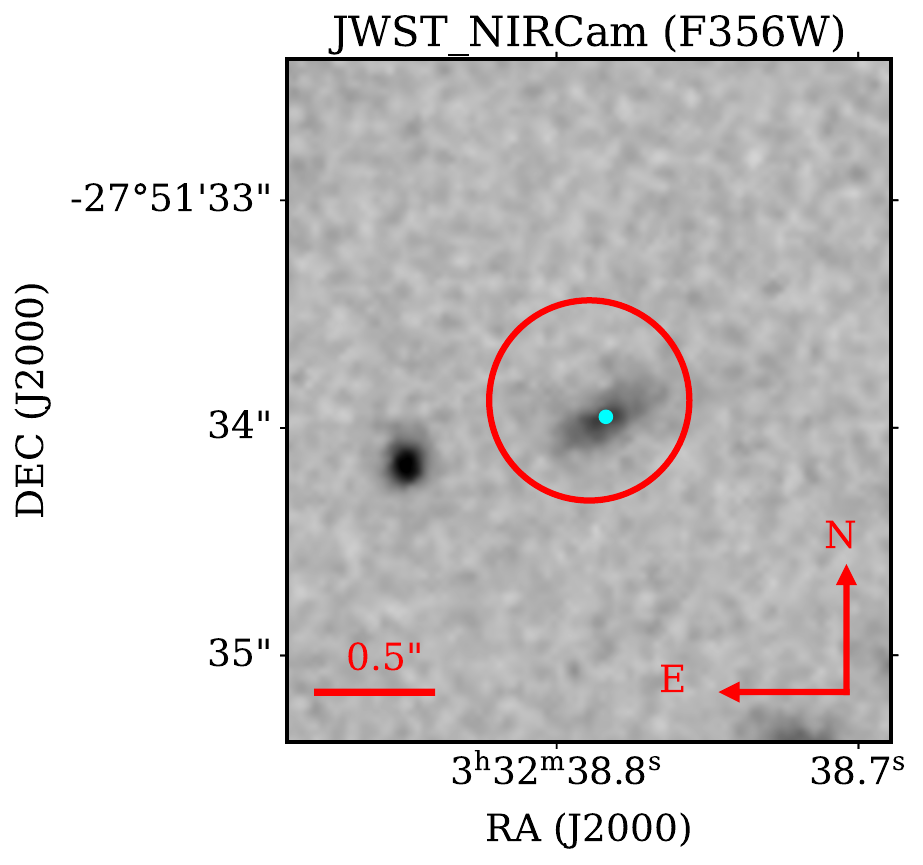}
    \includegraphics[scale=0.29]{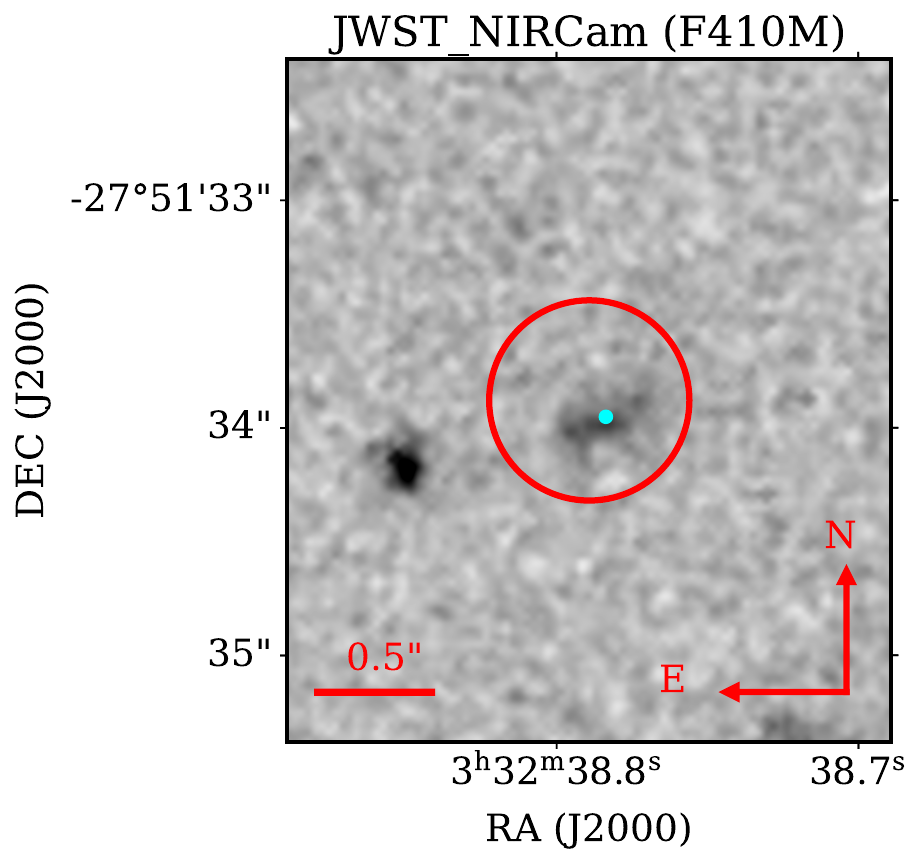}
    \includegraphics[scale=0.29]{XT1_F444W.pdf}
    \caption{The complete version of Fig.~\ref{fig:imaging_XRT_141001}. HST and JWST imaging of CDF-S XT1 at different filters (one per panel). The red circle depicts the 2$\sigma$ X-ray position uncertainty of CDF-S XT1, while the cyan dot shows the position of the host according to JADES.
    }
    \label{fig:imaging_XRT_141001_AP}
\end{figure*}

\begin{figure*}[h!]
    \centering
    \includegraphics[scale=0.29]{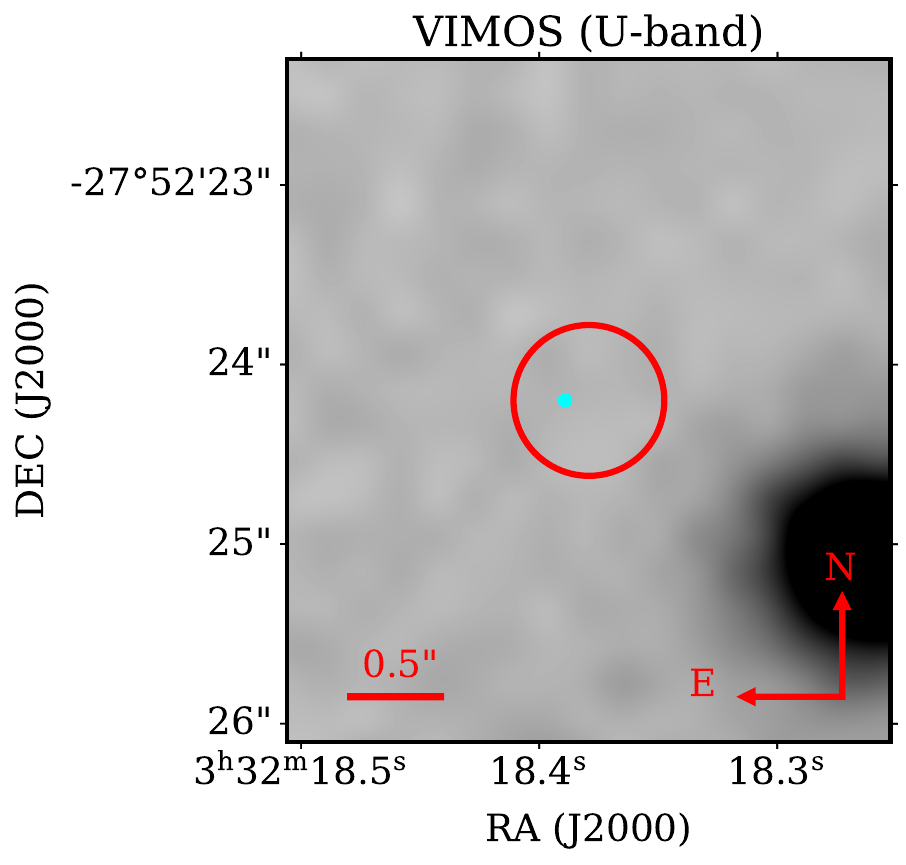}
    \includegraphics[scale=0.29]{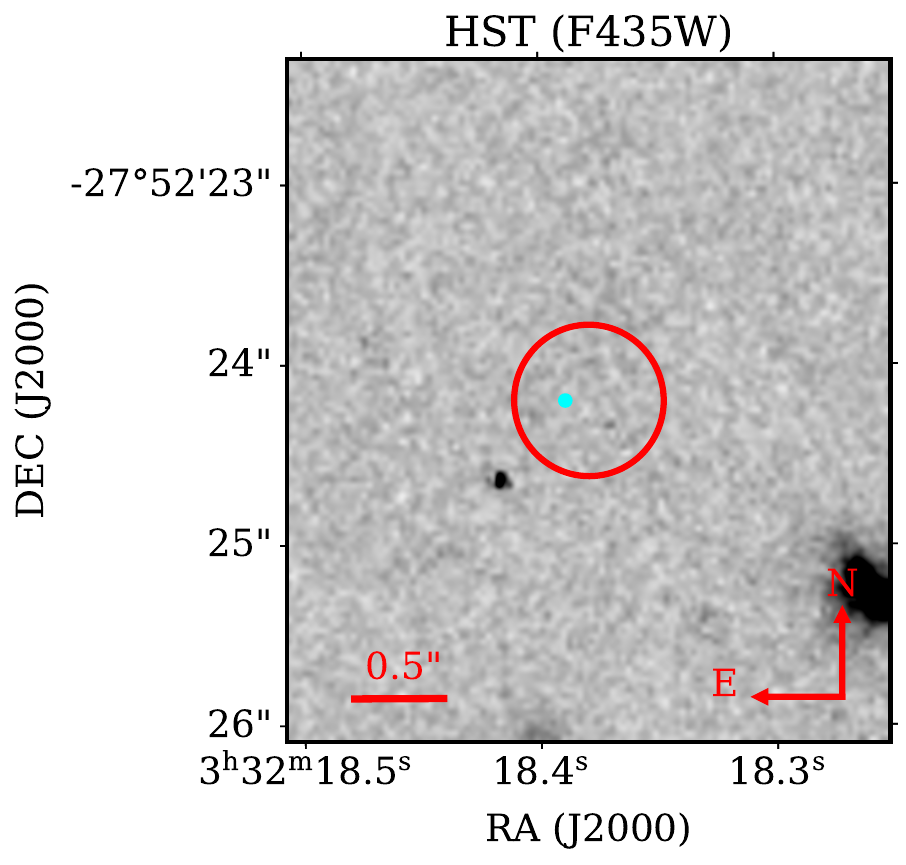}
    \includegraphics[scale=0.29]{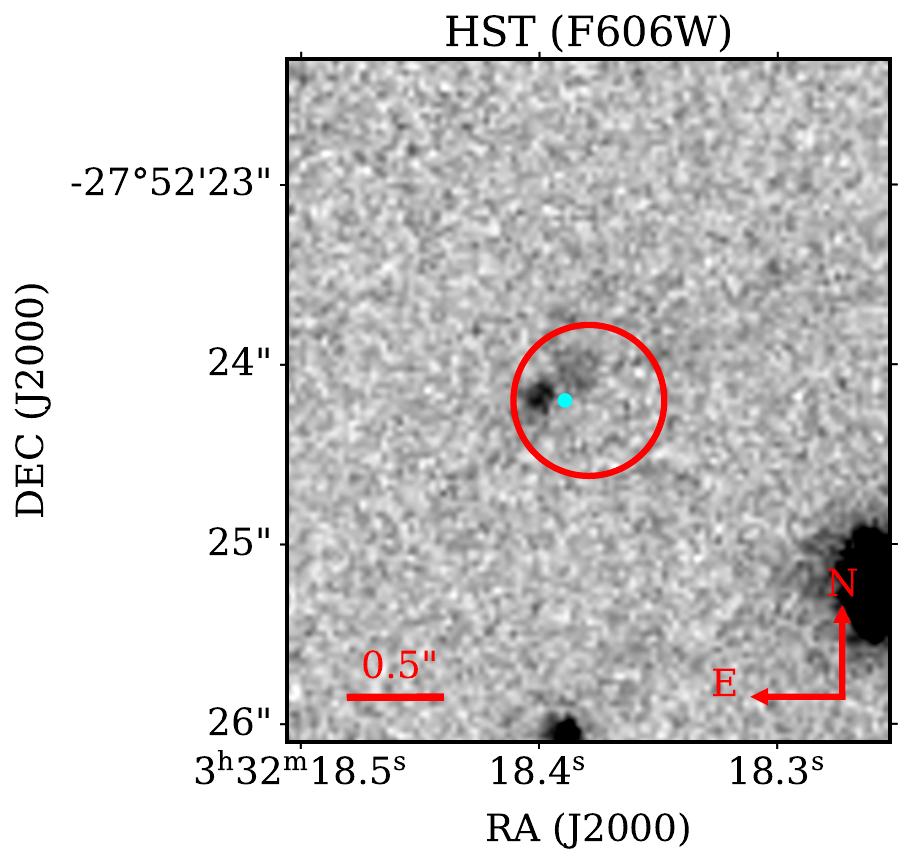}
    \includegraphics[scale=0.29]{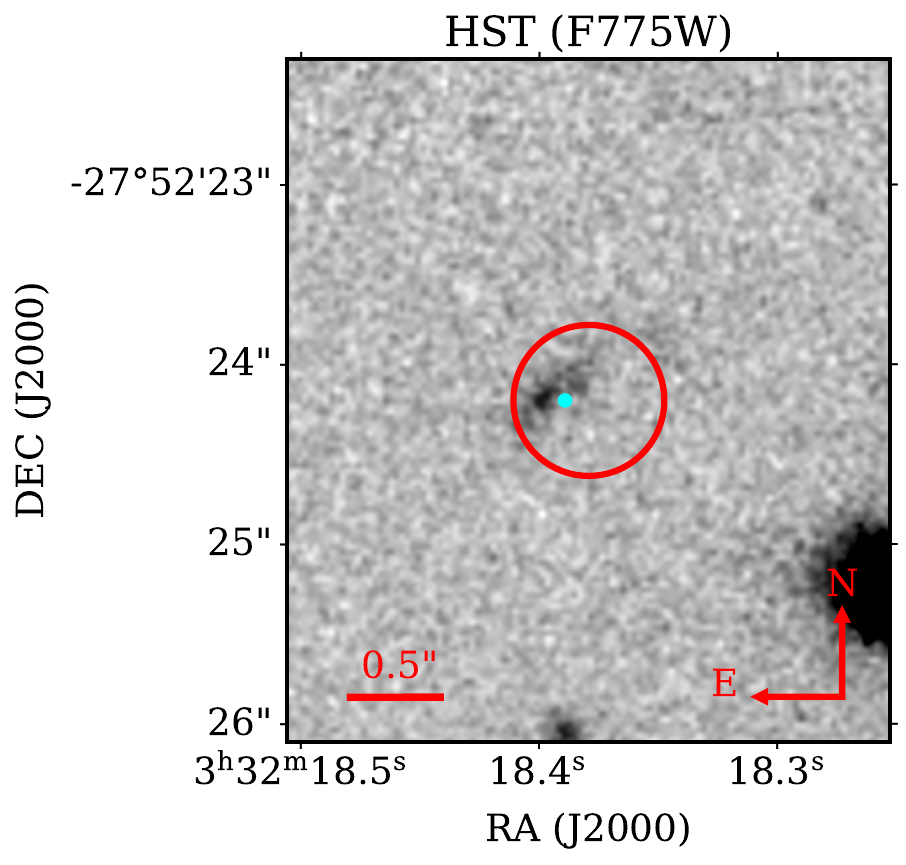}
    \includegraphics[scale=0.29]{XT2_F814W.pdf}
    \includegraphics[scale=0.29]{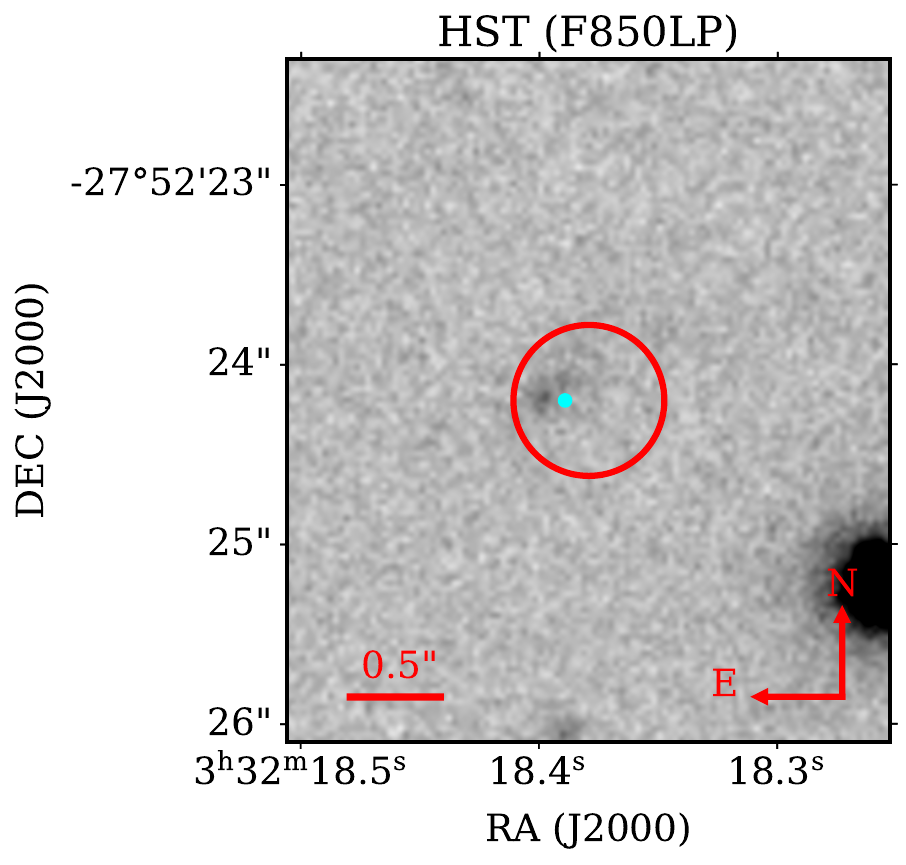}
    \includegraphics[scale=0.29]{XT2_F090W.pdf}
    \includegraphics[scale=0.29]{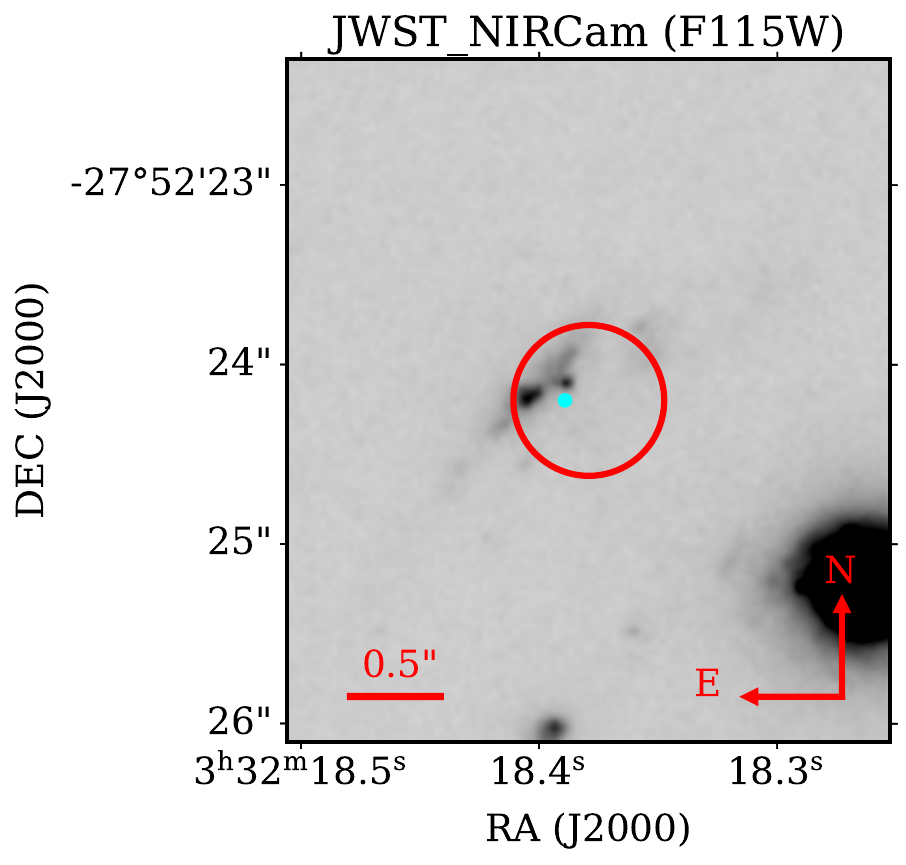}
    \includegraphics[scale=0.29]{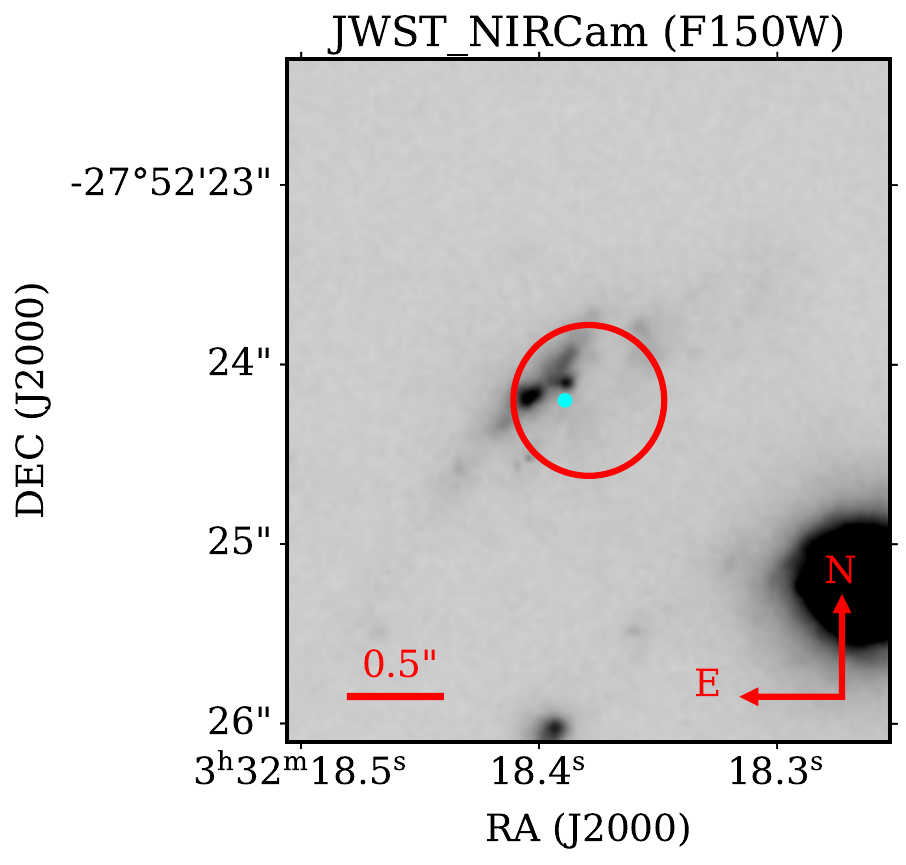}
    \includegraphics[scale=0.29]{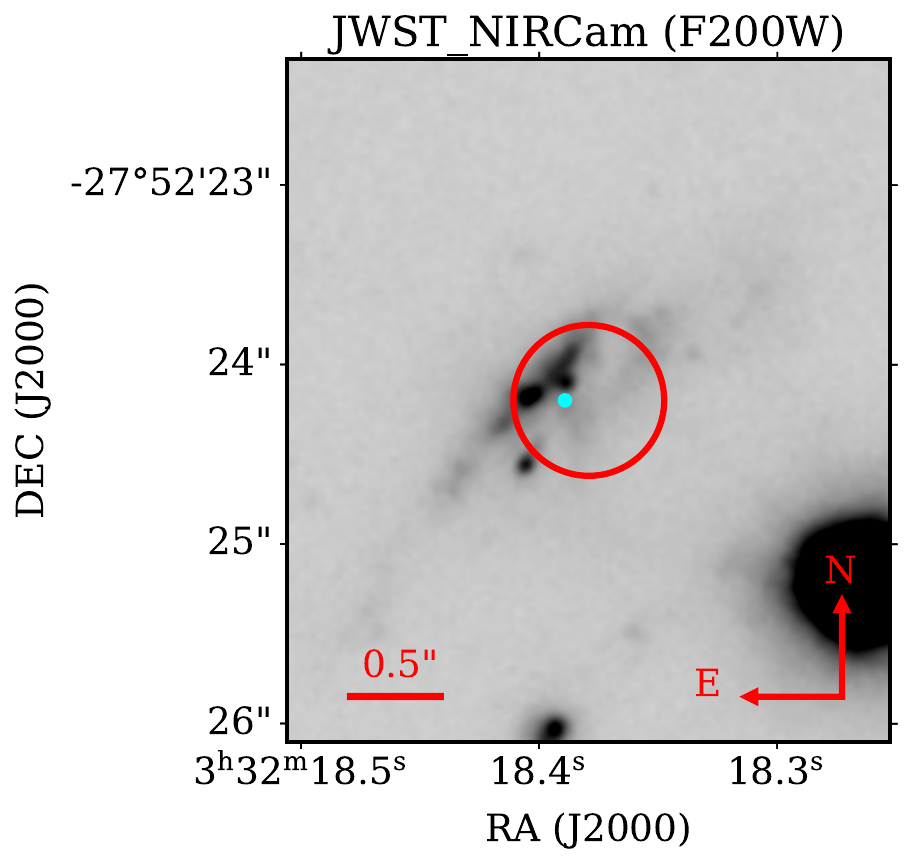}
    \includegraphics[scale=0.29]{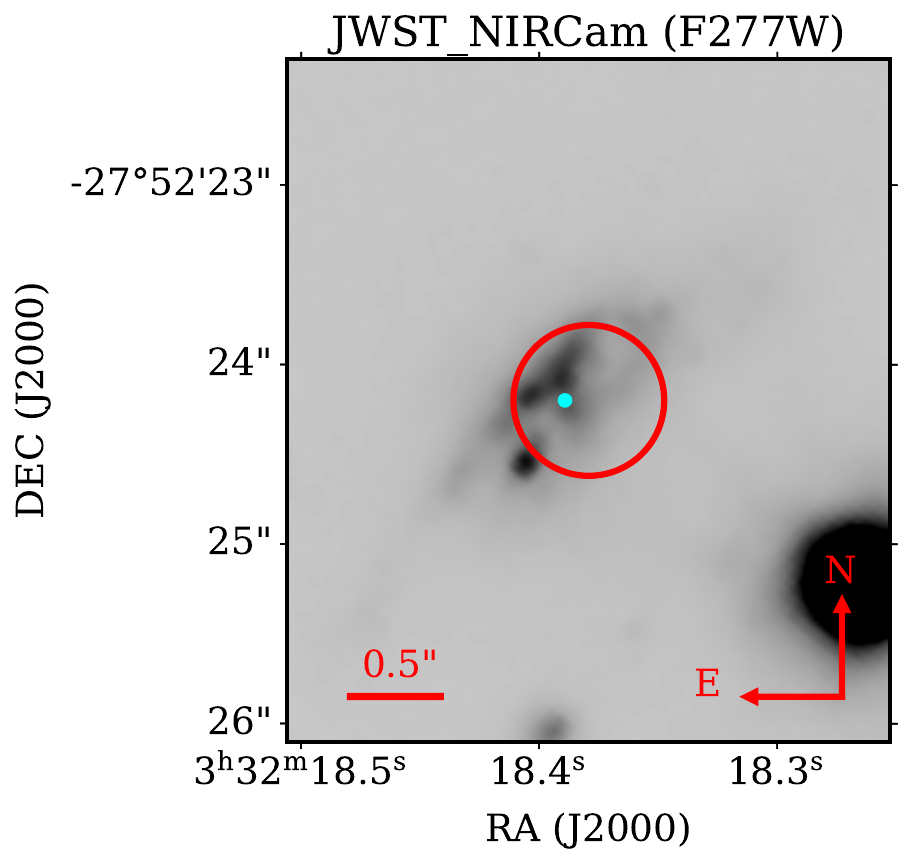}
    \includegraphics[scale=0.29]{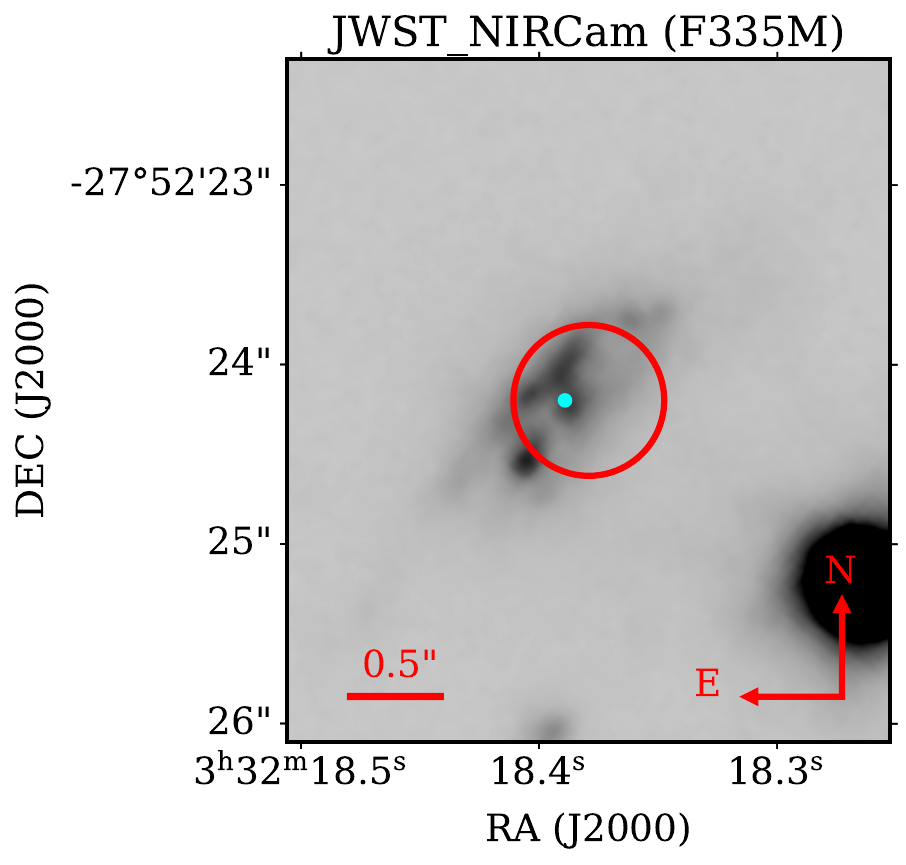}
    \includegraphics[scale=0.29]{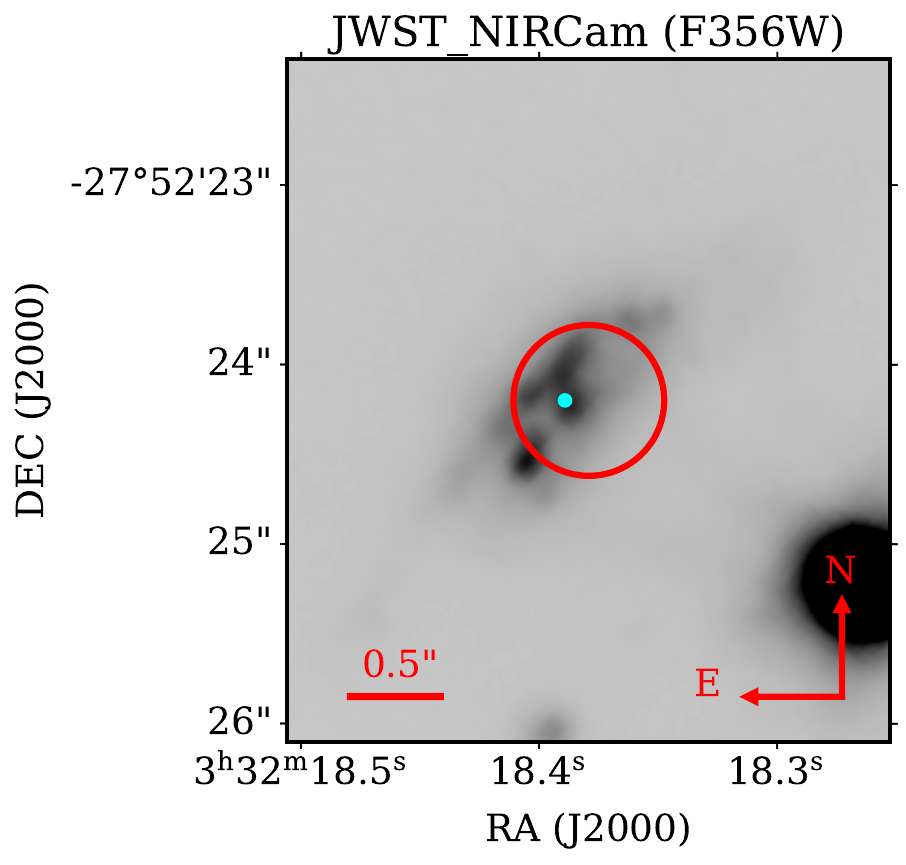}
    \includegraphics[scale=0.29]{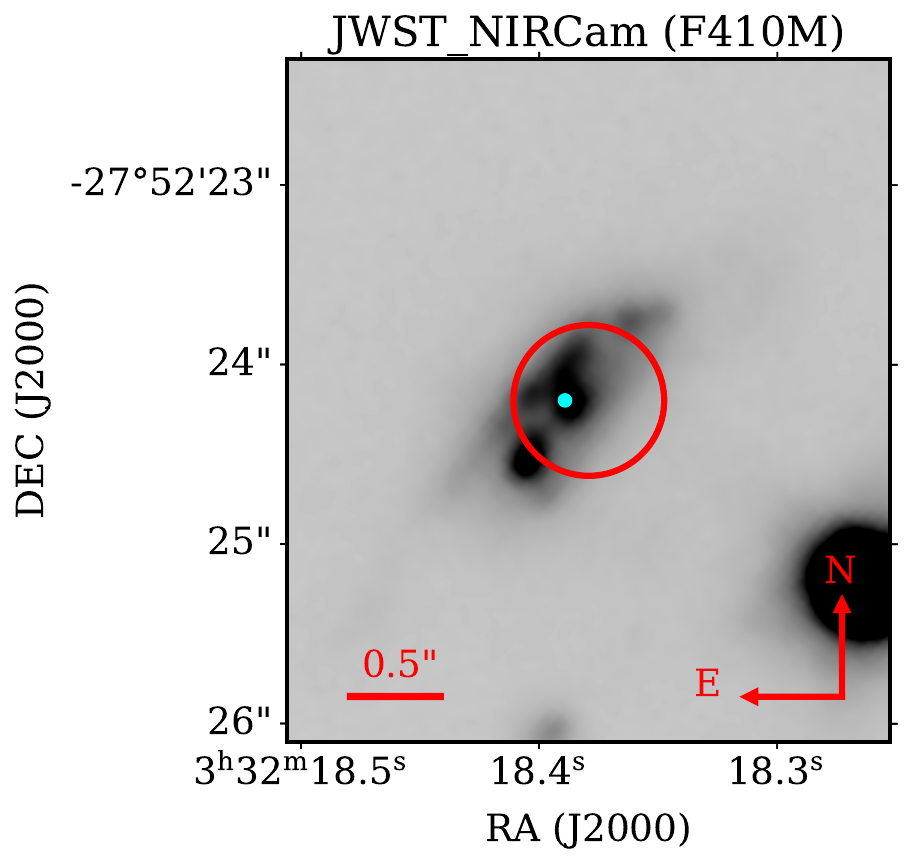}
    \includegraphics[scale=0.29]{XT2_F444W.pdf}
    \includegraphics[scale=0.29]{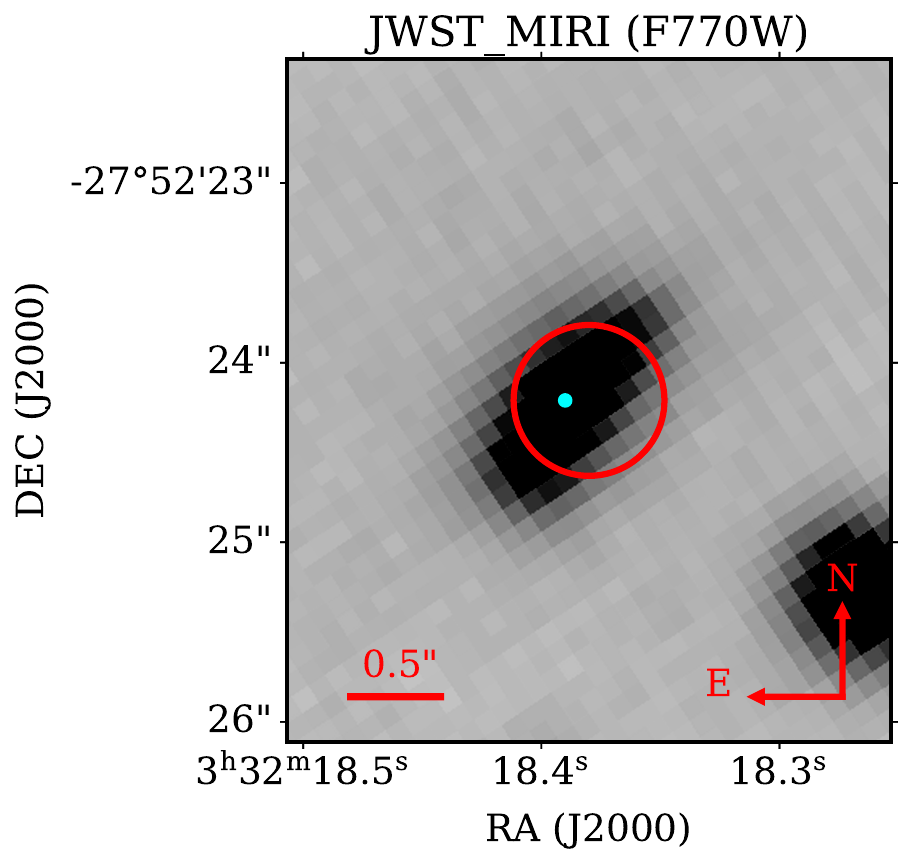}
    \includegraphics[scale=0.29]{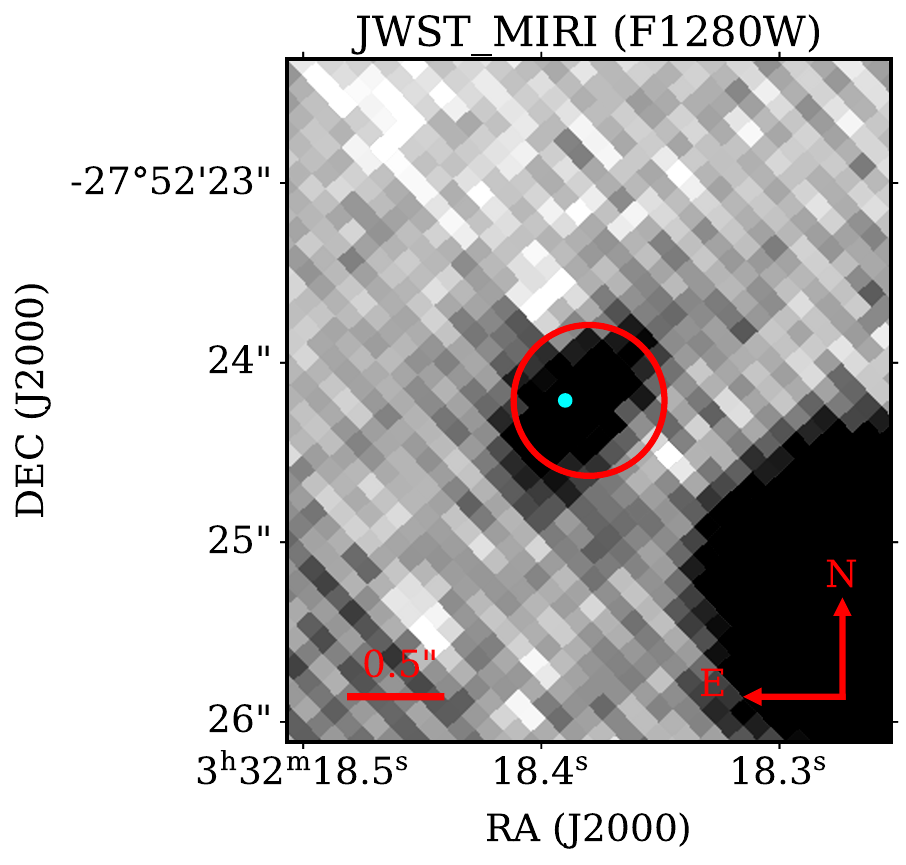}
    \includegraphics[scale=0.29]{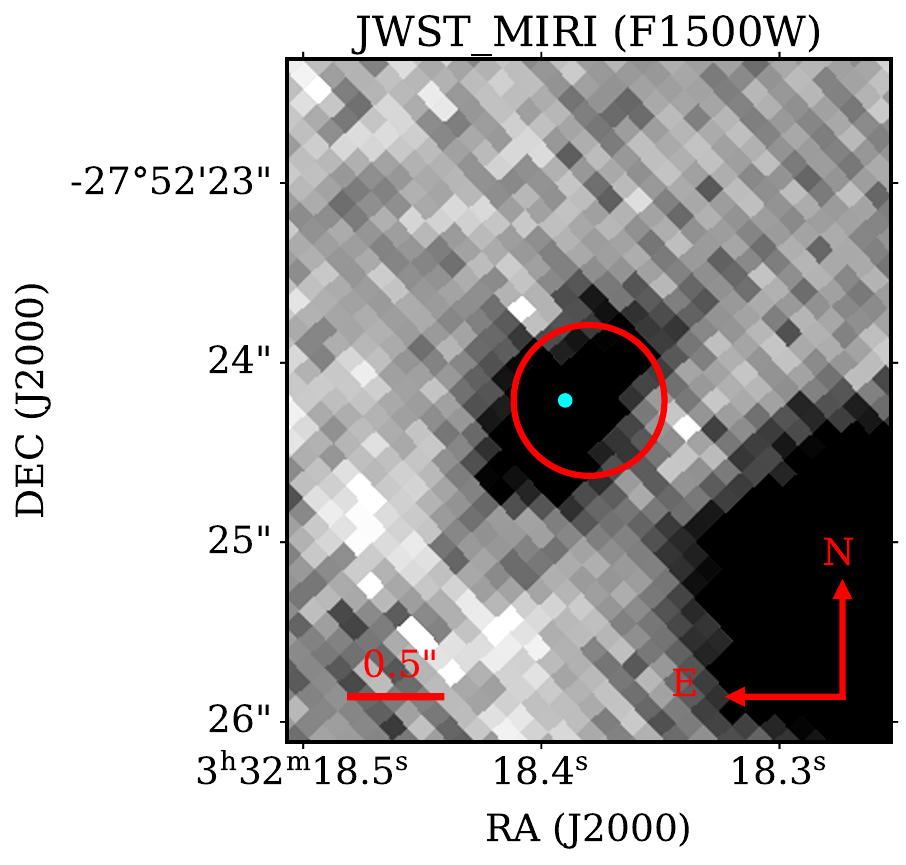}
    \caption{The complete version of Fig.~\ref{fig:imaging_XRT_150322}. HST and JWST imaging of CDF-S XT2 at different filters (one per panel). The red circle depicts the 2$\sigma$ X-ray position uncertainty of CDF-S XT2, while the cyan dot shows the position of the host according to JADES.}
    \label{fig:imaging_XRT_150322_AP}
\end{figure*}

\begin{figure}
    \centering
    \includegraphics[scale=0.75]{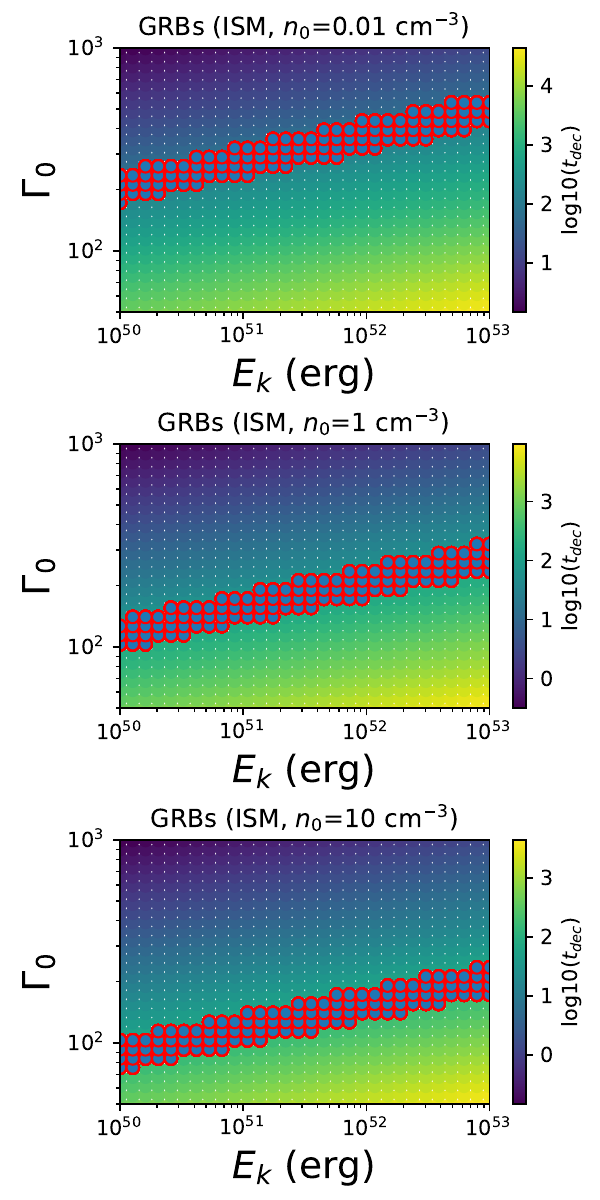}
    \caption{Parameter space exploration between the initial Lorentz factor, kinetic energy, and deceleration time (color coded) under the early afterglow scenario for three distinct uniform-density ISM regimes (from top to bottom: $10^{-2}$, $1$ and $10$~cm$^{-3}$. As explained in \S\ref{sec:deceleration}, to explore this possibility, we assume a plausible range of parameters for GRBs using Eq.~\ref{eq:003}. 
    The red markers represent the condition $t_{\rm dec}\approx t_{\rm X,peak}=110\pm50$~sec.}
    \label{fig:early_aft}
\end{figure}

We consider whether CDF-S XT1 could be related to an on-axis early afterglow, supported by the shape of its X-ray light curve, which matches some model predictions \citep[e.g.,][]{Sari1999}. Under this scenario, the early afterglow should come from the deceleration of the jet emission by the interaction with the circumburst medium. Furthermore, we assume that the X-ray emission does not have contamination from GRB prompt emission or flaring episodes from any central engine \citep{Ghirlanda2018}. The deceleration time, considering the fireball model and a constant-density hydrogen medium, is defined as \citep{Sari1999,Ghirlanda2018}:
\begin{equation}
    t_{\rm dec}(\rm ISM)\simeq185\left(\frac{E_{\rm k}}{10^{52}}\right)^{1/3}\left(\frac{\Gamma_0}{100}\right)^{-8/3}n_0^{-1/3}(1+z)\ {\rm sec.},
    \label{eq:003}
\end{equation}
where $E_{\rm k}$, $\Gamma_0$, $n_0$, and $z$ are the kinetic energy, initial Lorentz factor, number density of the ISM, and redshift, respectively. 
Here, the parameter space covered by LGRBs and SGRBs is $E_{\rm k}\sim10^{50}-10^{53}$~erg \citep{Piran2004,Fong2015,Berger2014}, $\Gamma_0\sim50-1000$ \citep{Racusin2011}, and an ISM number density of $n\sim10^{-2}-10$~cm$^{-3}$ \citep{Schulze2011,Berger2014,Salafia2022}. Figure~\ref{fig:early_aft} shows the color-coded deceleration time as a function of the initial Lorentz factor and kinetic energy, assuming three different ISM densities.

\section{redshift discrepancy for CDF-S XT2}\label{sec:redshift_discre}

\begin{figure*}
    \centering
    \includegraphics[scale=0.29]{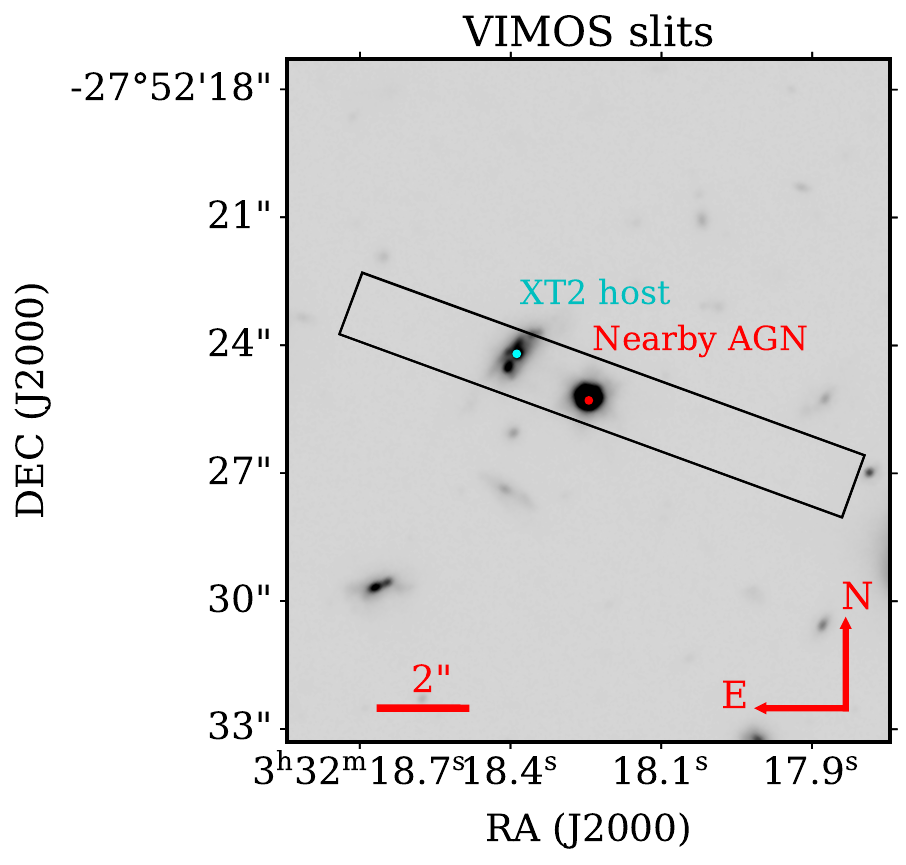}
    \includegraphics[scale=0.29]{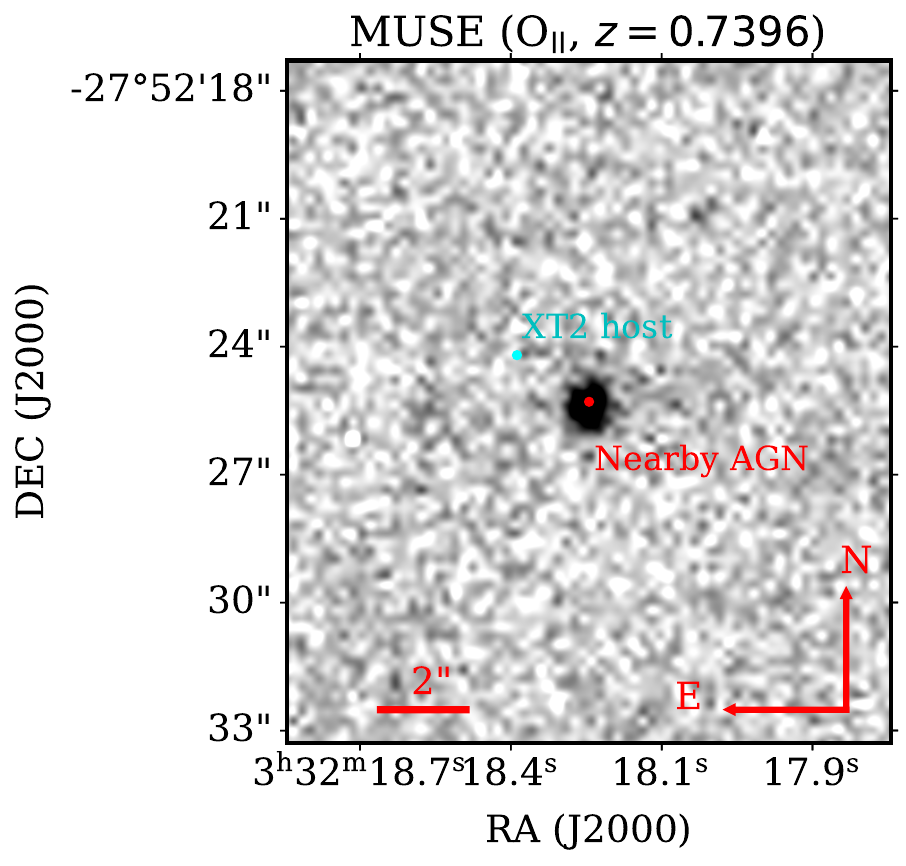}
    \includegraphics[scale=0.29]{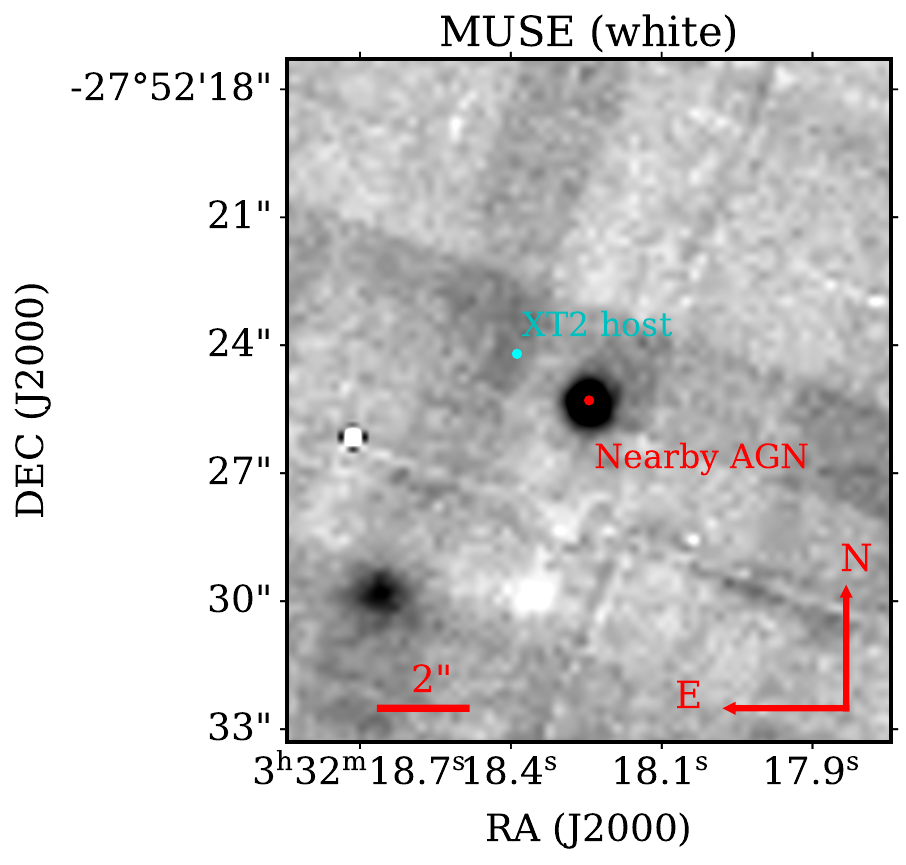}
    \includegraphics[scale=0.29]{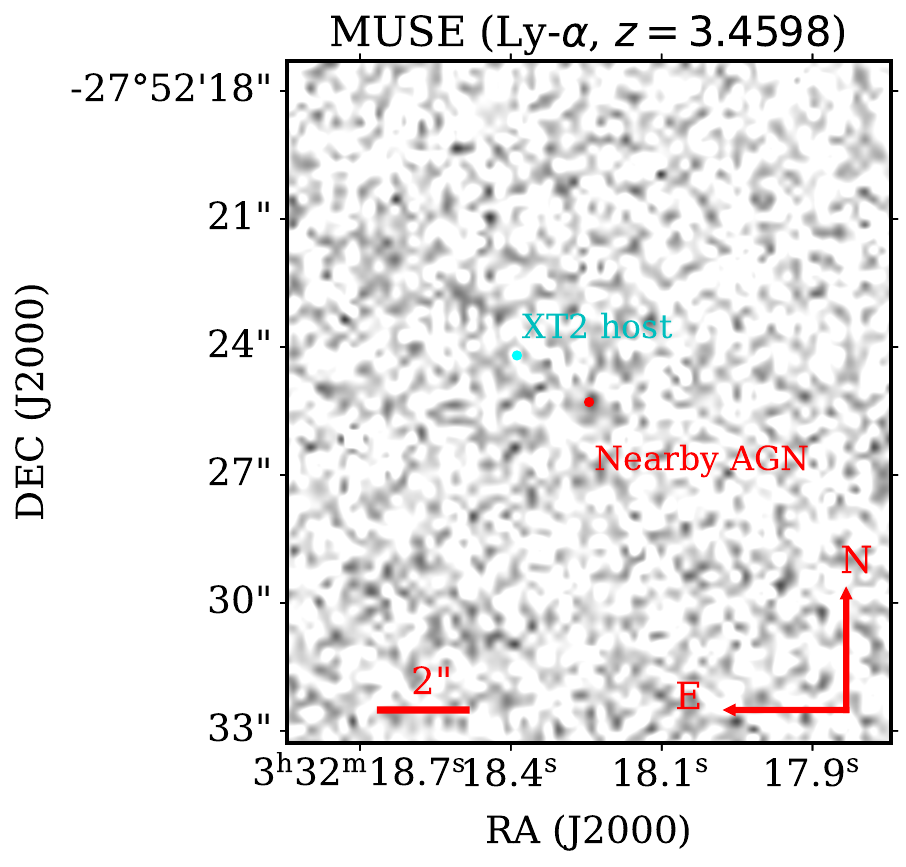}
    \caption{VIMOS slit-position and MUSE multiwavelength images of the field of CDF-S XT2. Left panel: JWST-F444W imaging superimposed with the VLT-VIMOS slits (black box) used by \citet{Balestra2010} for determining the redshift of XT2 and the nearby AGN CANDELS \#4210. 
    Left-center panel: MUSE collapsed imaging at [O{\sc ii}] emission line (i.e., $\lambda_{\rm cen}{=}6478$~\AA, width 10~\AA) of the field of XT2, where according to \citet{Balestra2010} the line [O{\sc ii}] of XT2's is detected by VLT-VIMOS. The AGN CANDELS \#4210 is clearly detected (consistent with a [O{\sc ii}] line at $z = 0.7396$). 
    Right-center panel: MUSE white-light image showing the position of the XT2 host and AGN CANDELS \#4210. The cyan and red dots depict the position of the XT2 host and AGN CANDELS \#4210, respectively.
    Right panel: MUSE collapsed imaging at the Ly-$\alpha$ emission line (i.e., $\lambda_{\rm cen}{=}1216$~\AA, width 10~\AA) of the field of XT2 at redshift $z_{\rm spec}{=}3.4598$. At these wavelengths, the XT2 host is not observed. The cyan and red dots depict the position of the XT2 host and AGN CANDELS \#4210, respectively.
    }
    \label{fig:slit_image}
\end{figure*}

\begin{figure*}
    \centering
    \includegraphics[scale=0.6]{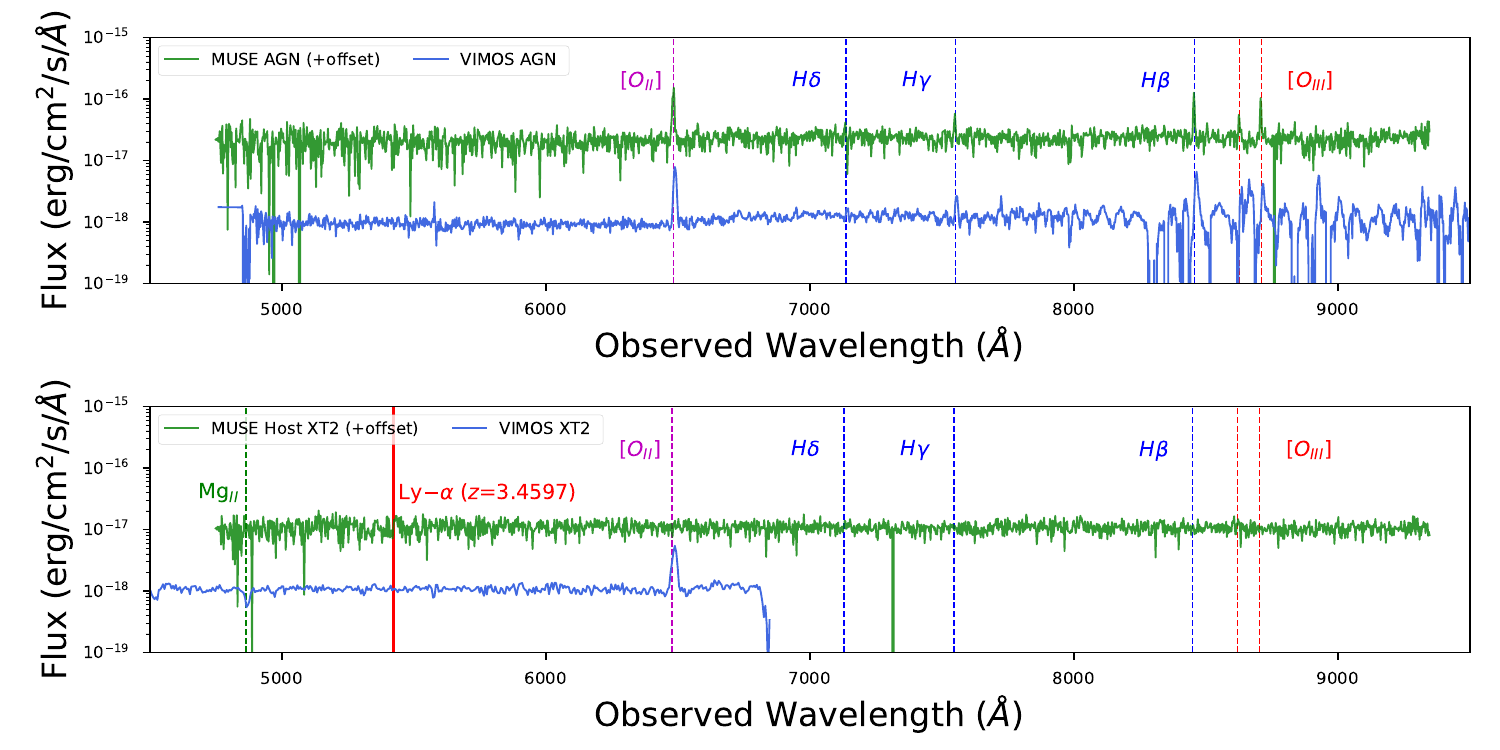}
    \caption{VLT/VIMOS (blue lines) and MUSE (green lines) spectra of the nearby AGN CANDELS \#4210 (top panel) and the host of XT2 (bottom panel) obtained by \citet{Balestra2010}. Moreover, both spectra show the most important emission lines (vertical dashed lines) used by \citet{Balestra2010} to determine the spectroscopic redshift of the AGN CANDELS \#4210 ($z=0.7396$; [O{\sc ii}], $H\beta$ and [O{\sc iii}]) and XT2's host ($z=0.7382$; [O{\sc ii}], and Mg{\sc ii}). On the other hand, the MUSE spectrum of the nearby AGN CANDELS \#4210 shows the [O{\sc ii}], $H\beta$ and [O{\sc iii}] emission lines, confirming the reported redshift of \citet{Balestra2010} ($z=0.7396$). However, the subtracted spectrum at the position of the XT2 host does not display the [O{\sc ii}] and Mg{\sc ii} lines reported by \citet{Balestra2010}. Finally, the MUSE spectrum of the XT2 host does not show the expected Lyman$\alpha$ line at redshift $z_{\rm spec}{=}3.4598$ (vertical red solid line).}
    \label{fig:old_spec_comp}
\end{figure*}

The redshift $z{=}0.7382$ reported by \citet{Balestra2010} for the host of CDF-S XT2 is clearly in conflict with the new redshift from {\it JWST}, shown in Fig.~\ref{fig:Spec_XRT_150322}. Here, we investigate the veracity of the previous redshift, which was regarded as of high quality. We began by examining the VLT-VIMOS (three exposures taken between 04:03:10.24 and 04:10:12.50 on 2004-11-17; PI Cesarsky) slit position and slit images to ascertain how the data reported by \citet{Balestra2010} were taken.\footnote{The raw slit imaging data is available in the ESO Observational Raw Data Archive: http://archive.eso.org/eso/eso\_archive\_main.html} Notably, the slits in the three images are all  1\farcs0$\times$ 12\farcs0, centered on an elliptical galaxy (CANDELS \#4210) hosting an X-ray AGN at $z{=}0.7396$ (see Fig.~\ref{fig:slit_image}, left panel). This galaxy lies $\approx$2$"$ to the southwest of XT2's identified host, which also falls within the slit, and was extracted as a secondary target by \citet{Balestra2010}. Given the proximity of the two objects, there is potential for contamination. The VLT-VIMOS spectra reported by \citet{Balestra2010} of the CANDELS \#4210 and host of CDF-S XT2 are shown in Fig.~\ref{fig:old_spec_comp}, as well as the most important emission lines used to measure their redshifts.\footnote{The VIMOS/VLT data are bias subtracted, flat-field corrected, wavelength calibrated, sky subtracted, and flux calibrated, and made available in the ESO Science Portal: http://archive.eso.org/scienceportal/home}

To concretely confirm the contamination scenario, we examine this location with the deep MUSE dataset (Wisotzki PI) from \citet{Inami2017} and \citet{Herenz2017}. We obtained the reduced (ESO MUSE pipeline v2.8.6) and calibrated MUSE data from the ESO archive\footnote{This data cube is bias subtracted, flat-field corrected, wavelength calibrated, sky subtracted, flux calibrated, stacked, and made available in the ESO Science Portal.}.
Moreover, we use the Zurich Atmosphere Purge \citep[ZAPv2.1;][]{Soto2016} on the data cubes, which is a high precision subtraction tool to improve the already performed sky subtraction\footnote{ZAP uses principle component analysis combined with filtering to construct a sky residual for each spaxel which is subtracted from the original data cube.}.
We improved the astrometry of the MUSE data cube by detecting sources in the collapsed white-light image and aligning them with matched objects in the CANDELS catalog \citep{Guo2013}. Using \texttt{PyMUSE} \citep{Pessa2018}, we then created a narrow-band image centered at $\lambda_{\rm cen}{=}6478\pm5$\AA~ to probe the rest-frame [O{\sc ii}] emission line at $z{=}0.738-0.740$ (Figure~\ref{fig:slit_image}) and a white-light image (Figure~\ref{fig:slit_image}). CANDELS \#4210 is robustly detected in both, while the host of XT2 is not visible in either. The absence of an emission line at the position of XT2's host supports the contamination scenario. Finally, we extracted MUSE spectra at the positions of CANDELS \#4210 and the host of CDF-S XT2 using the file viewer package \texttt{QFitsView} \citep{Ott2012} considering a circular aperture of 5~pixels, subtracting a background spectrum extracted in a region free of known sources in the JWST image. Figure~\ref{fig:old_spec_comp} depicts these spectra (shown in green), where the [O{\sc ii}] emission line $\lambda_{\rm cen}{=}6478$\AA~reported by \citet{Balestra2010} is not found, despite the MUSE data having sufficient sensitivity to detect it at moderate significance. Thus, we conclude that the reported redshift of $z{=}0.7382$ is not supported, confirming the result obtained by the JWST spectra.


\begin{table}
    \centering
    \caption{\emph{HST}-ACS/WFC3 and \emph{JWST}-NIRCam photometric data of the FXT XT1.}
    \scalebox{0.8}{
    \begin{tabular}{ccccc}
    \hline\hline
    Telescope & Instrument & Filter & AB mag. & Flux ($\mu$Jy)  \\ 
    (1) & (2) & (3) & (4) & (5) \\ \hline
    \emph{Blanco} & MOSAIC II & $U$ & $27.60{\pm}0.71$ & $0.033{\pm}0.022$ \\
    \emph{VLT} & VIMOS & $U$ & $28.81{\pm}0.45$ & $0.011{\pm}0.005$ \\
    \emph{HST} & ACS &  F435W & $ 28.51 {\pm} 0.58 $ & $ 0.014 {\pm} 0.008 $ \\
     \emph{HST} & ACS &  F606W & $ 27.67 {\pm} 0.2 $ & $ 0.031 {\pm} 0.006 $ \\
     \emph{HST} & ACS &  F775W & $ 27.35 {\pm} 0.3 $ & $ 0.042 {\pm} 0.012 $ \\
     \emph{HST} & ACS &  F814W & $ 27.71 {\pm} 0.18 $ & $ 0.030 {\pm} 0.005 $ \\
     \emph{HST} & ACS &  F850LP & $ 27.97 {\pm} 0.71 $ & $ 0.024 {\pm} 0.015 $ \\
     \emph{HST} & WFC3 &  F125W & $ 27.05 {\pm} 0.3 $ & $ 0.055 {\pm} 0.015 $ \\
     \emph{HST} & WFC3 &  F140W & $ 26.27 {\pm} 0.39 $ & $ 0.113 {\pm} 0.041 $ \\
     \emph{HST} & WFC3 &  F160W & $ 26.73 {\pm} 0.3 $ & $ 0.074 {\pm} 0.021 $ \\ 
\hline
     \emph{JWST} & \emph{NIRCam} &  F090W & $ 27.51 {\pm} 0.17 $ & $ 0.036 {\pm} 0.006 $ \\
     \emph{JWST} & \emph{NIRCam} &  F115W & $ 27.31 {\pm} 0.14 $ & $ 0.043 {\pm} 0.005 $ \\
     \emph{JWST} & \emph{NIRCam} &  F150W & $ 26.94 {\pm} 0.09 $ & $ 0.061 {\pm} 0.005 $ \\
     \emph{JWST} & \emph{NIRCam} &  F182M & $ 26.58 {\pm} 0.11 $ & $ 0.085 {\pm} 0.008 $ \\
     \emph{JWST} & \emph{NIRCam} &  F200W & $ 26.69 {\pm} 0.06 $ & $ 0.076 {\pm} 0.004 $ \\
     \emph{JWST} & \emph{NIRCam} &  F210M & $ 26.63 {\pm} 0.14 $ & $ 0.081 {\pm} 0.010 $ \\
     \emph{JWST} & \emph{NIRCam} &  F277W & $ 26.84 {\pm} 0.07 $ & $ 0.067 {\pm} 0.005 $ \\
     \emph{JWST} & \emph{NIRCam} &  F356W & $ 26.82 {\pm} 0.06 $ & $ 0.068 {\pm} 0.004 $ \\
     \emph{JWST} & \emph{NIRCam} &  F410M & $ 26.81 {\pm} 0.09 $ & $ 0.068 {\pm} 0.006 $ \\
     \emph{JWST} & \emph{NIRCam} &  F444W & $ 26.73 {\pm} 0.06 $ & $ 0.074 {\pm} 0.004 $ \\
    \hline\hline
    \end{tabular}}
    \tablefoot{The HST and JWST data were taken from \citet{Eisenstein2023b}, while the VLT and Blanco data were taken from \citet{Guo2013}.}
    \label{tab:XT1_photometry}
\end{table}

\begin{table}
    \centering
    \caption{\emph{HST}-ACS/WFC3, \emph{JWST}-NIRCam and \emph{JWST}-MIRI photometric data of the FXT XT2.}
    \scalebox{0.8}{
    \begin{tabular}{ccccc}
    \hline\hline
    Telescope & Instrument & Filter & AB mag. & Flux ($\mu$Jy)  \\ 
     (1) & (2) & (3) & (4) & (5) \\ \hline
     \emph{Blanco} & MOSAIC II & $U$ & ${>}26.63$ & ${>}0.081$ \\
     \emph{VLT} & VIMOS & $U$ & ${>27.97}$ & ${>}0.024$ \\
     \emph{HST} & ACS &  F435W & $26.58{\pm}0.34$ & $0.085{\pm}0.027$ \\
     \emph{HST} & ACS &  F606W & $ 25.64 {\pm} 0.21 $ & $ 0.201 {\pm} 0.039 $ \\
     \emph{HST} & ACS &  F775W & $ 24.99 {\pm} 0.24 $ & $ 0.366 {\pm} 0.079 $ \\
     \emph{HST} & ACS &  F814W & $ 24.97 {\pm} 0.15 $ & $ 0.372 {\pm} 0.050 $ \\
     \emph{HST} & ACS &  F850LP & $ 25.21 {\pm} 0.35 $ & $ 0.301 {\pm} 0.096 $ \\
     \emph{HST} & WFC3 &  F105W & $ 24.99 {\pm} 0.37 $ & $ 0.367 {\pm} 0.126 $ \\
     \emph{HST} & WFC3 &  F125W & $ 24.38 {\pm} 0.17 $ & $ 0.642 {\pm} 0.102 $ \\
     \emph{HST} & WFC3 &  F140W & $ 24.4 {\pm} 0.42 $ & $ 0.632 {\pm} 0.246 $ \\
     \emph{HST} & WFC3 &  F160W & $ 24.0 {\pm} 0.14 $ & $ 0.910 {\pm} 0.119 $ \\ \hline
     \emph{JWST} & \emph{NIRCam} &  F090W & $ 24.85 {\pm} 0.03 $ & $ 0.418 {\pm} 0.012 $ \\
     \emph{JWST} & \emph{NIRCam} &  F115W & $ 24.51 {\pm} 0.02 $ & $ 0.570 {\pm} 0.012 $ \\
     \emph{JWST} & \emph{NIRCam} &  F150W & $ 24.01 {\pm} 0.01 $ & $ 0.903 {\pm} 0.011 $ \\
     \emph{JWST} & \emph{NIRCam} &  F200W & $ 23.26 {\pm} 0.01 $ & $ 1.803 {\pm} 0.011 $ \\
     \emph{JWST} & \emph{NIRCam} &  F277W & $ 22.49 {\pm} 0.01 $ & $ 3.652 {\pm} 0.009 $ \\
     \emph{JWST} & \emph{NIRCam} &  F335M & $ 22.46 {\pm} 0.01 $ & $ 3.753 {\pm} 0.013 $ \\
     \emph{JWST} & \emph{NIRCam} &  F356W & $ 22.37 {\pm} 0.01 $ & $ 4.089 {\pm} 0.007 $ \\
     \emph{JWST} & \emph{NIRCam} &  F410M & $ 22.08 {\pm} 0.01 $ & $ 5.359 {\pm} 0.010 $ \\
     \emph{JWST} & \emph{NIRCam} &  F444W & $ 21.91 {\pm} 0.0 $ & $ 6.255 {\pm} 0.009 $ \\ \hline
    \emph{JWST} & \emph{MIRI} & F770W & $21.38 {\pm} 0.01 $ & $10.194 {\pm} 0.094$ \\
    \emph{JWST} & \emph{MIRI} & F1280W & $21.65 {\pm} 0.01 $ & $7.909 {\pm} 0.073$ \\ 
    \emph{JWST} & \emph{MIRI} & F1500W & $21.38 {\pm} 0.01 $ & $10.177 {\pm} 0.094$ \\ \hline
    \hline
    \end{tabular}}
    \tablefoot{Data were taken from \citet{Eisenstein2023b}, while the VLT and Blanco data are the $5\sigma$ magnitude limit for the $U-$band photometry of \citet{Guo2013}.}
    \label{tab:XT2_photometry}
\end{table}

\begin{table}
    \centering
    \caption{Parameters considered for fitting HST+JWST photometric data of CDF-S XT1 and XT2 host galaxies.}
    \scalebox{0.8}{
    \begin{tabular}{llc}
    \hline \hline
    Parameter & Definition & Prior \\ 
    (1) & (2) & (3)\\ \hline
    $z^a$ & Redshift & $\mathcal{U}{\sim}[0.5,4.0]$  \\
    $t_{\rm age}$~(Gyr) & Age of the galaxy at the time of observation & $\mathcal{U}{\sim}[0.1,15.0]$ \\
    $\tau$~(Gyr) & e-folding time of exponential SFH & $\mathcal{U}{\sim}[0.1,10.0]$ \\
    $\log(M_F/M_\odot)$ & Total mass formed & $\mathcal{U}{\sim}[6.0,13.0]$ \\
    $\log(Z/Z_\odot)$ & Stellar metallicity & $\mathcal{U}{\sim}[-1.0,0.5]$ \\
    $\log(U)$ & Nebular component & $\mathcal{U}{\sim}[-4.0,-1.0]$ \\
    $A_{\rm V,dust}$ & Dust attenuation & $\mathcal{U}{\sim}[0.0,4.0]$ \\    
    \hline 
    \end{tabular}}
    \tablefoot{The packages used for the fitting process are \texttt{Bagpipes} \citep{Carnall2018} and \texttt{Prospector} \citep{Leja2017}. \emph{Columns 1 and 2:} SFH parameters and their definitions, respectively. \emph{Column 3:} prior distribution per parameter.\\
    $^a$ The redshift as a free parameter is valid only for CDF-S XT1, while for CDF-S XT2, its redshift is fixed to $z=3.4598$.}
    \label{tab:SED_model}
\end{table}

\end{document}